\documentclass[useAMS,usenatbib]{mn2e}
\usepackage{epsfig,rotate,graphicx}
\usepackage[fleqn]{amsmath}
\usepackage{subfigure}
\usepackage{lscape}

\newcommand{\p}{\partial}
\newcommand{\mnras}{MNRAS}
\newcommand{\apj}{ApJ}
\newcommand{\aap}{A\&A}
\newcommand{\apjl}{ApJL}

\newcommand{\adot}{\dot{a}}

\newcommand{\be}{\begin{equation}}
\newcommand{\ee}{\end{equation}}
\newcommand{\gtrsim}{\;\raisebox{-.8ex}{$\buildrel{\textstyle>}\over\sim$}\;}
\newcommand{\lesssim}{\; \raisebox{-.8ex}{$\buildrel{\textstyle<}\over\sim$}\;}

\newcommand{\apjs}{{\it ApJS, }}

\newcommand{\nat}{{\it Nature, }}

\newcommand{\anas}{{\it A\&AS, }}

\title[Type III migration]{ Type III migration in a low viscosity disc}

\author[Lin and Papaloizou]{Min-Kai Lin$^1$ \thanks{E-mail: mkl23@cam.ac.uk} and John C. B. Papaloizou$^1$ \thanks{E-mail: J.C.B.Papaloizou@damtp.cam.ac.uk}\\
$^1$ Department of Applied Mathematics and Theoretical Physics,
University of Cambridge, Centre for Mathematical Sciences,\\
\  \ Wilberforce  Road, Cambridge, CB3 0WA, UK \\}

\begin{document}

\maketitle

\begin{abstract}
We study the type III migration of a Saturn mass planet in low viscosity
discs. The planet  is found to experience cyclic episodes of rapid 
decay in orbital radius, each amounting to a few
Hill radii. We find this to be due to the scattering of large-scale vortices present in the
disc. The origin and role of vortices in the context of type III migration is
explored. 
It is shown through numerical
simulations and semi-analytical modelling that spiral shocks induced
by a sufficiently massive planet will extend close to the planet's
orbital radius as well as being global prominent features. The production of vortensity across shock tips
results in thin high vortensity rings 
with a characteristic width of the local
scale height. For planets with masses equal to and above that of Saturn, the rings are
co-orbital features extending all the way around the orbit.
Linear stability analysis show such vortensity rings are dynamically
unstable. There exists unstable modes that are localised about local vortensity minima
which coincide with gap edges. Simulations show that vortices are
non-linear a outcome. %% The gravitational torque exerted by
%% the vortex on the planet has significant effect on the nature of
%% migration.

We used hydrodynamic simulations to examine vortex-planet
interactions. Their effect is present in
discs with kinematic viscosity less than about an order of magnitude smaller
than the typically adopted value of
$\nu=10^{-5}\Omega_pr_p(0)^2$, where $r_p(0)$ and $\Omega_p$ are the initial
orbital radius and angular velocity of the planet respectively.
We find that the magnitude of viscosity affects the nature of type III
migration but not the extent of the orbital decay. The role of vortices as
a function of initial disc mass is also explored and it is found that the amount of
orbital decay during one episode of vortex-planet interaction is
independent of initial disc mass. We incorporate the concept of the
co-orbital mass deficit in the analysis of our results and link it to the presence of
vortices at gap edges. 

\end{abstract}


%set the number of sectioning levels that get number and appear in the contents
%\setcounter{secnumdepth}{3}
%\setcounter{tocdepth}{3}

%\frontmatter
%\include{Dedication/dedication}
%\include{Acknowledgement/acknowledgement}
%\include{Abstract/abstract}

%\tableofcontents
%\listoffigures
%\printglossary  %% Print the nomenclature
%\addcontentsline{toc}{chapter}{Nomenclature}

%\mainmatter

%%% Thesis Introduction --------------------------------------------------
%\chapter{Introduction}
%\ifpdf
%    \graphicspath{{Introduction/Figs/PNG/}{Introduction/Figs/PDF/}{Introduction/Figs/}}
%\else
%    \graphicspath{{Introduction/Figs/EPS/}{Introduction/Figs/}}
%\fi
\section{Introduction}
%% \begin{itemize}
%% \item observations of hot jupiters, early theory, type i and type ii.
%% \item recent analytical work, review work
%% \item simulations of type iii, discovery. how type iii works. basic
%%   ideas. recent simulations of type iii (peplinski).
%% \item large-scale disc structures, vortices in discs (refs for fixed
%%   orbit). recent studies on effect of migration (on type i, li 2009 ou
%%   2007)
%% \item purpose of current work, structure of paper
%% \end{itemize}
The importance of planet migration
was realised with the discovery of the first
exoplanet with an orbital
period of 4 days \citep{mayor95}. Such planets, classified as `hot Jupiters'
orbit so close to their host stars it is difficult to understand
their formation in situ. It is thought they form further out then
migrate inwards due to
torques from the gaseous protoplanetary disc.\\
\indent Tidal interactions between a protoplanet and protoplanetary disc were in fact
studied well before observations of exoplanets \citep{goldreich79,goldreich80}. In
these early studies, the protoplanet is treated as a small perturbation,
the linearised hydrodynamic equations are then solved for
the disc response. Density waves launched at Lindblad resonances generally lead to decay
of the planet's semi-major axis. This scenario is referred to as type I migration \citep{ward97}.
Recent semi-analytical treatments include \cite{tanaka02} for
isothermal discs and \cite{paardekooper09b} for a non-isothermal
equation of state. 
For planet masses above that of Jupiter, linear theory ceases to apply and
the non-linear disc response produces a surface density
gap about the planet \citep{lin86}. In this migration mode, now called type
II, the planet's orbital radius is locked with the disc viscous
evolution. That is, it drifts towards the central star along with disc
material, while residing inside the gap. 
%\cite{ward97} has presented a
%unified theory of type I and type II migration.

Recently, a new migration mode, called runaway or
type III migration, was introduced by
\cite{masset03} and further discussed by
\cite{artymowicz04} and \cite{papaloizou05}.  Type III migration is a self-sustaining mechanism with
short migration timescales $\leq O(10^2)$ orbits. A series of detailed numerical studies
was presented by \cite{peplinski08a,peplinski08b,peplinski08c} ,  who considered a Jupiter-mass planet and 
  included corrections for self-gravity and thermal
  effects. In their standard
  case, the planet's semi-major axis is reduced by a factor of $\sim3$
  within 50 orbital periods. They also concluded that type III migration is strongly dependent on the flow inside the Hill radius
  of the planet.

Type III migration is also applicable to intermediate mass planets comparable
to Saturn in massive discs. The planet opens a partial gap which, in
the case of a migrating planet, allows a net drift of disc material across
the planet's orbital radius through executing U-turns at the end of
horseshoe orbits. The change in the fluid elements' orbital radius
implies a torque on the planet that is proportional to its migration
rate. The migration rate also increases with the co-orbital mass
deficit $\delta m$, which is proportional to the difference between surface
density of the co-orbital region and that of the orbit-crossing flow.

 Most
previous disc-planet simulations have included large enough
viscosity to suppress instabilities resulting in a smooth
behaviour. Even numerical viscosity is sufficient to
suppress instabilities. For example, the studies of
\cite{peplinski08a,peplinski08b,peplinski08c} did not display instabilities even 
with zero physical viscosity. However, they employed a Cartesian code 
which, as pointed out by \cite{valborro07}, is probably too diffusive
to allow steep gradients to form and become unstable. 

In this work, we study a modified form of type III migration,
 which is non-smooth,  induced by large-scale
vortices originating from instabilities. The vortex forms
outside the Hill sphere and flows
across the co-orbital region. Changing
the planet softening length or adopting the equation of state of
\cite{peplinski08a}, which modifies flow inside the Hill sphere, is
found not to significantly affect the vortex-planet interaction
described in this paper.

The linear theory of instabilities in inviscid 
accretion discs associated with ring structures, sharp edges and vortensity
extrema is well-known
\citep{papaloizou84,papaloizou89,lovelace99,li00} and
vortices develop in the non-linear regime \citep{li01}. In the context of
protoplanetary discs, steep gradients arise from gap edges associated
with sufficiently massive planets. \cite{valborro07} performed
simulations of a Jupiter-mass planet held on fixed orbit to show the
formation of large-scale vortices at gap edges.  The migrating
case was considered by \cite{ou07} for a Neptune-mass planet. The
authors find non-smooth migration associated with non-axisymmetric surface
density enhancements near the gap edge. In this paper we examine
this effect in more detail for a Saturn mass planet undergoing type
III migration.

This paper is organised as follows. Section \ref{modeleq} 
describes the set of equations that we used to model the disc-planet system and in
Section \ref{rings} we describe the formation of vortensity rings using
numerical simulations and a semi-analytical model. We then study
their dynamical stability in Section \ref{Dynstab}. In 
Section \ref{mig_visc} we present hydrodynamic simulations of type III migration as
a function of viscosity, high-lighting the effect of vortices at low
viscosity. We analyse the inviscid case in detail in Section ref{inviscid} and
consider varying the disc mass {\bf and planet mass.} Finally in Section \ref{conc} 
we summarise our results.
%%% ----------------------------------------------------------------------

%%% Local Variables: 
%%% mode: latex
%%% TeX-master: "../thesis"
%%% End: 

%\chapter{Ring structures in disc-planet interactions}\label{rings}
%\ifpdf
%    \graphicspath{{Rings/Figs/PNG/}{Rings/Figs/PDF/}{Rings/Figs/}}
%\else
     \graphicspath{{Rings/Figs/EPS/}{Rings/Figs/}}
%\fi
\section{Basic equations and model}\label{modeleq}
We consider a planet of mass $M_p$ orbiting a central star 
of mass $M_*.$  We adopt a cylindrical coordinate system $(r,\phi,z)$
where $z$ is the vertical coordinate increasing in the direction normal to the disc plane
for which the unit vector is ${\hat{\bf k}}.$ 
We integrate vertically to obtain a
  two-dimensional flat disc model, for which
the governing hydrodynamic equations
in a frame uniformly rotating with angular velocity $\Omega_p{\hat{\bf k}}$ are the continuity 
equation
\begin{align}\label{continuity}
\frac{D\Sigma}{Dt}=-\Sigma\nabla\cdot\mathbf{u},
\end{align}
and the equation of motion
\begin{align}\label{momentum}
\frac{D\mathbf{u}}{Dt}+2\Omega_p\hat{\mathbf{k}}\wedge\mathbf{u} = 
-\frac{1}{\Sigma}\nabla P - \nabla\Phi_{\mathrm{eff}},
\end{align}
where $D/Dt$ is the total derivative and the vertically integrated pressure  $P=c_s^2(r)\Sigma$ with
$c_s=h(GM_*/r)^{1/2}.$ 
Here $\Sigma$ is the surface density, ${\bf u}$ is the velocity and the effective
gravitational and centrifugal  potential is
given by
\begin{align}
\Phi_{\mathrm{eff}} = &-\frac{GM_*}{r} - \frac{GM_p}{\sqrt{r^2+r_p^2
+2rr_p\cos(\phi-\phi_p)+\epsilon^2}}\notag\\
&-\frac{1}{2}\Omega_p^2r^2 + \frac{GM_p}{r_p^2}\cos{(\phi-\phi_p)}\notag\\
&+\int \frac{G\Sigma(r^\prime,\phi^\prime)}{r^{\prime2}}\cos{(\phi-\phi^\prime)}r^\prime dr^\prime d\phi^\prime
\end{align}
\citep{masset02}. The last two terms on RHS are indirect terms accounting for acceleration of the primary.
In the above, the cylindrical coordinates of the planet that is confined to remain in the disc plane
are $(r_p,\phi_p,0)$
 and the  softening length $\epsilon = 0.6H(r_p)$
 where  $H(r)=hr$ is the disc semi-thickness and $h=0.05$ the constant
 aspect ratio. All numerical and analytical work are based on the
   two-dimensional equations, but $H$ and $\epsilon$
   accounts for vertical structure.

 \subsection {Vortensity conservation}\label{secvort}
The vortensity  being the ratio of the  $z$ component of the  vorticity  (hereafter just  called the vorticity, $\omega$)
 to the surface density is
defined to be 
\begin{equation}\label{vortensity}
\eta =  \frac{\omega}{\Sigma} \equiv \frac{\omega_r+2\Omega_p}{\Sigma},
\end{equation}
where $\omega_r=\hat{\mathbf{z}}\cdot\nabla\wedge\mathbf{u}$ is the relative
vorticity seen in the rotating frame.
 It is well known that In barotropic flows without shocks,
it follows from (\ref{continuity}) and (\ref{momentum}) that the  vortensity
$\eta = \omega/\Sigma$  is
conserved for a  fluid particle. 
When an isothermal equation of state with variable sound speed as is used 
here is adopted, vortensity is no longer strictly conserved.
However, the sound speed varies on a global scale
so that  when   phenomena are considered on a local scale,  vortensity is 
 conserved to a good approximation in the absence of shocks.

\subsection{Hydrodynamic simulations}\label{sec2}
Our work is based on numerical simulations and specific parameters
depend on the problem at hand, as is described in the following
sections. Here we present the general set up.

For convenience we adopt dimensionless units such that the
orbital radius of the planet, $r_p(t)$, is initially $r_p(0)=2$. 
Time is expressed in units of the initial orbital period $P_0=2\pi/\Omega_p$, where
$\Omega_p=\sqrt{GM_*/r_p^3(0)}$ is the planet's initial Keplerian angular
velocity. The unit of mass is taken to be the central mass $M_*$ which
for working purposes may be considered to be $1M_{\odot}.$ The disc
has an  initially uniform surface density
$\Sigma=\Sigma_0\times10^{-4}$.  Some simulations include a constant
kinematic viscosity $\nu=\nu_0\times 10^{-5}$ in physical units  of
$r_p^2(0)\Omega_p$, with $\nu_0$ being a dimensionless
constant. We considered $\Sigma_0=1$---9 and $\nu_0=0$---1.     
%% In these units the 
%% the radial computation domain  for the simulations described in this
%% section is  $r=[1,3]$.
%%  For the simulations described in this section the planet is held fixed
%%  on its initial circular orbit with Keplerian angular velocity
%% $\Omega_p=\sqrt{GM_*/r_p^3}$ which also defines the rotation of the coordinate system.   % defines the dimensionless time we adopt. 
%% The planet mass is increased gradually from zero such that it attains its full
%% value after  five orbital periods.

We use the FARGO code to solve for the disc response
\citep{masset00a,masset00b}. FARGO evolves the hydrodynamic
equations in global cylindrical co-ordinates centred on the central
object. We used a non-rotating frame for the simulations described
here. Indirect terms arising from the non-inertial frame are included
in the potential. The code uses a finite-difference scheme with van Leer upwind advection
, similar to the
ZEUS code \citep{stone92} but it employs a modified algorithm for
azimuthal transport that allows for large time steps. 
We take $2\pi$-periodic boundary condition in
azimuth and wave damping boundary conditions at disc boundaries
\citep{valborro06}. We remark that vortensity
ring formation occurs as a result of  shocks near the planet and therefore is
unaffected these boundary conditions.

In the case where the planet is allowed to migrate, a fifth order
Rungke-Kutta method was used to integrate its equation of
motion. {\bf Simulations ran with halved time-step show
very similar results to the corresponding standard set-up.  
In particular, the extent of planet migration by vortex-scattering and
the number of events of scattering is the same. We believe the RK5
integrator is sufficiently accurate to capture this interaction, which
can be regarded as a two-body problem. }

\section{Ring structures in disc-planet interactions}\label{rings} 
In this section we  present hydrodynamic simulations to
show that narrow  surface density rings are brought about as  a consequence
of the fact that highly peaked 
vortensity  (being the ratio of the vorticity to the surface density) 
 rings are produced by flow through quasi-steady shocks
located close to  the planet. For
sufficiently massive planets, the associated vortensity generation occurs as fluid
elements execute a horseshoe turn in the co-orbital region. 
Focusing on such
cases, we construct  a simple model that enables an estimate of the shock
location to be made together with rate of the vortensity
generation as material flows through it. Our model is in good agreement with
numerical simulations and confirms the process of
vortensity generation within a planet's co-orbital region  which later
plays an important role in modifying  the flow structure there.

\subsection{ Vortensity generation by the shocks 
induced by a Saturn mass planet}\label{fiducial}
\indent We first consider the case with $M_p=2.8\times10^{-4}M_*$ corresponding
to a Saturn mass planet around a solar mass star.  Here, the planet potential is switched on over $5P_0$.
The computation
domain is $r=[1,3]$ and the
resolution was $N_r\times N_\phi=1024\times3072$,  corresponding to 
radial and azimuthal grid spacings of   $\Delta r\simeq
0.02r_h$ and $r\Delta\phi\simeq0.05r_h$ respectively, where
$r_h=(M_p/(3M_*))^{1/3}r_p$ is the planet's Hill radius. The planet
is held on fixed circular orbit and the disc has density $\Sigma_0=1$ with
no explicit viscosity ($\nu=0$).

\begin{figure}
\center
\includegraphics[width=1.0\linewidth]{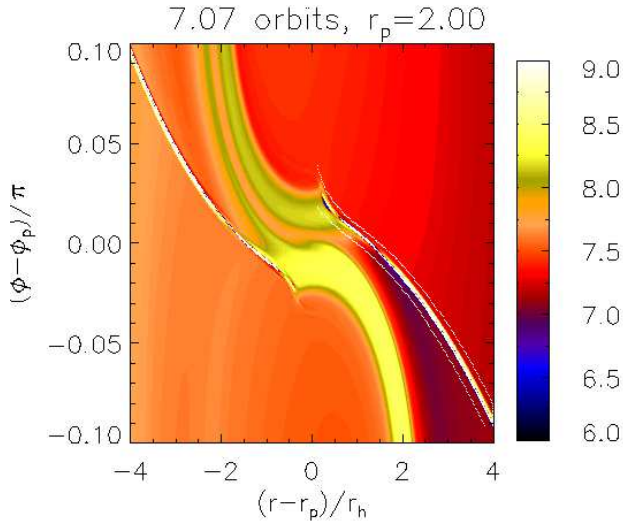}\caption{Vortensity
generation and destruction across shocks induced by a Saturn-mass
planet in an inviscid disc. The plot shows $\ln{\eta} =\ln{(\omega/\Sigma)}$.
Regions of increased vortensity are clearly visible as half horseshoes
above and below the planet. The increase occurs as material passes through
parts of the shock fronts that extend  into  the co-orbital region.}
\label{vortxy_020} 
\end{figure}

Fig.\ref{vortxy_020} shows the vortensity field at
$t=7.07$ close to the planet. Vortensity is
generated/destroyed as material passes through the two spiral shocks.
 For the outer shock, vortensity
generation occurs for fluid elements executing a
horseshoe turn ($|r-r_p|\lesssim r_h$) while  vortensity is reduced for
fluid that passes by the planet, but the change is smaller in magnitude
in the latter case. The
situation is similar for the inner shock, but some post-shock material
with increased vortensity continues to pass by the planet. This stream begins at
$r\simeq -r_h$ but such feature is absent at the outer
shock, suggesting a lack of symmetry about $r=r_p$, possibly resulting from
the non-uniform vortensity background being $\propto r^{-3/2}$.
Note that although a pre-shock fluid element that would be  on a horseshoe trajectory,
may in fact pass by the planet after crossing the shock, it is
clear that vortensity rings originate from passage through shock fronts interior
to the  co-orbital  region that would correspond to the horseshoe region
for free particle motion.

The streams of high vortensity eventually move around the whole orbit outlining the
entire co-orbital region. Fig. \ref{vortxy_020} shows  that they are generated
along a small  part of the shock front  of length  $\sim r_h$. This  results in
thin rings with  a similar  radial width.
The fact that they originate from horseshoe
material can  enhance the contrast  as post-shock
inner disc horseshoe material is mapped  from $r-x$ to  $r+x$ 
to become
adjacent to post-shock outer disc material passing by the planet.

\begin{figure}
\center
\includegraphics[width=1.0\linewidth]{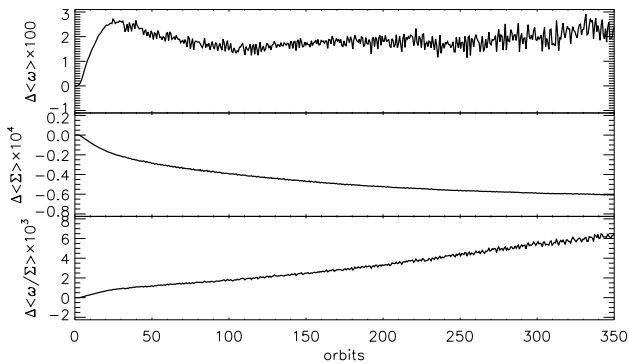}\caption{Long-term
  evolution of co-orbital properties for a Saturn-mass planet held on
  fixed orbit. Vorticity, density and vortensity 
  perturbations are expressed   relative to their  values at
  $t=0$. \bf{Angle brackets denote an average over the co-orbital region defined by the annulus $r=[r_p-2.5r_h,r_p+2.5x_s]$.}
  \label{coorb_plot_0}}
\end{figure}

We also show in Fig. \ref{coorb_plot_0} long-term evolution of
average co-orbital properties from a corresponding low resolution run
($N_r\times N_\phi = 256\times768$). The co-orbital region is taken to be
the annulus $[r_p- x_s, r_p+x_s]$. We fix $x_s=2.5r_h$, as is typically
measured from hydrodynamic simulations
\citep{artymowicz04b,paardekooper09} for intermediate or massive planets. In Appendix
\ref{xslimit} we show that in the particle dynamics limit
$x_s \lesssim 2.3r_h$, comparable to the value adopted above.

Vorticity generation occurs within $t\lesssim25$ orbits,
after which it remains approximately steady.  For a Jupiter-mass planet the time taken to reach this state is about
$50$ orbits, but most of the vorticity generation also occur in the first $\sim 25$ orbits. It is important to note
that subsequent vortensity increases {\bf is} in narrow rings and fluctuations are due to
instabilities associated with the rings. The average surface density falls more gently as
the planet opens a gap, which is a requirement for the rings to be
self-supported. Consequently, we observe a rise in co-orbital
vortensity. Fig. \ref{coorb_plot_0} reflects modification of co-orbital properties
on dynamical time-scales due to shocks, so we expect migration mechanisms which depend on
co-orbital structure to be affected.


\subsection{ Location of the  vortensity generation region for different planet masses}
\begin{figure}
\center
\includegraphics[width=1.0\linewidth]{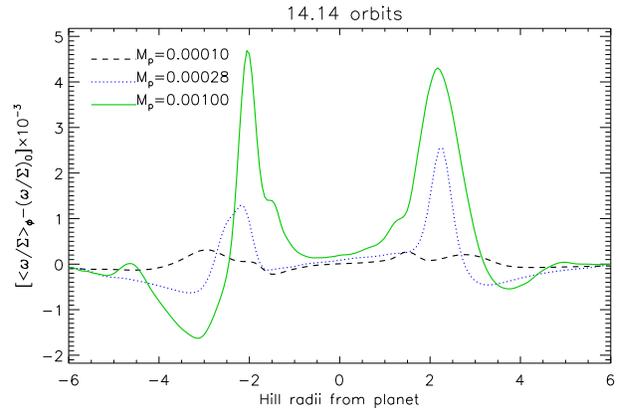}\caption{Azimuthally averaged vortensity
  perturbation for different planet masses.} \label{compare2_020}
\end{figure}

The process of  formation of vortensity rings discussed 
here differs from  that observed
in \cite{koller03} and \cite{li05}, where the rings are generated by shocks outside
the co-orbital region. This is because the authors used a
smaller planet mass. 
To illustrate  the effect of reducing the mass, we ran
simulations with $N_r\times N_\phi = 256\times768$ for $M_p\times10^{4}=1,\,2.8$ and
$10$. Fig. \ref{compare2_020} compares azimuthally averaged
vortensity perturbations induced in each case after $14.14$ orbits. 
The intermediate and high mass cases are
qualitatively similar, having $\Delta(\omega/\Sigma)$ maxima at
$r-r_p\sim\pm 2r_h$ and minima at $r-r_p\sim3 r_h$. The magnitude of
$\Delta(\omega/\Sigma)$ increases with $M_p$ because higher masses
induce higher Mach number shocks. As the half-width of the horseshoe region
is $x_s=2.5 r_h$ for such masses, vortensity rings are co-orbital features.

The low mass $M_p=10^{-4}$ case has much smaller
$|\Delta(\omega/\Sigma)|$. There is a vortensity maximum at $r-r_p=-3r_h$
but nothing of similar magnitude at $r-r_p>0$. 
 \cite{paardekooper09}
found the co-orbital half-width, in the limit of zero softening
for low mass planets, to be:
\begin{align}
x_s=1.68(M_p/h)^{1/2}r_p.
\end{align}
For $M_p=10^{-4},\,h=0.05$, $x_s=2.33r_h$.
Non-zero softening gives a smaller $x_s$. Hence, vortensity
rings for low mass planets occur outside the  co-orbital region as is
confirmed in our simulations and they are very much weaker. 

%%%%%%%%%%%%%%%%%%%%%%%%%%%%%%%%%%%%%%%%%%%%%%%%%%%%%%%%%%
%% HIGH RESOLUTION RESULTS FOR VORTENSITY GENERATION V.S. M_P
%%%%%%%%%%%%%%%%%%%%%%%%%%%%%%%%%%%%%%%%%%%%%%%%%%%%%%%%%%
%% \delta m = -5\times10^{-5} for M_p = 10^{-4}
%% \delta m = -2\times10^{-4} for M_p = 2.8\times10^{-4}
%% \delta m = +5\times10^{-4} for M_p = 10^{-3}
%% the 1d vortensity plot is very similar to low resolution case (picture
%% at /data/stardisk/mkl23/hyades/vorticity/compare2_040.ps). in the
%% above, only radial derivative part of voriticy is used. the low
%% resolution stuff presented above used full expression for
%% vorticity, there is little difference between the two. low
%% resolution was needed to examine long term evolution/equilibrium
\subsection{ A semi-analytic model for co-orbital 
vortensity generation by shocks}\label{model}
\indent We now  construct a semi-analytic model describing the process of
shock-generation of vortensity.  In this way we gain an understanding
of the physical processes involved and later how they lead to 
strong torques and fast migration of the planet and under what conditions
simulations can represent them accurately.

More specifically we model the outer
spiral shock in Fig. \ref{vortxy_020}. To do this we need the  pre-shock flow field,  the shock front
location  and then to evaluate the vortensity change undergone  by
material as it passes through the shock.
We now consider these in detail.

\subsubsection{Flow field}
\indent We first simplify the geometry by adopting the shearing box
approximation \cite[e.g.][]{paardekooper09}. As we are concerned with
the flow near the planet, we consider a local Cartesian co-ordinate
system $(x,y)$ co-rotating with the planet at angular
speed $\Omega_p$ and with origin at its centre of mass. Without the planet, the velocity field is
Keplerian $\mathbf{u}=(0,-3\Omega_p x/2)$. 
In order to deal with the pre-shock flow we make the assumption that pressure effects can be ignored when
 compared to the planet potential.  This ballistic  approximation is appropriate for 
 a slowly varying  supersonic flow as is expected to be appropriate for the pre-shock
 fluid. The   velocity $\mathbf{u}$  satisfies the local form of
 equation (\ref{momentum}) with  $P=0$ and the effective
 potential
 \begin{align}
\Phi_{\mathrm{eff}} =  - \frac{GM_p}{\sqrt{x^2+y^2+\epsilon^2}}-\frac{3}{2}\Omega_p^2x^2
\end{align}
and thus follows from particle dynamics. The indirect terms are neglected in this approximation. 
We write a fluid particle's trajectory in a steady state flow  in the form
 of  $ x=x(y)$ and the corresponding velocity field
 as $\mathbf{u}=\mathbf{u}(y).$
 Noting that  on a particle trajectory we have 
 $\frac{D}{Dt}=u_y\frac{d}{dy}$,  it follows from the local form of (\ref{momentum}) 
 that on a  particle trajectory we have the following system of three simultaneous
 first order differential equations:
\begin{align}
\label{semi-an}&\frac{du_y^2}{dy}=-4\Omega_pu_x-\frac{2M_py}{(x^2+y^2+\epsilon^2)^{3/2}}\equiv Q,\\
&\frac{d}{dy}\left(u_xu_y^2\right)= u_y\left[3\Omega_p^2x-\frac{M_px}{(x^2+y^2+\epsilon^2)^{3/2}}\right]
+u_xQ\notag \\
&\phantom{\frac{d}{dy}\left(v_xv_y^2\right)=}+2\Omega_pu_y^2,\\
&\frac{d}{dy}\left(xu_y^2\right)=xQ+u_xu_y.
\label {semi-an1}\end{align}
We  use these to solve for the state vector
$\mathbf{U}(y)=[u_y^2,\,u_xv_y^2,\,xu_y^2]$ in $x>0$ for a particular
particle. The boundary condition is 
\begin{align}
(u_y,\, u_x,\, x)\to(-3\Omega_p x_0/2, \, 0, \, x_0) \,\,\mathrm{as}\,\,
y\to\infty , 
\end{align}
where $x=x_0$ is the particle's unperturbed path. 
The totality of paths obtained by considering a continuous range of $x_0$
then constitutes the flow field.  Having obtained the velocity field, we 
use  vortensity conservation to obtain $\Sigma$. The surface density is then
\begin{equation}
\Sigma(x,y) = \frac{2\Sigma_0}{\Omega_p}(\omega_r+2\Omega_p),
\end{equation}
where the unperturbed absolute vorticity is
$\omega =\Omega_p/2.$  Numerically, we
calculated the relative vorticity  by relating it to the circulation through 
\begin{equation}
\omega_r\simeq\frac{1}{\Delta A}\oint\mathbf{u}\cdot\mathbf{dl}
\end{equation}
where the integration is taken over a closed loop about the point of
interest and $\Delta A$ is the enclosed area. This avoids
errors due to  numerical
differentiation on the uneven grid in $(x,y)$  generated by solving the above system.
\subsubsection{Location of the shock front}
\indent The physical reason for shock formation is the fact that the
planet presents an obstacle to the flow. Where the relative flow is subsonic,  the
presence of this obstacle can be communicated to the fluid via sound
waves emitted by the planet. In regions where the relative flow is supersonic,  the fluid is unaware
of the planet via sound waves (but  the planet's
gravity is felt).  We estimate the location of the boundary separating these two regions
by  specifying an appropriate characteristic curve or ray defining a sound wave.  
This is a natural location for shocks. Applying this
idea to Keplerian flow, \cite{papaloizou04} obtained a good match
between the predicted theoretical shock front and the wakes associated with a low mass
planet. For general velocity field $\mathbf{u}$, the characteristic
curves satisfy the equation 
\begin{align}\label{shock_front}
\frac{dy_s}{dx} =
\frac{\hat{u}_y^2-1}{\hat{u}_x\hat{u}_y-\sqrt{\hat{u}_x^2+\hat{u}_y^2-1}},
\end{align}
where $\hat{u}_i= u_i/c_s$. The  positive sign of the square root
has been chosen so that the curves have negative slope
in the domain $x>0$   with $u_y<0.$  As the fluid flows
from super-sonic ($y>y_s$) to sub-sonic ($y<y_s$), fluid located at 
 $y=y_s$  begins
to know about the planet through pressure waves \citep{papaloizou04}. In
Keplerian flow, the rays defining the shock fronts originate from the  sonic points at
$x=\pm2H/3,\,y=0$. In a  general flow,  sonic points where  ($|\mathbf{u}|=c_s$),
at which the rays may start,
lie on  curves and can occur for $x<2hr_p/3$.  The starting
sonic point for   solving  equation  ( \ref{shock_front}) that we eventually  adopted has
 $x=0$ and the value of  $y >0$ that is furthest possible
from the planet.

\subsubsection{Vorticity and vortensity changes across a shock}
The jump in absolute vorticity $[\omega]$ is readily obtained by
resolving the fluid motion parallel and perpendicular to the shock
front  \citep[eg.][]{kevlahan97}. As we do not solve the energy equation,
 shocks are locally  isothermal and our expression differs from
those of  \citet{kevlahan97}, accordingly a brief derivation of $[\omega]$ is
presented in  Appendix B. The result for a steady shock is
\begin{align}\label{vorticity_jump}
[\omega]=-\frac{(M^2-1)^2}{M^2}\frac{\p u_\perp}{\p S}+(M^2-1)\omega
-\left(\frac{M^2-1}{u_\perp}\right)\frac{\p c_s^2}{\p S},
\end{align}
where $u_\perp$ is the pre-shock  velocity component perpendicular to the shock front, 
 $M=u_\perp/c_s$ is perpendicular Mach number
and $c_s=hr^{-1/2}$ is the sound speed. $\p/\p S$ is the derivative
along the shock. It is  important to note that for (\ref{vorticity_jump}) to hold,
 the direction of increasing $S,$ the direction of positive  $u_{\perp},$
 and the vertical direction should form
a right handed triad. We take increasing $S$ as moving away from 
the planet. The local
isothermal equation of state produces the last term on RHS of
equation  (\ref{vorticity_jump}). Because of the slow variation of $c_s$  
its contribution is not important in this application. \\ 
\indent The vortensity jump $[\omega/\Sigma]$ follows immediately from
equation (\ref{vorticity_jump}) as
\begin{align}\label{vortensity_jump}
\left[\frac{\omega}{\Sigma}\right]=-\frac{(M^2-1)^2}{\Sigma M^4}\frac{\p
  u_\perp}{\p S}-\left(\frac{M^2-1}{\Sigma M^2u_\perp}\right)\frac{\p c_s^2}{\p S},
\end{align}
which essentially reduces to the expression derived by \cite{li05} if
$c_s=\mathrm{constant}$ (and a sign change due to different
convention). The sign of  $[\omega/\Sigma]$ depends mainly on the gradient of
$u_\perp$ (or $M$) along the shock. As in our case  $u_\perp<0,$ the width of the  increased vortensity
rings produced is determined by  the length along the shock where $|M|$ is
increasing. Note that $[\omega/\Sigma]$ does not 
depend explicitly on the  pre-shock vortensity, unlike the absolute vorticity jump.

\subsubsection{Comparison of the results of the semi-analytic model with numerical simulations}
\begin{figure}
\center
\includegraphics[width=1.0\linewidth]{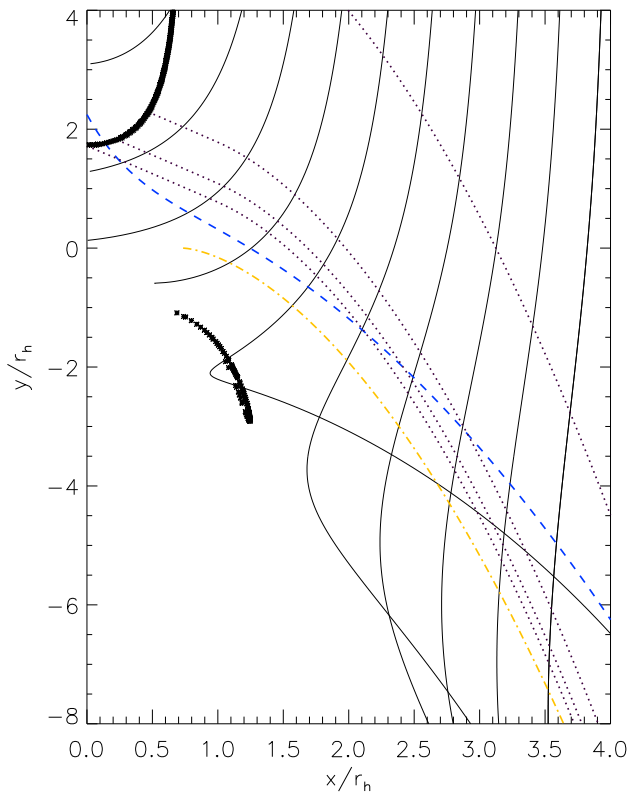}\caption{Solid lines: particle paths from the reduced zero-pressure
momentum equations (equations \ref{semi-an} - \ref{semi-an1}); thick lines: curves composed of sonic points
 on which $|\mathbf{u}|=c_s$;
dotted lines: theoretical shock fronts; dash-dot line: solution to equation (\ref{shock_front}) for Keplerian flow; dashed line: 
polynomial fit to simulation shock front. The actual shock front begins  at a sonic point 
around $x=0.2r_h$.}
\label{shock_curve2_}
\end{figure}

We  now compare the results of hydrodynamic simulations
with those obtained from the model.  
First we illustrate the particle paths that constitute
the flow field together with the  shock
fronts obtained by assuming coincidence with the characteristic curves
that are obtained from the semi-analytic model 
in Fig. \ref{shock_curve2_}.
A polynomial fit to the simulation shock front is also
shown.
 Particle paths cross for $x > r_h,\,y < -2r_h$
 so  that the neglect of pressure becomes invalid.  Accordingly  the
solution to equation (\ref{shock_curve2_})  should not be trusted in this
region.  Fortunately, vortensity generation occurs at a distance from the planet that is  within 
$r_h$, where pre-shock particle paths do not cross.

In Fig. \ref{shock_curve2_}, the estimated shock location is
qualitatively good and tends to the Keplerian solution further away. 
The important feature is that the shock can
extend close to $x=0$, across horseshoe orbits. If the flow were
purely Keplerian there could be no significant vortensity generation
close to $x=0$ because the flow becomes sub-sonic
there. Furthermore, only circulating fluid can be shocked in that
case. Shock-generation of vortensity inside the co-orbital region is
only possible for sufficiently massive planets that induce large enough
non-Keplerian velocities.

\begin{figure}
\center
\includegraphics[width=\linewidth]{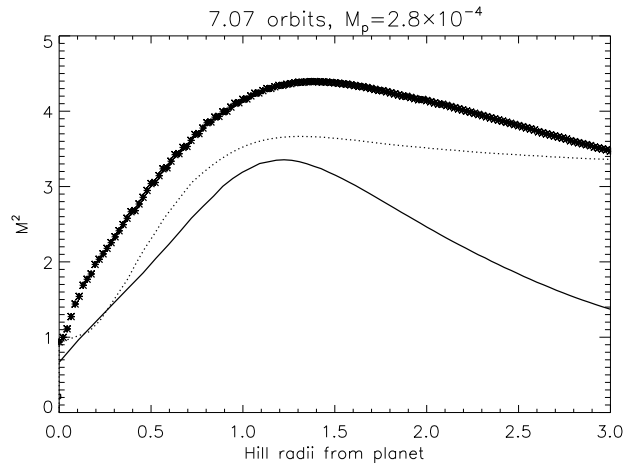}\caption{Perpendicular
  Mach number squared $M^2$ along the outer spiral shock  illustrated in
  Fig. \ref{vortxy_020}. 
Asterisks: simulation data; dotted line:  $M^2$ obtained 
  from the simulation shock
  and the semi-analytic flow; solid line: $M^2$
  obtained from the  semi-analytic shock front and flow.}
\label{sc3_machsq_020}
\end{figure}

\begin{figure}
\center
\subfigure[\label{mp2.8}]{\includegraphics[width=\linewidth]{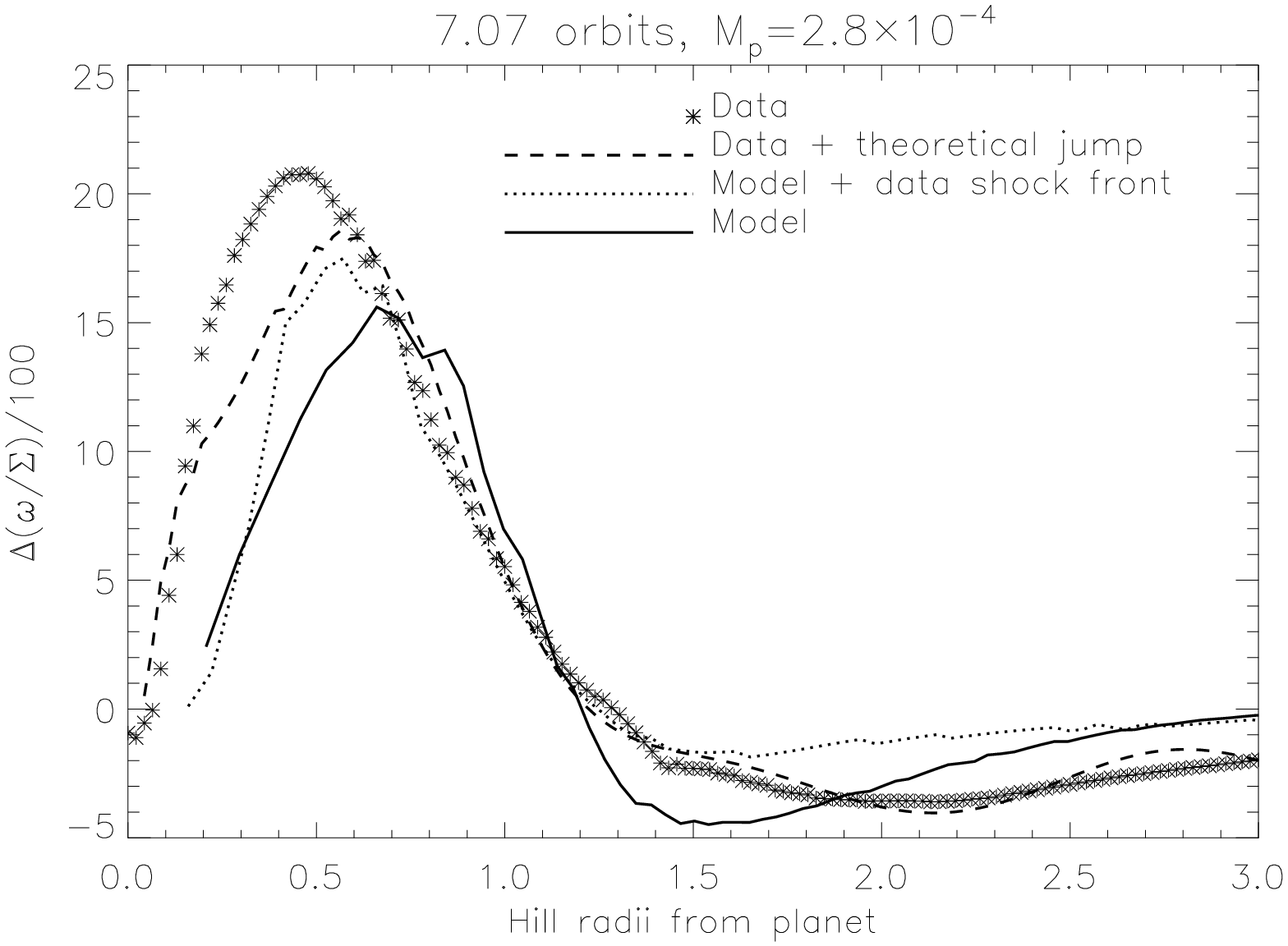}}
\subfigure[\label{mp2}]{\includegraphics[width=\linewidth]{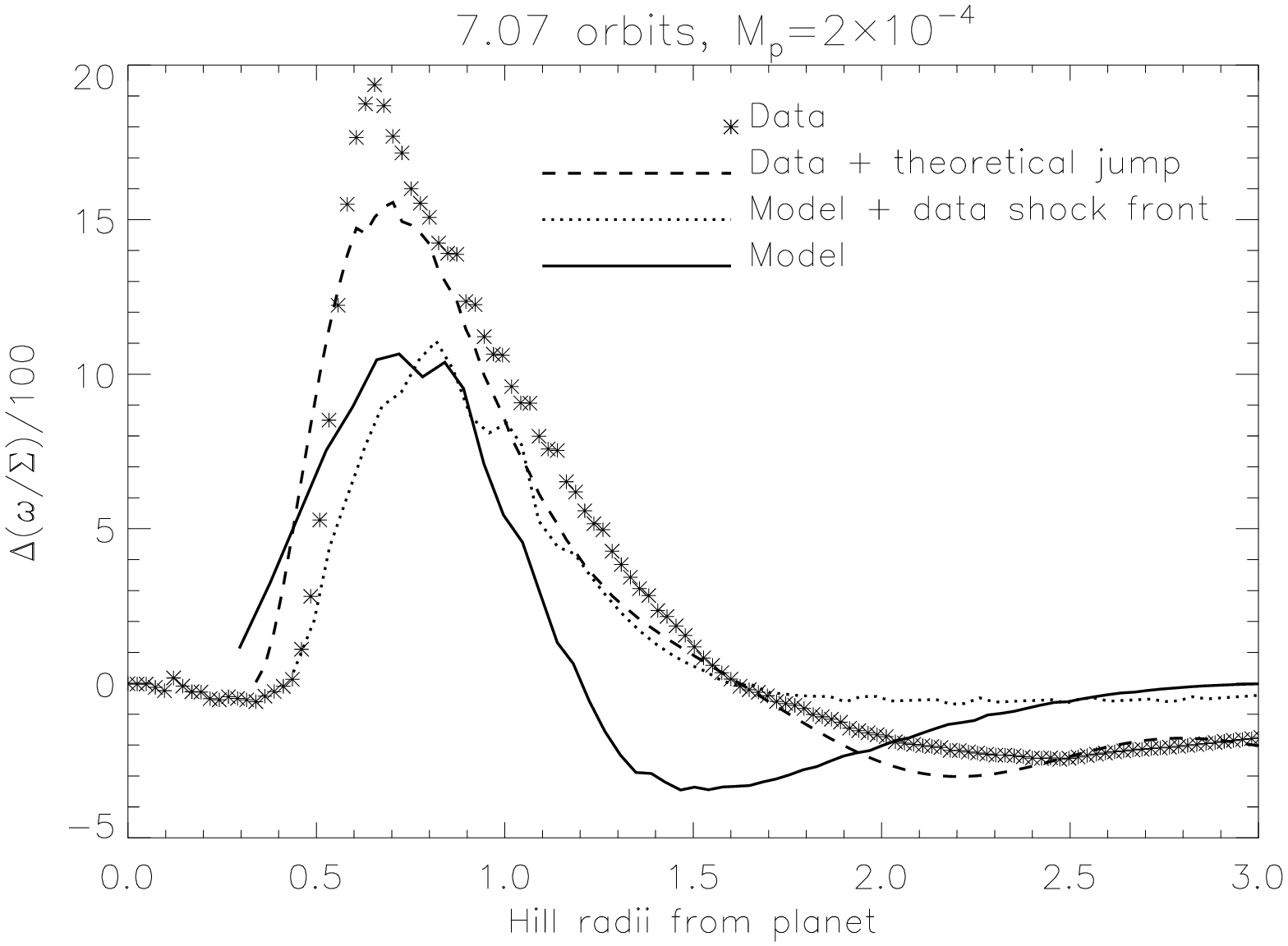}}
\caption{Semi-analytic  and actual vortensity jumps across a shock in the
  co-orbital region. Asterisks are measured from the simulations. The
  dashed lines were obtained by using pre-shock simulation data coupled
  with the jump condition  specified by equation( \ref{vortensity_jump}). The  dotted lines 
  were obtained using the semi-analytic flow field together with
  the location of the  shock front obtained from the simulation.  The  solid
  curves  correspond to the semi-analytic model for both the flow field and shock front.}\label{vort_jump}
\end{figure}

\indent The key quantity determining vortensity generation is the
perpendicular Mach number. Fig. \ref{sc3_machsq_020} compares $M$ from
simulation and model.  Although the semi-analytic model gives a shock  Mach number
that is somewhat smaller than found from the simulation, all curves have $|M|$
increasing from $x=0\to 1.3r_h$ which is the important domain for vortensity
generation.  Thus equation (\ref{vortensity_jump})
 implies  vortensity generation in all cases. %% Note that in the significant domain
%%   $x \in [0\to 1.3r_h]$ the Mach numbers deviate by less than 50\%.
Fig. \ref{mp2.8} illustrates various  combinations of
semi-analytic modelling  and simulation data that have been used to
estimate the  vortensity jump
across the shock.
The qualitative similarities  
between the various curves confirms that
vortensity generation  occurs  within co-orbital
material about one Hill radius away from the planet. 
It is shocked as it undergoes horseshoe turns.
 
Assuming that material is mapped from $x\to -x$ as it switches to
the inner leg of its horseshoe orbit,  we expect the outer spiral
shock to produce a vortensity ring peaked at $r-r_p\sim -0.5r_h$ of
width $O(r_h)$ ($=O(H)$), with a similar discussion
applying to the inner shock which is expected to produce
a vortensity ring peaked at $r-r_p\sim 0.5r_h$ . Of course, as a
fluid element moves away from the U-turn region, $|r-r_p|$ increases,
but it remains on a horseshoe orbit. Thus,
thin vortensity rings are natural features of the co-orbital
region for such planet masses.

We have  also tested our model for $M_p=2\times10^{-4}$ in
Fig. \ref{mp2}.  There is still  a good qualitative 
match between the simulation and  the model;
even though  lowering $M_p$  makes  the zero-pressure
approximation, adopted to determine the semi-analytic flow field, less good. 
Decreasing $M_p$ shifts vortensity generation
away from the planet in the semi-analytic model, as is also observed in 
the hydrodynamic simulation
(Fig. \ref{compare2_020}). In this case, there is no vortensity
generation in $r-r_p<0.5r_h$ but vortensity rings are still
co-orbital (with  peaks at $\sim0.7r_h$).

We remark that steep
vortensity gradients are associated with 
dynamical instabilities \citep[e.g.][]{papaloizou85} and this is
explored in more detail below. 
\section{Dynamical stability of vortensity rings}\label{Dynstab}
Having established the origin of the vortensity rings
and the mechanism
for producing them, 
we go on to  study the linear stability of the shock-modified
protoplanetary disc model described above.
This is an important issue as instability can lead to their breaking
up into mobile non-axisymmetric structures which can
affect the migration of the planet significantly.

The linear stability 
of inviscid barotropic  axisymmetric
rings  and discs without a planet 
is well developed \citep{papaloizou84,papaloizou85,papaloizou89}.
The extension to a 
non-barotropic  equation of state was  undertaken more recently
 \citep{lovelace99,li00}. This work  indicates that  sharply peaked
  surface density or  vortensity features in the  disc  are  associated with dynamical
instabilities. It is directly relevant to the types of configuration
that we have found to be produced by the disc planet interactions
described above. For example 
\cite{valborro07} considered the linear instability of gap edges associated
with a Jupiter mass planet
and connected it to  vortex formation there. 

%Here we  consider a different
% parameter   for which highly peaked vortensity features appear in
%the coorbital (horseshoe)  region. 
Here we consider the stability of the partial gap opened by a Saturn-mass planet,
for which type III migration is expected in a massive disc. 
In order to be able to do this
we need to be able to define an appropriate background  equilibrium 
axisymmetric structure
to perturb. We show that this is possible. 
\subsection{Basic background  state}
\begin{figure}
\center
\includegraphics[width=\linewidth]{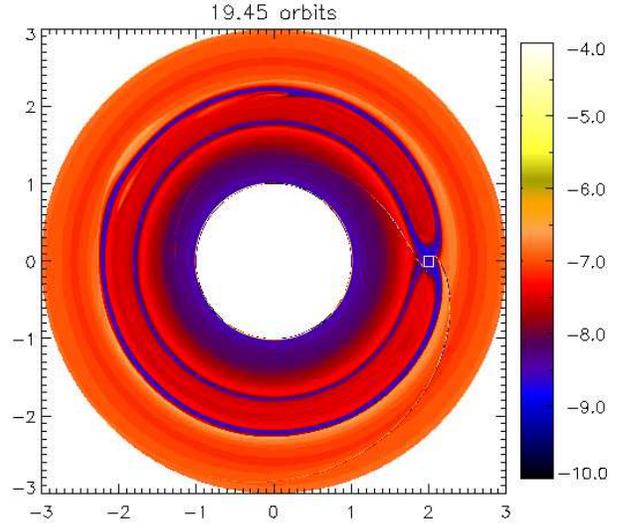}
\caption{A contour plot of $\ln{(\Sigma/\omega)}$ when the vortensity ring like structures
have formed in the co-orbital region. These are almost axisymmetric apart from in the
region very close to the planet. The centre of the small square marks the planet position.
\label{vort_055}}
\end{figure}

 The
basic state   should be axisymmetric and time independent with no radial velocity
($\p/\p\phi=\p/\p t=u_r=0$). To set this up we  suppose the
vortensity profile $\eta(r)$ is known (e.g. via shock modelling as above) 
and that accordingly we have 
\begin{align}\label{vortensity_definition}
\eta(r) = \frac{1}{r\Sigma}\frac{d}{dr}\left(r^2\Omega\right),
\end{align}
The radial momentum equation gives
\begin{align}\label{hydrostatic}
r\Omega^2 = \frac{1}{\Sigma}\frac{dP}{dr} + \frac{GM_*}{r^2}
\end{align}
Using these together with the locally isothermal equation of state,
  $P=h^2GM_*\Sigma/r,$ $h$ being the constant aspect ratio, we obtain a single equation
for $\Sigma$ in the form
\begin{align}\label{basic_density}
h^2\frac{d^2\ln{\Sigma}}{dr^2} =& \frac{h^2-1}{r^2}  +
\frac{2\Sigma\eta}{\sqrt{GM_*}} \left(\frac{1-h^2}{r}+h^2\frac{d\ln{\Sigma}}{dr}\right)^{1/2}\notag\\
&-\frac{2h^2}{r}\frac{d\ln{\Sigma}}{dr}
\end{align}
We solve equation (\ref{basic_density}) for $\Sigma$  with  $\eta(r)$ 
taken as an azimuthal average from the fiducial
simulation described in (\S\ref{fiducial}) at a time at which vortensity rings have
developed. These structures are essentially axisymmetric  
apart from in the close neighbourhood of the planet. They are illustrated
 in Fig. \ref{vort_055}. The boundary conditions is $\Sigma=\Sigma_0=1$ at
$r=1.1$, and $r=3.$  These conditions are consistent with the fact that
 shock-modification of the surface density profile only
occurs near the planet ($r=2$). 

\begin{figure}
\center
\subfigure[$\Sigma$\label{custom_dens}]{\includegraphics[scale=0.33]{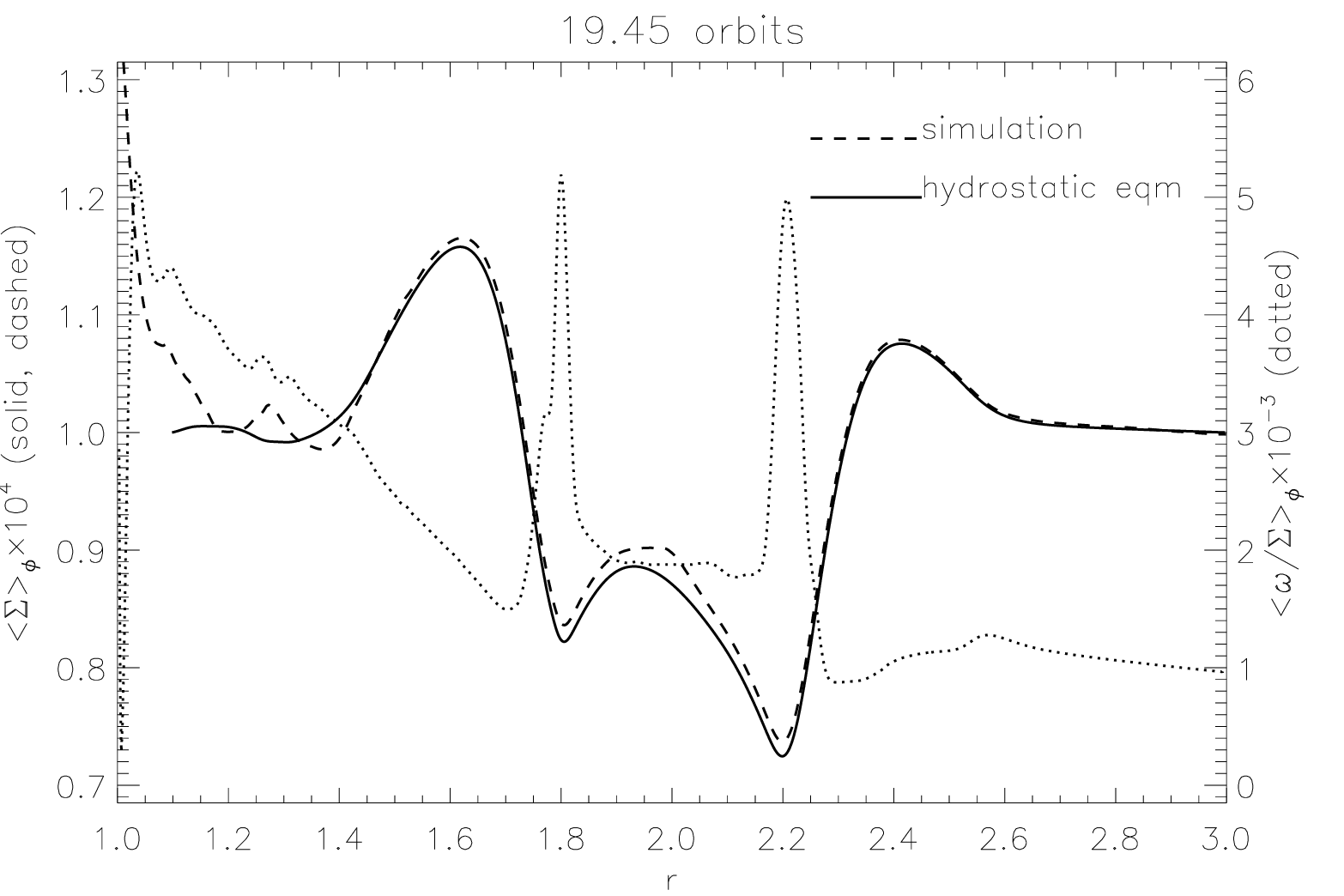}}
\subfigure[$\Omega/\Omega_k$\label{custom_urad}]{\includegraphics[scale=0.33]{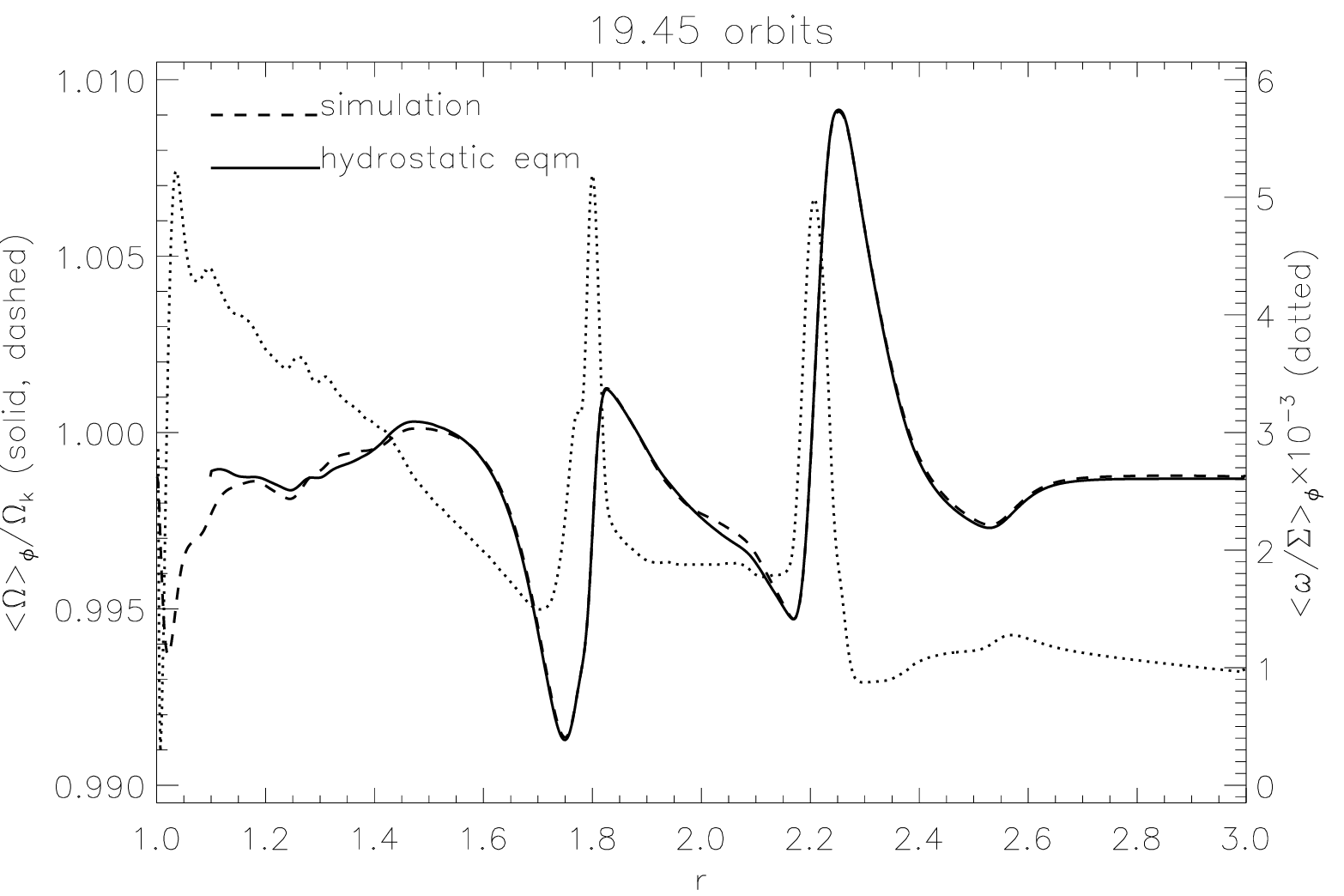}}
\caption{A comparison of the axisymmetric surface density and
  angular velocity (scaled by Keplerian speed) profiles 
obtained by solving equation  (\ref{basic_density}), with the same quantities obtained by azimuthally averaging
the simulation data over $2\pi$.  
The azimuthally averaged vortensity profile from simulation data is also shown (dotted).} The co-orbital region is $r=[1.77,2.22]$.
\end{figure}

A comparison  between hydrostatic surface density  
and angular velocity  profiles
 obtained by solving equation  (\ref{basic_density}) with those obtained  as azimuthal averages from
 the corresponding simulation is illustrated in
Fig. \ref{custom_dens} and Fig. \ref{custom_urad}.  The agreement is generally very good  
indicating that the adoption  of the basic axisymmetric state  defined by the simulation vortensity profile,  together with equation  (\ref{hydrostatic}),  for stability analysis
should be a valid procedure.  Fig. \ref{custom_dens} shows that  vortensity rings reside 
in the horseshoe region just
inside the gap. At  a surface density extrema equation  (\ref{basic_density})
implies that
\begin{align*}
h^2\frac{d^2\ln{\Sigma}}{dr^2} = \frac{h^2-1}{r^2} +
2\Sigma\eta\sqrt{\frac{1-h^2}{GM_*r}}.
\end{align*}  
Since $h\ll1$, if $\eta$ is sufficiently
large then $d^2\Sigma/dr^2 > 0$. Hence vortensity maxima will coincide with
surface density minima.
 In our basic state, local minima and maxima are separated by $O(H)$ ($\simeq
0.1$).
 As vortensity rings originate from
spiral shocks, equation  ( \ref{basic_density}) demonstrates the  link
between shocks and gap formation. Given the double-ringed vortensity profile $\eta(r)$, which may be
estimated by modelling shocks as described above, we can solve equation (\ref{basic_density}) for the axisymmetric
surface density profile $\Sigma(r)$. We will then find that in order for the rings to be in hydrostatic equilibrium, a gap in the surface density must be 
present around the planet's orbital radius, which is in between the vortensity peaks. Sufficiently massive planets induce 
strong shocks, implying larger vortensity maxima, and hence deeper surface
density minima or gaps. 
\subsection{Linearised equations}
\indent Having established the ring-basic state, we perform linear analysis to determine its stability. 
The analysis is for a two-dimensional disc and isothermal perturbations.  This is much simpler than the
 non-barotropic 2D disc studied by \cite{li00}, who considered adiabatic perturbations and hence 
 involved the energy equation, omitted in the present work.  This is so that  linear analysis is consistent with hydrodynamic simulations presented above and later. The 3D barotropic disc was studied by \cite{papaloizou87} where edges 
and the role of the vortensity were first considered.  We obtain 
the governing equation for locally isothermal
perturbations  by linearising the  continuity equation and the equation of motion as seen in the non-rotating
frame
\begin{align}\label{euler}
&\frac{\p\Sigma}{\p t} + \nabla\cdot{(\Sigma\mathbf{u})} = 0\\
&\frac{\p\mathbf{u}}{\p t} + \mathbf{u}\cdot\nabla\mathbf{u} =
-\frac{1}{\Sigma}\nabla P -\nabla\Phi,\label{momeq}
\end{align}
where $P=c_s^2\Sigma$   with $c_s^2=h^2GM_*/r.$ The total potential $\Phi$ is assumed
fixed. We set
%\begin{align}
%(\Sigma, u_r, u_\phi) \to (\Sigma, 0, u_\phi)+(\delta\Sigma(r), \delta u_r(r), 
%\delta u_\phi(r))\times\exp{i(\sigma t + m\phi)},\\
%\end{align}
\[
\begin{pmatrix}
\begin{matrix}
\Sigma \\
u_r\\
u_\phi
\end{matrix}
\end{pmatrix}
\to 
\begin{pmatrix}
\begin{matrix}
\Sigma \\
0\\
u_\phi
\end{matrix}
\end{pmatrix}
+
\begin{pmatrix}
\begin{matrix}
\delta\Sigma(r) \\
\delta u_r(r)\\
\delta u_\phi (r)
\end{matrix}
\end{pmatrix}
\times \exp{i(\sigma t + m\phi)}
\]
where, the first term on the right hand side corresponds to
the basic background state, 
 $\sigma$ is a complex frequency and $m,$ the azimuthal mode number,  is a positive
integer. 
The linearised  equation  of motion (equation (\ref{momeq})) gives
\begin{align}
\delta u_r &= -\frac{c_s^2}{\kappa^2 - \bar{\sigma}^2}
\left(i\bar{\sigma}\frac{dW}{dr} + \frac{2im\Omega W}{r}\right)\label{velocities1}\\
\delta u_\phi &= \frac{c_s^2}{\kappa^2 - \bar{\sigma}^2}
\left(\Sigma\eta\frac{dW}{dr} + \frac{m\bar{\sigma}}{r}W\right)\label{velocities2}
\end{align}
where
$W \equiv \delta\Sigma/\Sigma$
is the fractional density perturbation,
$\kappa^2 = 2\Omega\Sigma\eta$
is the epicycle frequency  expressed in terms of the  vortensity $\eta,$ and
$\bar{\sigma} \equiv \sigma +m\Omega(r) = \sigma_R + m\Omega(r) + i\gamma$ 
is the Doppler-shifted frequency with  $\sigma_R$ and $\gamma$ being real. 

Substituting  equation  (\ref{velocities1})---(\ref{velocities2}) into
the linearised continuity equation
\begin{align}
i\bar{\sigma} W = -\frac{1}{r\Sigma}\frac{d}{dr}\left(r\Sigma\delta
  u_r\right) - \frac{im}{r}\delta u_\phi
\end{align} 
yields  a  governing equation for $W$  of the form
%\begin{align}\label{goveq}
%$$ 
% \frac{d}{dr}\left(\frac{\Sigma}{\kappa^2 -\bar{\sigma}^2}\frac{dW}{dr}\right)
%+\hspace{4cm} $$
\begin{align}\label{goveq}
%\frac{d}{dr}\left(\frac{\Sigma}{\kappa^2 -
%    \bar{\sigma}^2}\frac{dW}{dr}\right)+
&\frac{d}{dr}\left(\frac{\Sigma}{\kappa^2 -\bar{\sigma}^2}\frac{dW}{dr}\right)
\notag\\
&+ \left\{\frac{m}{\bar{\sigma}}\frac{d}{dr}
 \left[\frac{\kappa^2}{r\eta(\kappa^2-\bar{\sigma}^2)}\right]
     -\frac{r\Sigma}{GM_*h^2}
 -\frac{m^2\Sigma}{r^2(\kappa^2-\bar{\sigma}^2)}\right\}W = 0.
\end{align}
This  equation leads to  an eigenvalue problem 
for the complex eigenvalue $\sigma.$ Co-rotation 
resonance occurs at $\bar{\sigma} = 0$ which requires
$\gamma =0$ for it to be on the real axis. 
Then for the equation  to remain regular,  the gradient of the  terms in
square brackets must vanish at co-rotation. This results in the requirement that
the gradient of $\eta r$ vanish there. Because the sound speed varies
with radius, this is slightly different from
the condition that the gradient of $\eta$ should vanish which applies
in the barotropic strictly isothermal case \citep{papaloizou84,papaloizou85,papaloizou89}. However, because  $\eta$ varies vary rapidly in the region of interest and 
 the modes we consider are  locally confined in radius, this difference is of no
essential consequence.
 Lindblad resonances occur when
$\kappa^2-\bar{\sigma}^2 = 0$, but as is well known, and can be seen from
formulating a governing equation for $\delta u_r$ rather than $W$,
these do not result in a singularity.

\subsubsection{Simplification of the governing ODE}
\indent It is useful to simplify equation  (\ref{goveq})  to gain further
insight. To to this we consider  modes localised around the co-rotation circle
 such that the condition 
$\kappa^2\gg|\bar{\sigma}^2|$ is satisfied. Beyond this region
the mode amplitude is presumed to be exponentially small.
 Now, recognising  that $GM_*h^2/r = c_s^2$, the
ratio of the second to last to last  term in equation  (\ref{goveq}) is
\begin{align*}
\frac{r^2\kappa^2}{m^2c_s^2}\sim\frac{1}{m^2h^2}.
\end{align*}
For a thin disc $h\ll1$, so if we consider $m=O(1)$ this ratio is
large and the last term in equation  (\ref{goveq})  can be neglected. This is
 also motivated by the fact that only low $m$ modes  are
observed in simulations. 
 Doing this and replacing $\kappa^2-|\bar{\sigma}^2|$ by $\kappa^2$,
 equation  (\ref{goveq}) reduces to the simplified form
\begin{align}\label{goveq2}
\frac{d}{dr}\left(\frac{rc_s^2\Sigma}{\kappa^2}\frac{dW}{dr}\right)
  +\left\{\frac{m}{\bar{\sigma}}\frac{d}{dr}
 \left[\frac{c_s^2}{\eta}\right]
     -r\Sigma\right\}W = 0.
\end{align}
We comment that in this form  equation  \ref{goveq2} is valid
for any fixed $c_s$  profile.
\subsubsection{Necessity of extrema}
 Multiplying equation  \ref{goveq2}
by $W^*$ and integrating between $[\,r_1,r_2]$ assuming, consistent
with a sharp exponential decay, that  $W=0$ or
$dW/dr=0$ at these  boundaries we find that
\begin{align}\label{neccessary}
\int^{r_2}_{r_1}\frac{m}{\bar{\sigma}}\left(\frac{c_s^2}{\eta}\right)^\prime|W|^2dr
=\int^{r_2}_{r_1}r\Sigma|W|^2 dr + \int^{r_2}_{r_1}\frac{r\Sigma
  c_s^2}{\kappa^2}|W^\prime|^2 dr
\end{align}
where derivatives are indicated with a prime. Since the  right hand side is real, 
the imaginary part
of the left and side must vanish. For general complex $\sigma$ this implies that
\begin{align}\label{vortmin}
\gamma\int^{r_2}_{r_1}\frac{m}{(\sigma_R+m\Omega)^2+\gamma^2}
\left(\frac{c_s^2}{\eta}\right)^\prime|W|^2dr
=0.
\end{align}
 Thus for a growing mode ($\gamma\neq0$) to exist we need
$(c_s^2/\eta)^\prime =0$ at some $r$ in $[\,r_1,r_2]$. 
For
our disc, $c_s^2$ varies on a scale  $O(r)$, but $\eta$ varies on  a scale  $H\ll r$, thus given
that the range of relative 
 variation of the vortensity  $\eta$ is of order unity, we infer that this quantity
 needs to  have stationary points in order  for there to be unstable 
 modes. In the barotropic case \cite{papaloizou89} have shown that vortensity
maxima are stable while minima are associated with instabilities. 
 We expect these conclusions to  apply here
also  because of the local nature of the modes of interest.  We can approximate 
$c_s^2\sim\mathrm{constant}$,  or equivalently  adopt a barotropic equation of
state locally without changing the character of the problem.
 Referring back to Fig. \ref{custom_dens} it is clear
that  our basic state satisfies the necessary criterion for
instability.

\subsection{Numerical solution of the eigenvalue problem}
 We  have solved the eigenvalue problem
for the  full equation  (\ref{goveq}) using a 
shooting method  that employs an adaptive Runge-Kutta integrator 
and a  multi-dimensional Newton method \citep{press92}. 
 For low $m$ ($\leq 3$) , unstable modes mainly comprise an evanescent disturbance
 near co-rotation (or the vortensity minimum at $r=r_0$) and the simple  boundary condition $W = 0$ applied
at the inner boundary $r/r_p=0.55$ and the outer boundary $r/r_p=1.5$ produces  
good results. 
\begin{figure}
\center
\subfigure[Inner edge]{\includegraphics[width=0.75\linewidth]{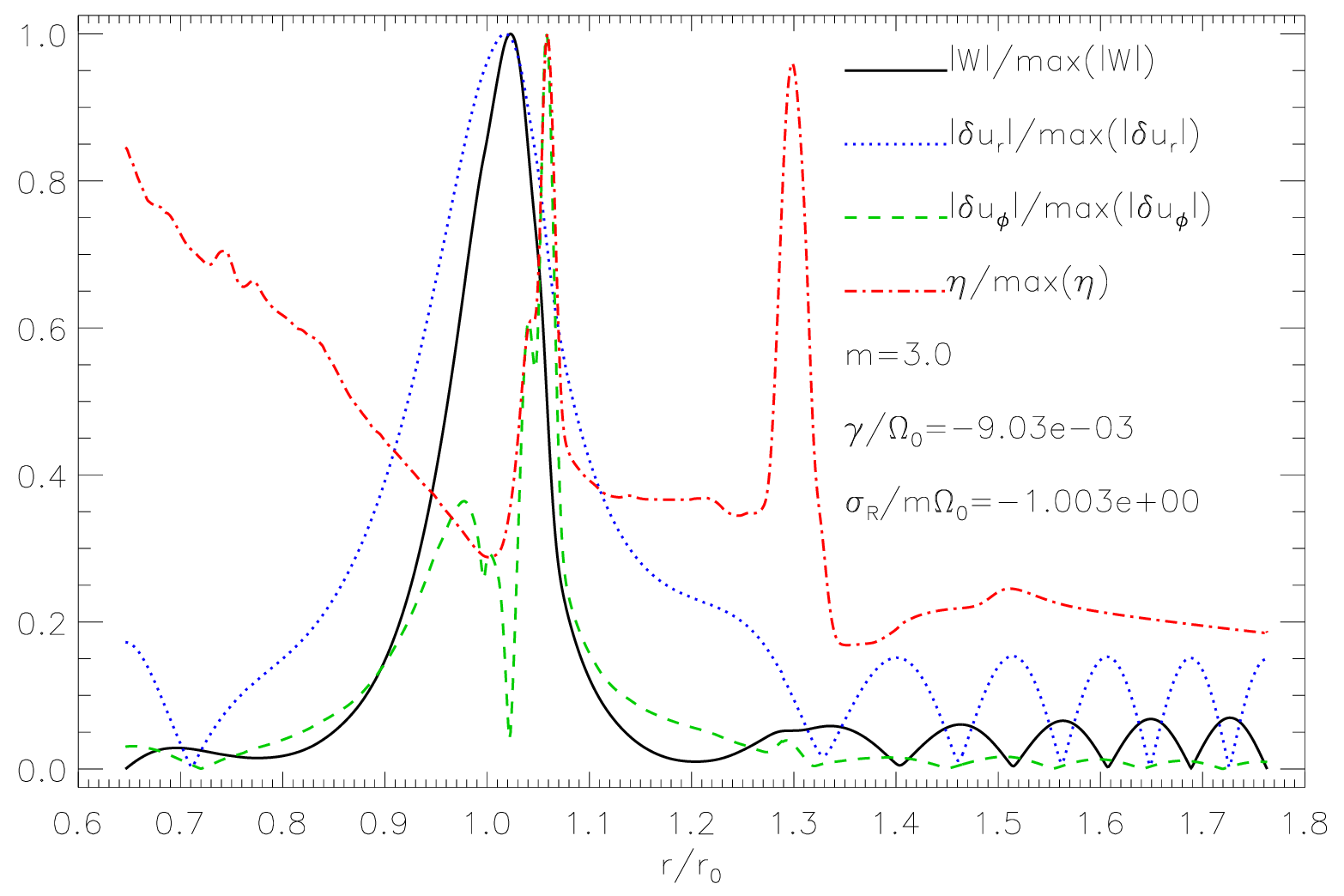}}
\subfigure[Outer edge]{\includegraphics[width=0.75\linewidth]{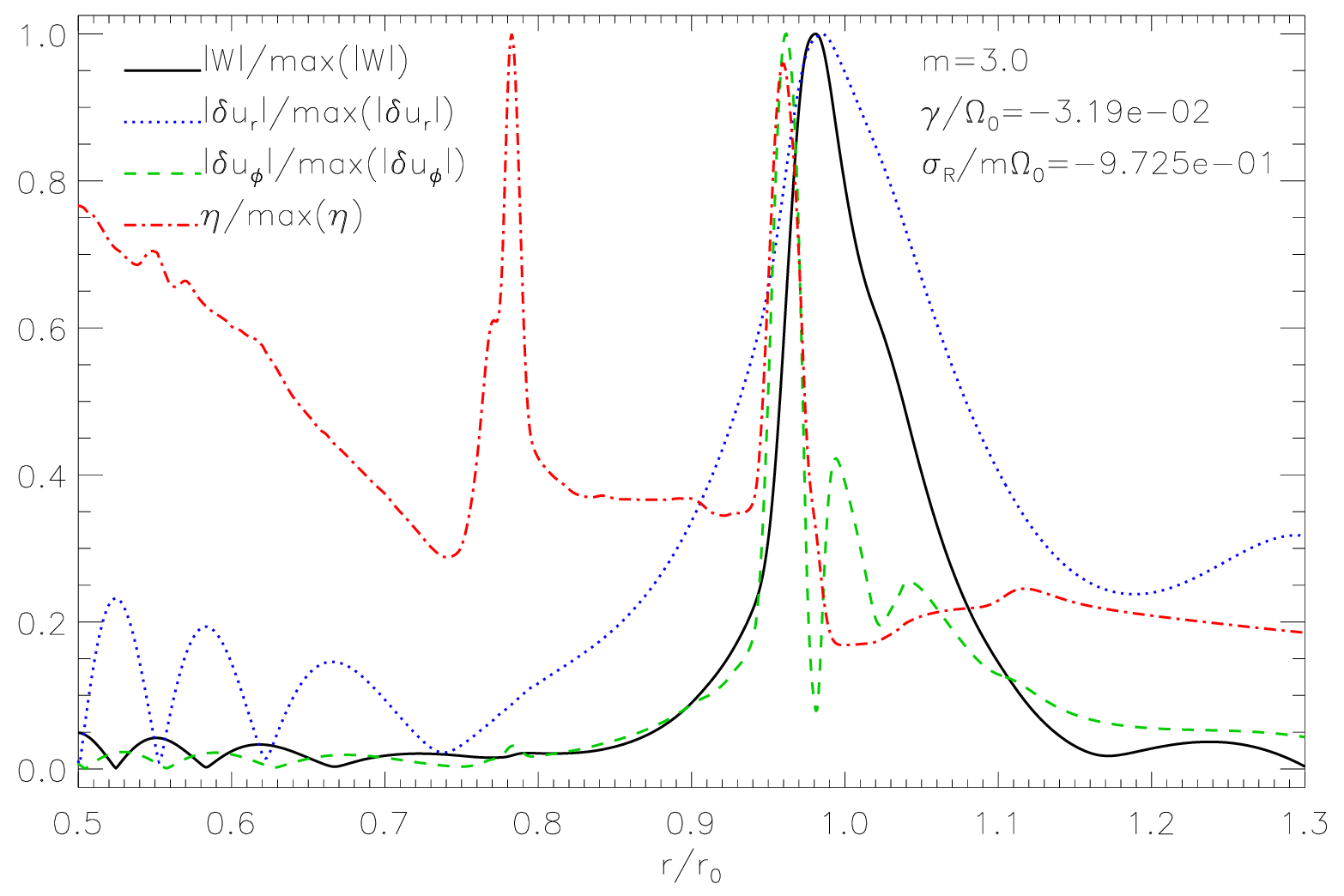}}
\caption{$m=3$ eigenmodes obtained  with the  boundary condition $W=0.$  The
perturbed quantities are plotted as $|W|$ (solid, black), $|\delta u_r|$ (dotted,blue), 
$|\delta u_\phi|$ (dashed, green) and scaled by their maximum
values in $r=[1.1,3.0]$. The background vortensity profile is also shown (dash-dot, red). $r_0=1.7,\,2.3$
correspond to the inner and outer vortensity minima, respectively.\label{s1h005}}
\end{figure}
As  $m$ increases,  the  Lindblad resonances approach  $r=r_0$
and a  significant portion of the mode is wave-like 
requiring  the application of outgoing  radiation boundary conditions.   
We determined these  using the  WKBJ  approximation (see eg.  \cite{korycansky95}). 
\subsubsection{Eigenmode calculations}
%We have determined unstable eigenmodes using the above procedures for 
%$1 \le m \le 8. $
We now discuss some example solutions to illustrate the instability of gap edges. 
We recall that the simulations indicate the ultimate  dominance of small $m$ values.
One class of mode is associated with the inner vortensity ring while another
is associated with the outer ring.

 As a typical example of the behaviour that is found for low $m,$ each type of eigenmode  for  $h=0.05$ and  $m=3$ 
is shown in Fig. \ref{s1h005}. The background vortensity profile is also shown.
The instabilities ($\gamma<0$) are associated with the  vortensity minima at the inner
and outer  gap edge, as has also been 
observed by \cite{li05} in simulations. 
  the modes are evanescent around co-rotation and the  vortensity peaks 
behave like wall through which the instability scarcely penetrates.
 The mode decays away from $r=r_0$. For $m=3$
Lindblad resonances occur at $r_L/r_0=0.76,\,1.21$  from which 
waves travelling away from co-rotation are emitted. 
 However,
the oscillatory amplitude is at most $\simeq 20\%$ of that at $r=r_0$. Hence for  low-$m$ 
the dominant effect of the instability will be due to perturbations near co-rotation. 
Increasing $m$ brings $r_L$ even closer to $r_0$, 
waves  then propagate through the gap. 
 The growth time-scale of the inner mode with $m=3$ is  $\sim14P_0,$.  
The outer mode has a growth rate that is about three times faster.  Very similar
results and growth rates were found  for $m=1$  and $m=2$. Since the instability
grows on dynamical time-scales, we expect non-linear interaction of vortices 
to occur within few tens of orbits and to affect planet migration
if the latter were also on similar or longer time-scales.

After the onset of linear instability and the formation of several vortices,
it has been observed that non-linear effects 
cause them to eventually 
merge into a single vortex \citep{valborro07}.  
This eventually interacts with the planet. 
In the fiducial simulation,
rapid migration begins at $55P_0$ 
which is compatible with the characteristic  growth 
times found  from linear theory. 

Fig. \ref{s1h005}  indicates that 
the outer edge is more unstable than inner edge. 
The vortensity peaks are of similar
height, but the inner minimum in  $\eta$  is less pronounced 
than outer because of the background profile. In this sense the outer edge
profile is more extreme, and hence less stable.

\begin{figure}
\center
\includegraphics[width=0.75\linewidth]{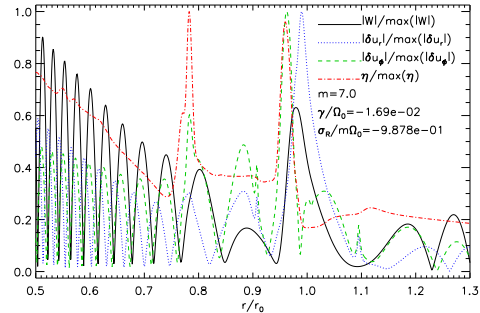}
\caption{$m=7$ eigenmodes associated with the outer edge, obtained
  with WKBJ boundary condition. The
perturbed quantities are plotted as $|W|$ (solid, black), $|\delta u_r|$ (dotted, blue), $|\delta u_\phi|$ (dashed, green) and scaled by their maximum
values in $r=[1.1,3.0]$. The background vortensity profile is also shown (dash-dot, red). $r_0=2.3$ is the radius of the outer vortensity minima.  
 \label{s1h005m7}}
\end{figure}

\indent For illustration, we show in Fig. \ref{s1h005m7} eigenfunctions
for the $m=7$ mode of the outer ring. The equivalent mode was not found for
the inner edge because high-$m$ are quenched. Radiative boundary conditions were 
adopted in this case. Although the WKBJ condition is the appropriate physical boundary condition
 , its application here is uncertain because the boundaries cannot be considered `far' from the gap.
However, solutions are actually not sensitive to 
boundary conditions as reported by \cite{valborro07}.  Note the two spikes in 
$\delta u_r$ and $\delta u_\phi$ at $r/r_0\simeq0.90,\,1.09$ which correspond to Lindblad resonances. These are not singularities as can be seen from $W(r)$ which is smooth there; other eigenfunctions were calculated from 
the numerical solution for $W$ and thus may be subject to numerical errors. We see that increasing $m$ increases the amplitude in the
wave-like regions of the mode, but the growth rate is smaller than for $m=3$. As the instability operates on dynamical time-scales, low-$m$ modes will dominate over high-$m$ modes, particularly through non-linear evolution and interaction of the former. 

We have also examined solutions with $h=0.03,\,0.04$ and 0.06. In general, as $h$ is lowered, $\gamma$ becomes more negative.  As the disc temperature is lowered with $h$, there are stronger shocks, and larger modifications to the disc profile, or steeper gradients.  Hence we expect edges to become more unstable. In the case of $h=0.03$, the rings become unstable (vortices form) before they reach an approximately axisymmetric state. 

Having understood the origin of vortices, we proceed to study
their effect on planet migration, in the type III regime. 
%\subsection{Comparison of inner/outer edges}
%\subsection{Growth rates and azimuthal wave number}
%\subsection{Application to disc-planet problem}
%\ifpdf
%  \pdfbookmark[2]{bookmark text is here}{And this is what I want bookmarked}
%\fi
% ------------------------------------------------------------------------
%%% Local Variables: 
%%% mode: latex
%%% TeX-master: "../thesis"
%%% End: 

%\include{Bump/bump}
%\chapter{Vortex-planet interactions}
%\ifpdf
%    \graphicspath{{Vortex/Figs/PNG/}{Vortex/Figs/PDF/}{Vortex/Figs/}}
%\else
     \graphicspath{{Vortex/Figs/EPS/}{Vortex/Figs/}}
%\fi
%\indent 
\section{Simulations of fast migration driven by Vortex-planet interaction}\label{mig_visc}
    We now  present hydrodynamic simulations of disc-planet
    interactions where the planet is free to migrate, so that $r_p=r_p(t)$.
     Vortices that form in the co-orbital region near
    the gap edge move through the co-orbital region
     and cause torques to be exerted on the planet.
     The interaction between
    such large-scale structures and the planet leads to non-monotonic  type III
    migration and is the focus of this section. 
      The type III torque increases with the co-orbital mass deficit $\delta m$:
\begin{align}\label{delta_m}
\delta m = 8\pi r_p B_p \left[x_sw(-x_s)-\int^0_{-x_s}w(x)dx\right]
\end{align}
\citep{masset03},
where $w=\Sigma/\omega$ and $B_p=\frac{1}{2r}\partial_r
(r^2\Omega)$ 
is the Oort constant evaluated at the planet radius $r_p$.
Hence the inverse vortensity $\Sigma/\omega$ will be central
to the discussion.\\
\indent The simulations described  below 
have computational domain $r=[\,0.4, 4.0\,]$
with resolution $N_r\times
N_\phi=768\times 2304$.
%% Some of the simulations are done with a constant kinematic viscosity $\nu.$ 
%% The viscosity is specified in units of $ \nu_0=10^{-5}r_p^2\Omega_p.$ 
The planet
is set in circular orbit after which 
the initial radial velocity is set to be 
 $v_r=-3\nu/2r.$ as expected for a steady accretion disc \citep{lyndenbell74}
 The initial surface density is chosen to be $\Sigma_0=7.0$, corresponding to a few times the
value appropriate to the minimum mass solar nebula in order to achieve rapid migration
when a typical viscosity $\nu_0=1$ is used
\citep{masset03,valborro07}. 
For most of these simulations the full planet potential is applied from
$t=0$. Similar results were obtained if the potential is switched on over 5 orbits, which
was employed in Section \ref{rings} when discussing formation of vortensity rings. In those cases, vortices were observed to form. Switching on
the planet potential over several orbits does not weaken the instability, vortex-planet interactions still occur (see below), although at a slightly
later time than if the planet is introduced at $t=0$.

\indent Type III migration is numerical challenging due to its 
dependence on flow
near the planet, one issue being the numerical resolution. \cite{dangelo05} reported
the suppression of type III migration in high resolution simulations. The main migration
feature discussed below is brief phases of rapid migration due to 
vortex-planet interaction, which does not depend on conditions very close to the planet.
Our preliminary experiments with resolutions of $N_r\times N_\phi=
192\times576,\,256\times768$ both show such behaviour, thus we believe
the higher resolution used below is sufficient to study this interaction. \\
\indent The locally isothermal equation of state implies no temperature changes
due to the planet. This could lead to mass accumulation in the Hill sphere 
and thus spurious torques from within. \cite{peplinski08a} used an equation
of state where temperature increases close to the planet. 
We have also considered this model where
the modification is applied to within the Hill sphere and
at $t=0$,  $c_s(r=r_p)$ is 18\% higher than the local isothermal value.
This again yields vortex-planet scattering, which is an indication that 
significant torque contribution during the interaction is not due to material 
inside the Hill sphere. Qualitatively, the same phenomenon is observed for
the case where the temperature increase is 38\%.\\
\indent Finally, there is the issue of softening. We set
$\epsilon=0.6H$ as before, but have considered softenings of 
$0.5H$ and $0.7H$ and both cases display vortex-planet interaction, 
although at earlier times when $\epsilon$ is lower. This is expected since
smaller softening has the same effect as increasing the planet mass, and thus
producing a more unstable disc. 
\subsection{Viscosity and vortices}
%\indent It is well-known that ring structures in accretion discs are
%subject to dynamical instabilities such as the Paploizou-Pringle
%instability \citep{papaloizou84,papaloizou85,papaloizou89, lovelace99, li00}
%and 
We have shown above that the ring structures formed by a Saturn mass
planet in our case are linearly unstable. 
It has been shown
through numerical simulations that such instabilities
are expected to lead to vortex production
\citep{li01} and this has been seen 
 when  steep surface density  gradients are present outside the co-orbital
 region of a lower mass planet
\citep{koller03, li05}. Non-monotonic migration
was observed by \cite{ou07} when a vortex is present inside the 
co-orbital region, but the role of the vortex was not analysed in
detail.

 Fixed-orbit simulations of
disc-planet interactions show that a dimensionless viscosity of order $10^{-5}$ suppresses
vortex  formation \citep{valborro07}.
 As a consequence  studies using viscous discs
typically yield smooth migration curves. 
We have confirmed this in our case  with numerical simulations.\\

\subsection{Dependence of the migration rate on viscosity}
\begin{figure}
 \centering
 \includegraphics[trim=0cm 0cm 6cm 4cm, clip, width=1.0\linewidth]{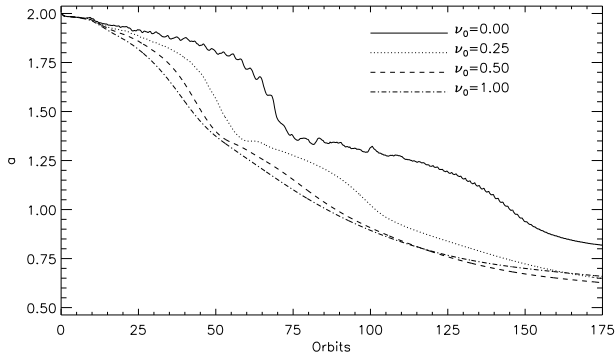}
 \caption{Type III migration as a function of viscosity \label{orbit2_visc}}
\end{figure}

%Here we focus on the interaction between vortices produced
%by instability of the vortensity rings  and  the planet.
We first study type III migration as a function of viscosity.
The effect of vortices appears at low viscosities. 
Fig. \ref{orbit2_visc} shows the orbital semi-major axis  $a(t)$
for viscosities {\bf $\nu_0=0$ to $\nu_0=1$}.  As the orbit is very nearly circular, $a(t)$ is always close
to the instantaneous orbital radius $r_p(t)$. 
 We considered a case with $\nu_0=10^{-3}$ which showed almost
  identical $a(t)$ curve to $\nu_0=0$. The numerical viscosity is thus
  of order $\nu=O(10^{-8})$ which is much smaller than the typically
  adopted physical viscosity of $\nu=10^{-5}$ in such simulations.

With the standard viscosity $\nu_0=1$, $a\to a/2$ in
less than $100P_0$, implying type III migration
\citep{papaloizou06}.  Comparing different $\nu_0$, $a(t)$ is indistinguishable for
$0<t\lesssim15$, since viscous time-scales are much longer than
the  orbital time-scale.  At $t=20$, $|\adot|$ increases with $\nu_0$.  In the
limit $\nu\to 0$, the horse-shoe drag
for a fixed orbit  is  $\propto \nu$
\citep{balmforth01,masset02}. However, in this case $\adot\neq 0$ there
is a much  larger  rate-dependent  torque
responsible for type III migration, for which explicit
dependence on viscosity has not been demonstrated analytically. 

  Migration initially accelerates inwards
($\adot,\,\ddot{a}<0$) and subsequently  slows  down at $r \sim 1.4$
(independent of $\nu_0$). For $\nu_0=1.0,\,0.5$ migration proceeds
 smoothly, decelerating towards the end of the simulation
 at which point the orbital radius has decreased by a factor of $\sim 2.7$ 
Migration curves  for  $\nu_0=1.0$ and $\nu_0=0.5$ are quantitatively similar. 
Lowering  $\nu_0$ further  enhances the deceleration at $r \sim 1.4$
until in the inviscid limit  the  migration stalls before eventually
restarting. 

Despite  
differences in detail, the overall extent of the
orbital decay in all of these cases  is similar. This is
expected in the model of  type III migration where the torque is 
due to circulating fluid material switching from $r_p-x_s\to
r_p+x_s$. 
In this model the extent of the  orbital decay should not depend on the nature of the
flow across $r_p$, but only on the amount of disc material
participating in the interaction, or equivalently the disc mass
and this does not depend on $\nu$.
 On the other hand, the flow may not be a smooth function
of time with migration proceeding  through  a series  of  fast and slow episodes
as observed in Fig. \ref{orbit2_visc}. Our argument is only valid
if migration proceeds via the type III mechanism. We have not explored
higher viscosities ($\nu$ of more than a few times $10^{-5}$) in detail but migration
time-scales of test cases indicate that in this regime, type III
migration is not operating.

\subsection{Stalling of type III migration }
\begin{figure}
 \centering
 \includegraphics[width=1.0\linewidth]{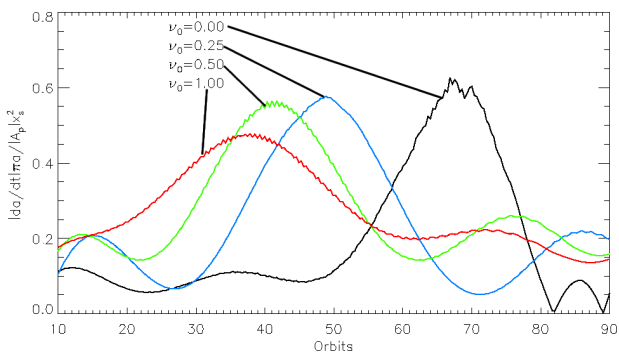}
 \caption{Evolution of $\chi$, the ratio of libration-to-migration time-scale as a function of viscosity.
 Libration time is measured at $r_p-r=x_s$ and migration time-scale is that across $x_s$. 
 \label{timescales_plot}}
\end{figure}

The issue discussed here  is what inhibits the growth of
$|\adot|$? Descriptions of (inward) type III migration usually assume that 
the libration time at $r_p-x_s$ is much less than
the time to migrate  across the co-orbital region \citep{masset03}. This implies that
\begin{align}\label{slow_migration}
\chi\equiv\frac{|\adot|\pi a}{|A_p|x_s^2}\ll 1,
\end{align}
where $A_p={1/2}({\p\Omega}/{\p r})$ at $r_p(t)$ and
$a$ is the changing semi-major axis. \cite{papaloizou06}
present a similar critical rate, but  with the same dependence on
$\Omega,\,a,\,x_s.$ 
If equation  (\ref{slow_migration}) holds, co-orbital material is trapped in  
libration on horse-shoe orbits and migrates with the planet. When
$\chi\gtrsim 1$ the horseshoe  region shrinks to a tadpole, and material is
trapped in libration about the L4 and L5 Lagrange points
\citep[as observed by][]{peplinski08b}.
This can tend to remove the co-orbital mass deficit $\delta m$ \footnote{We can 
  regard the process of changing from a partial gap extending for nearly
  the whole azimuth to one with a smaller azimuthal extent, as gap
  filling.}  which reduces the migration  torque.
   Comparing $\chi$ for
cases shown in Fig. \ref{timescales_plot}, it is clear that  migration
with $\chi \ll 1$  ( hereafter in this context termed slow
migration even though it may be much faster
than type II migration) does not always hold, with $\mathrm{max}(\chi)\sim0.6$ being
comparable for different $\nu_0.$ By following the evolution of a
passive scalar we checked that horse-shoe material no longer migrates
with the planet when $|\adot|$ is large. This
occurs for all $\nu$ but only the low viscosity cases exhibit
stalling. Hence, while horse-shoe material is lost due to
fast migration, it is not responsible for stopping it.\footnote{One may intuitively
  expect migration to be self-limited.}
  By examining the inviscid case in detail later, we show that
the stopping  of migration is due to the flow of a vortex across the co-orbital region,
where some of it becomes  trapped in libration.
%%%%%%%%%%%%%%%%%%%%%%%%%%%%%%%%%%%%%%%%%%%%%%%%%%%%%%%%%%%%%%%%%%%%%%%%%%%%%%
%% DIMENSIONLESS MEASURE OF MIGRATION RATE : RATE/CRITICAL RATE. v.s. NU_0
%%%%%%%%%%%%%%%%%%%%%%%%%%%%%%%%%%%%%%%%%%%%%%%%%%%%%%%%%%%%%%%%%%%%%%%%%%%%%%
%% critical rate adot_f given by papaloizou 06 in PPV. the figure is
%% located at /data/stardisk/mkl23/hyades/viscosity/timescales_plot_crit.ps
\subsection{The connection between the vortensity and fast migration}
\begin{figure}
  %\center
  \subfigure[\label{visc1d_evolA}]{\includegraphics[width=0.5\linewidth]{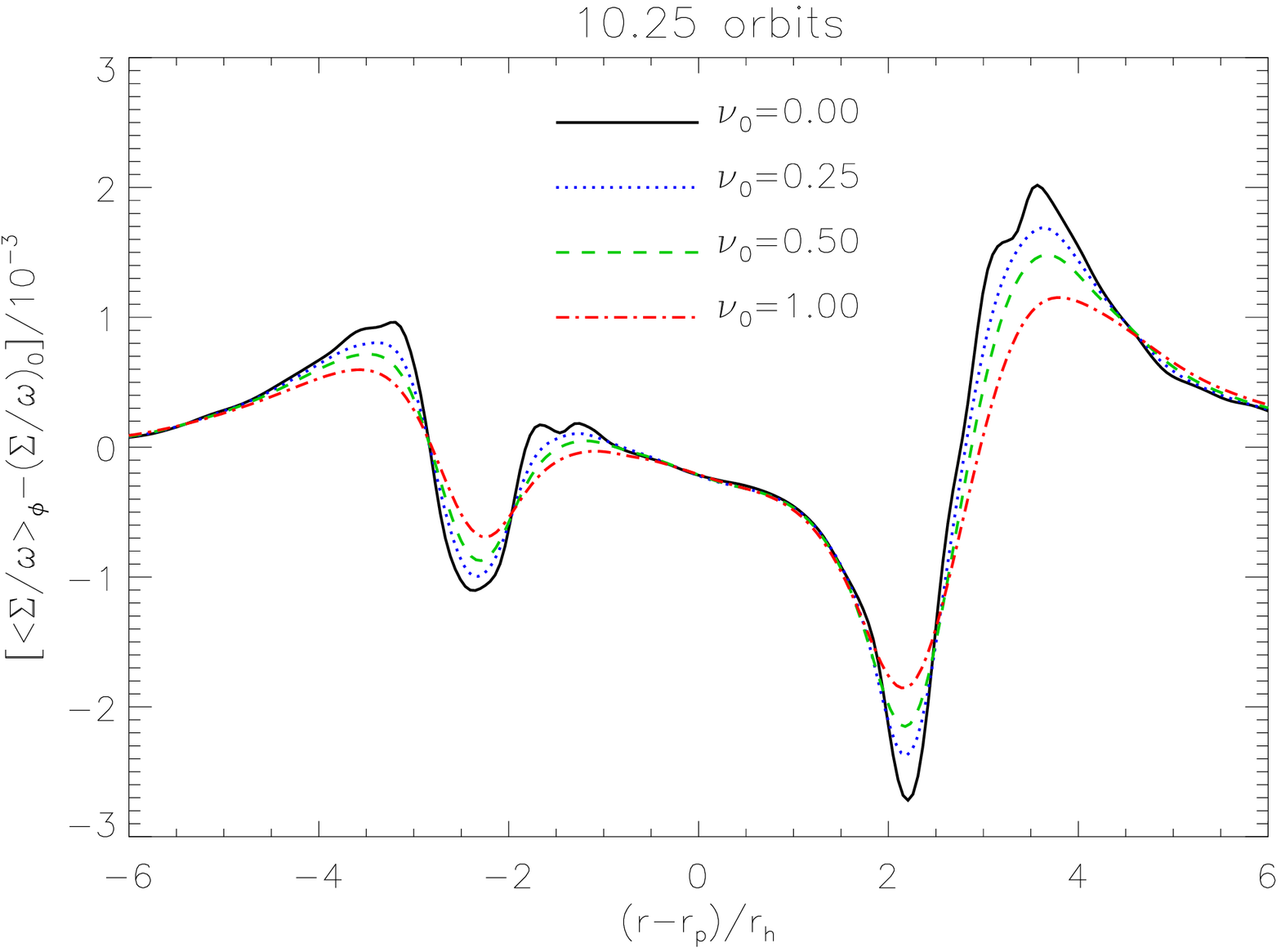}}\subfigure[\label{visc1d_evolB}]{\includegraphics[width=0.5\linewidth]{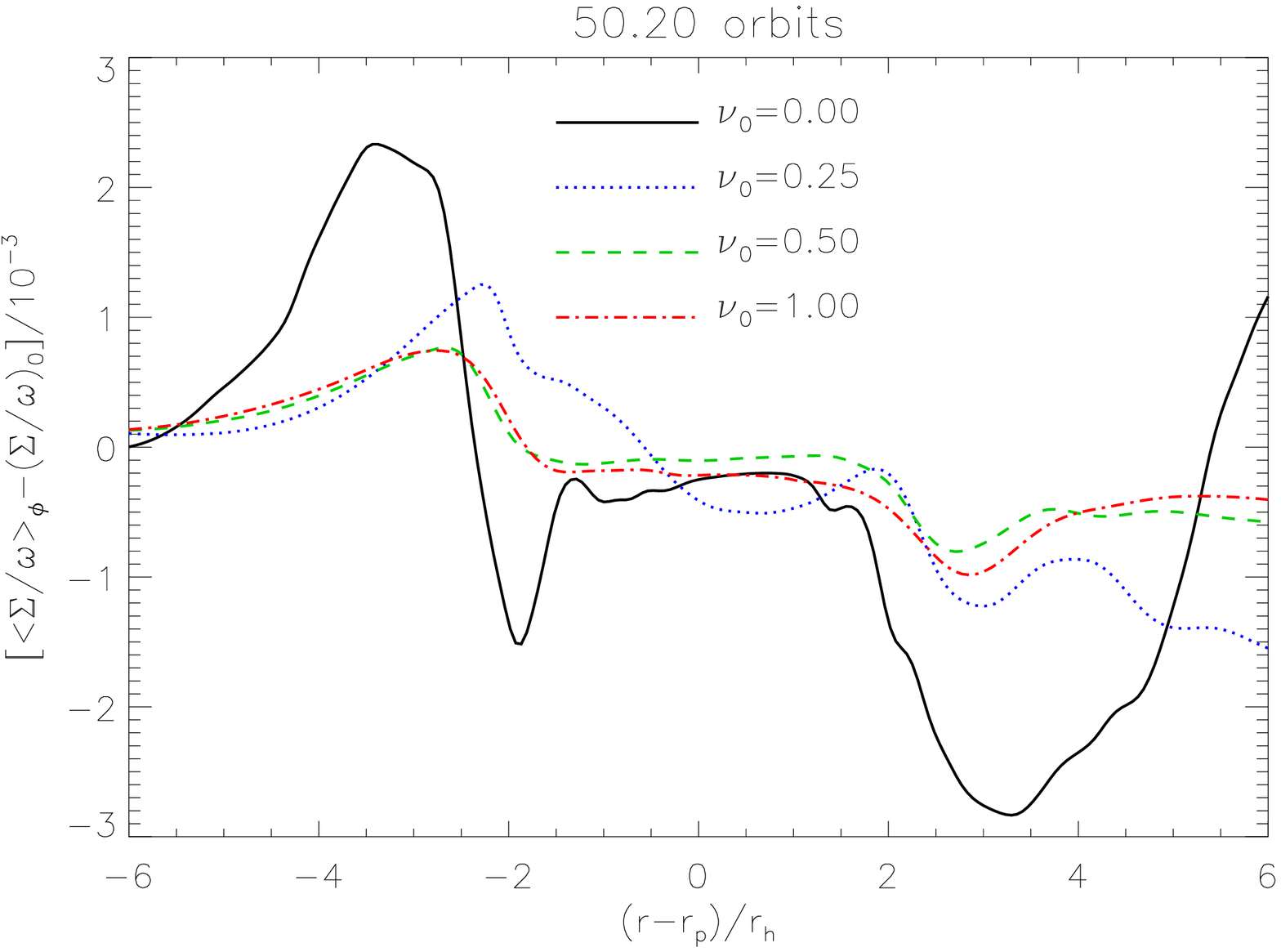}}\\
  \subfigure[\label{visc1d_evolC}]{\includegraphics[width=0.5\linewidth]{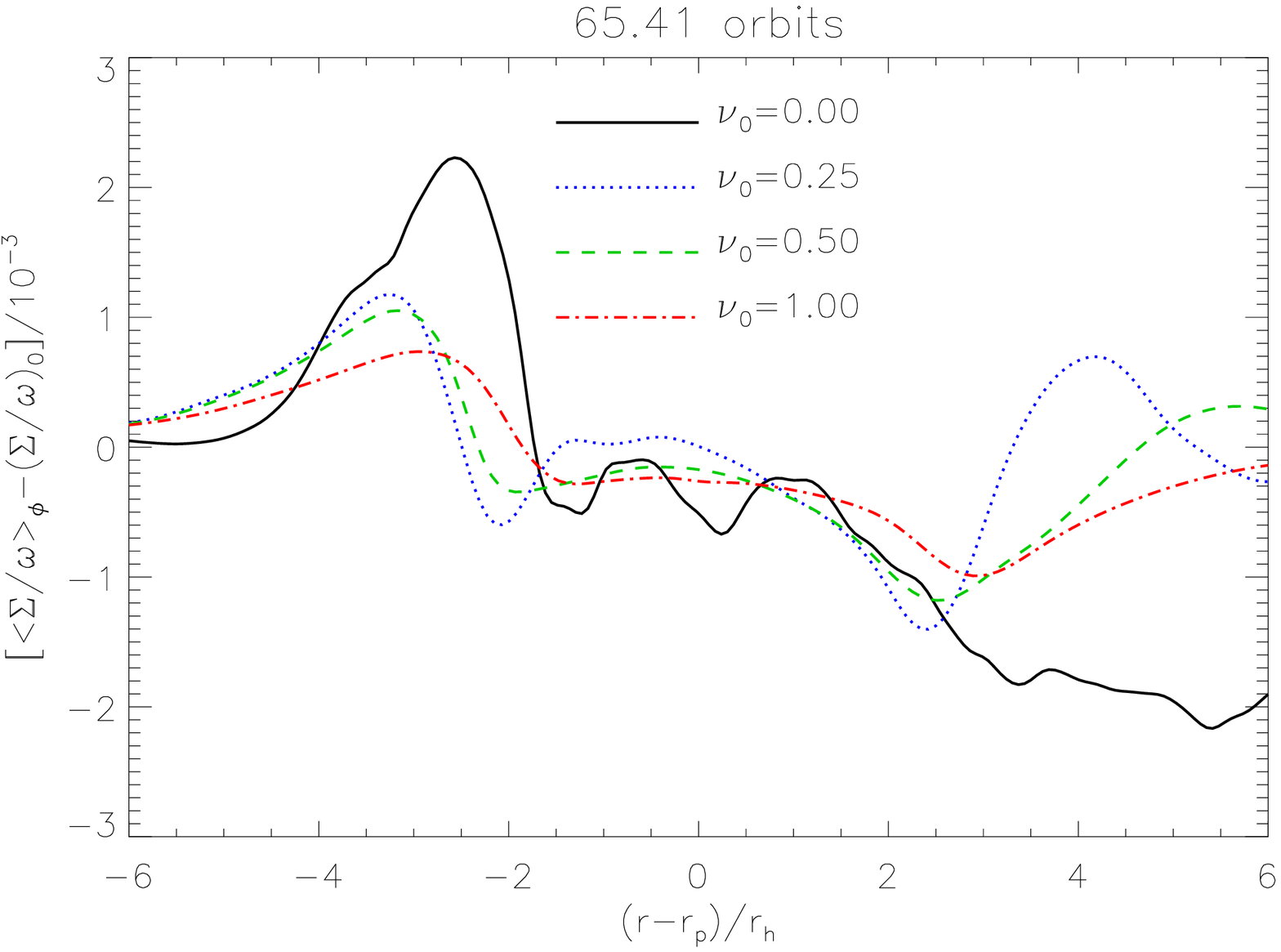}}\subfigure[\label{visc1d_evolD}]{\includegraphics[width=0.5\linewidth]{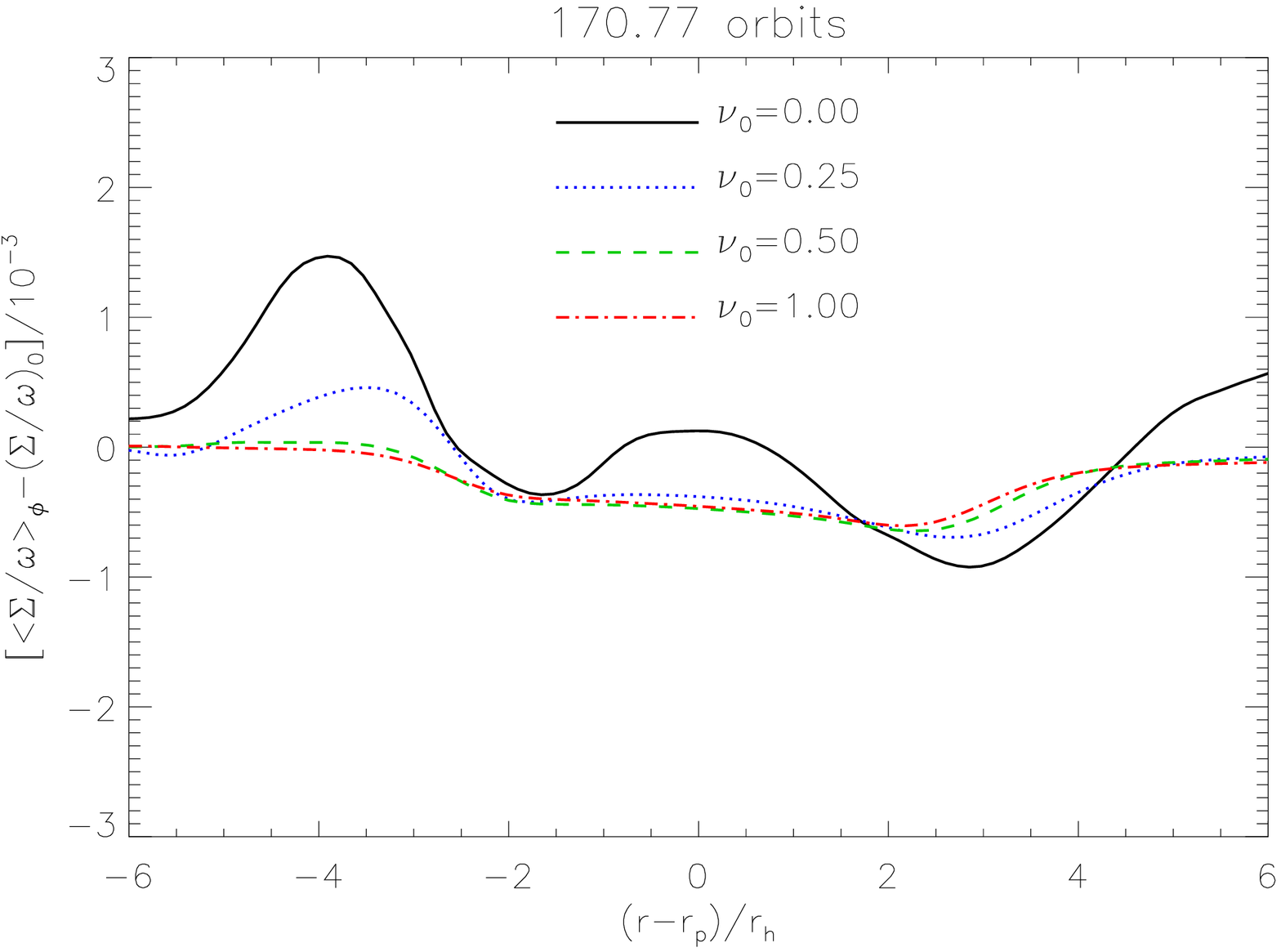}}
\caption{$\Sigma/\omega$-perturbation for different viscosities for the
 case $\Sigma_0 = 7$ and  $M_p=2.8\times10^{-4}$.  Here we use
  $\omega = r^{-1}\p_r(ru_\phi)$ to approximate the  vorticity.  This is valid since
  $|u_r|\ll|u_\phi|$. \label{visc1d_evol}} 
\end{figure}

The difference between the value
of the inverse of the vortensity  $\Sigma/\omega$ evaluated
in the co-orbital region and the value  associated with material that passes
from one side of the co-orbital region to the other defines the co-orbital mass
deficit $\delta m$ in \cite{masset03}.
It is often assumed that the vorticity is slowly varying so that
the difference in the values of inverse vortensity  reduces, to within a scaling
factor, simply to the difference in the values of the surface density.\\
\indent Although  \citet {masset03} assumed
steady, slow migration in the low viscosity limit, it is nevertheless
useful to examine its evolution in relation to the migration of the planet
(Fig. \ref{orbit2_visc}).
 Fig. \ref{visc1d_evol} shows the azimuthally averaged
$\Sigma/\omega$-perturbation following planet migration. Introducing
the planet modifies the co-orbital structure on orbital
time-scales. Vortensity rings develop
at $r-r_p\simeq\pm 2r_h$ for $t\lesssim10$
 (Fig. \ref{visc1d_evolA}).  We showed above that  this initial modification is due to spiral shocks
extending into the horse-shoe region  and also that  the ring-structure is
unstable to the production of vortices.\\ 
 \indent Increasing $\nu$ reduces the rings' amplitude,
but their  locations are unaffected. Taking the length-scale of interest as
$l=r_h\simeq 0.1$, for $\nu_0=1$ the viscous time-scale is $t_d=l^2/\nu\simeq 56P_0$.
hence at $t\lesssim10$ viscous diffusion is not significant even
locally. Thus ring-formation is not sensitive to the value of $\nu.$ 
We note the
correspondence between the similarity of  the $\Sigma/\omega$ profiles and
similarity in $a(t)$ for different $\nu$ in the initial phase. That 
is, the co-orbital disc structure determines migration \citep{masset03}.  
Dependence on the value of the  viscosity  is seen  beyond  $t=50$ (Fig. \ref{visc1d_evolB}),
producing much smoother (and similar) profiles for
$\nu_0=0.5$ and  $\nu =1.0.$\\
 \indent For a fixed orbit, we would expect profiles to
be smoothed on a viscous timescale.
When there is migration, we should also consider
advection of vortensity (in the planet frame).
Following the planet, the
perturbed profile can be smooth if the planet migrates through
the background without carrying its original co-orbital
material. This can be regarded as the loss of horseshoe material due to fast
migration. Any deviation from the background must then be due to local
vortensity generation/destruction  as the planet moves through
(e.g. shocks). \\
\indent In Fig. \ref{visc1d_evolB} only the inviscid case
is still in slow migration, and only this disc retains the
inner ring (low $\Sigma/\omega$). This  suggests that the
inner vortensity ring inhibits inward migration.  In terms of $\delta
m$, for $\nu\neq0$ the planet resides in a gap (co-orbital
$\Sigma/\omega$ is less than that at the inner separatrix, or $\delta
m>0$) whereas in the inviscid case $\delta m\sim 0$.\\
\indent Consider the
$\nu=0$ case. The outer vortensity ring has widened to $\sim 2r_h$ (c.f.
Fig. \ref{visc1d_evolA}). It is centred at 
$3r_h$ so that  co-orbital dynamics may not account for it. However,
the migration implies a flow of material across $r_p$ from the interior region.
 The increased region of
low $\Sigma/\omega$ exterior to the planet may be due to this flow.
 Notice the high-$\Sigma/\omega$ ring at $+4r_h$ in
Fig. \ref{visc1d_evolA} is no longer present in
Fig. \ref{visc1d_evolB} because this ring is not co-orbital and
therefore does not migrate with the planet.\\
\indent At $t=65$ (Fig. \ref{visc1d_evolC}) the $\nu_0=0.5$ and $\nu=1.0$
cases continue smooth rapid migration (Fig. \ref{orbit2_visc}) with qualitatively unchanged
profiles. 
For $\nu_0=0.25$, characteristic vortensity double-rings
re-develop after a stalling event at $t\sim60$
(Fig. \ref{orbit2_visc}). 
The peaks
and troughs of $\Sigma/\omega$ recover forms that are close to those in the initial phase
(Fig. \ref{visc1d_evolA}). At this time the inviscid case is in rapid migration.\\
%{\bf ****IS THE ABOVE TRUE**********. yes}
\indent Fig. \ref{visc1d_evolD} shows  the final
$\Sigma/\omega$-perturbation profiles . Viscous cases are in slow
migration, and have  much  smoother profiles. 
This can be due to
diffusion (since we are at late stage of evolution) and/or migration
across the background,  in both situations there is little disc material carried by the
planet. %that participated in previous fast-phases. 
The inviscid case is also in slow
migration but retains the double-ring structure. This indicates two
possible co-orbital configurations which slow down type III
migration.\\
\indent In this section we reviewed the disc
 vortensity evolution and its connection to the state of migration.
 A correlation between migration (fast, slow) and
co-orbital structure is apparent. The effective action of viscosity appears to be through its 
modification of  the disc structure, rather than  associated viscous torques  acting on
the co-orbital region. 

\subsection{Evolution of the co-orbital region}
\begin{figure}
  \center
  \subfigure[$\nu_0=0.5$\label{nu0d5}]{\includegraphics[scale=.4]{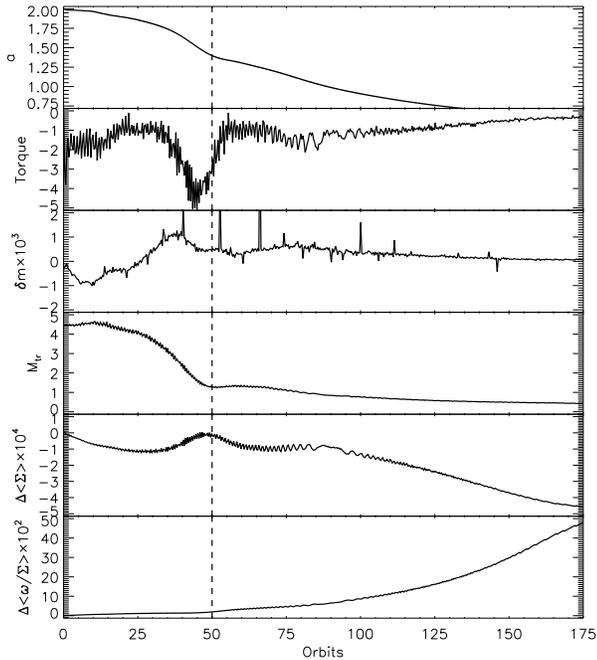}}
  \subfigure[$\nu_0=0$\label{nu0}]{\includegraphics[scale=.4]{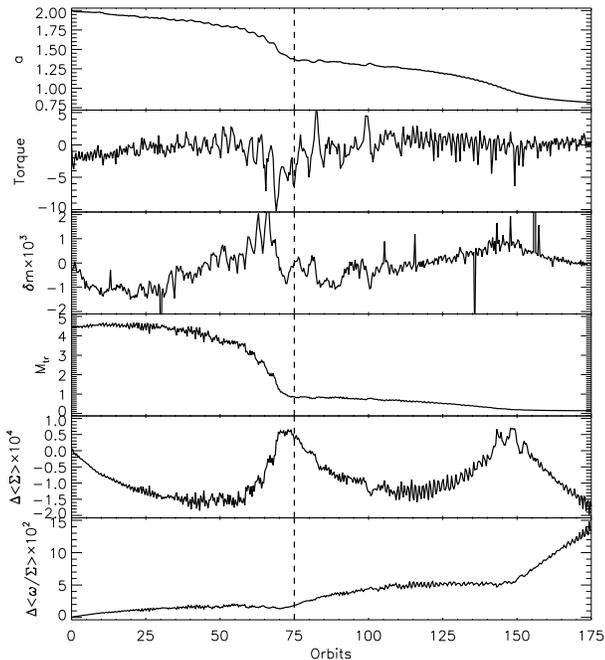}}
\caption{Effect of viscosity  on co-orbital evolution. 
The time evolution is illustrated for the model with $\nu=0.5$ (upper panel)
 and for the inviscid model (lower panel).  From top to bottom of each panel the evolution of
 the orbital radius, $a(t),$ the  co-orbital mass deficit $\delta m$, the  torque, the tracer mass
  $M_\mathrm{tr}$,  the surface density $\langle\Sigma\rangle$ and the vortensity
  $\langle\frac{\omega}{\Sigma}\rangle$ are plotted. Angle brackets denote space
  averaging over the annulus $[r_p- 2.5r_h,r_p+2.5r_h$ and $\Delta$ denotes perturbation
  relative to $t=0$. The vertical line indicates stalling of  migration.
  \label{coorb_plot}}  
\end{figure}

 In Fig. \ref{coorb_plot} we illustrate the  evolution of the following
quantities associated with the co-orbital region that are related to the migration
torque induced on the planet: 
\begin{enumerate}
\item The  gravitational torque acting on the planet due to fluid with  $|r-r_p| \le 2.5r_h$,
  excluding fluid within $r_h$ of the planet. This includes co-orbital material and
  orbit-crossing fluid, the latter being responsible for the type III 
  torque.  
\item The  co-orbital mass deficit $\delta m$ is computed from
  azimuthally averaged, one-dimensional disc profiles.  We
  took the separatrices to be at $|r-r_p| = 2.5r_h.$   
\item $M_\mathrm{tr}$, the mass of a passive scalar initially placed
  such that  $|r - r_p|= 2 r_h$ \footnote{Due to the uncertainty in the horse-shoe half-width
  $x_s$ we put the tracer  well within the region defined by
  $x_s=2.5r_h$ to ensure it  is
  co-orbital.}. Note that  $M_\mathrm{tr}=\mathrm{const.}$ if it migrates with
  the planet.
\item The average density $\langle\Sigma\rangle$ and vortensity
  $\langle\omega/\Sigma\rangle$ of the region $|r-r_p| < 2.5r_h.$
\end{enumerate}

The evolution of viscous and inviscid cases  is  qualitatively  similar up to the  stalling
indicated by vertical lines in Fig. \ref{coorb_plot}.
 Prior to this
there is  a rapid migration phase associated with large a negative torque
which we find  originates from
material crossing the planet orbit. Co-orbital torques
are oscillatory and the period/amplitude is longer/larger for  $\nu_0=0$
than for $\nu_0=0.5.$  During rapid migration phases, the inviscid torque is twice
as negative than in the viscous case. While the torque in a viscous disc
remains negative, torques in an inviscid disc can be positive
due to the formation of  large-scale vortices.  Note that the torque
does not originate from within the Hill sphere since it is excluded
from the summation. 

%\subsection{Flow through the coorbital region}
Migration is slow until
sufficient difference builds up between co-orbital and circulating flow at which point 
there is  a sudden flow-through the co-orbital region. At the same time
there is significant loss of the original horse-shoe material
($M_\mathrm{tr}$ decreases by $\sim 80\%$). The flow-through is
reflected in $\nu_0=0.5$
($\nu_0=0$) by a $\sim 17\%$ (36\%) increase in
$\langle\Sigma\rangle$ from $t=25\to50$ ($t=50\to75$). In the case of zero viscosity,
this material is in the form of a high surface density vortex.
 We note from
Fig. \ref{nu0}  that there is a repeated episode where there is a fall
followed by a rise  in
$\langle\Sigma\rangle.$ However,  in the late stages of evolution of
 the case with $\nu_0 =0.5,$
$\langle\Sigma\rangle$ decreases monotonically and the
migration is slower.

%As expected, and can be seen in Fig. \ref{coorb_plot}   phases of rapid migration 
%are associated with a large positive co-orbital mass deficit.

The co-orbital mass deficit is initially negative\footnote{
Ring structure in the vicinity of the separatrix means that the sign of $\delta 
m$ is sensitive to the adopted value of $x_s.$  Here we regard
$\delta m$ as a representation of gap depth, so we fix $x_s=2.5r_h$.} 
as the rings develop. It subsequently  increases resulting in
the onset of  type III migration  and is most positive during  the following rapid
migration phase, with  peak values $\delta m\times10^3 \sim1,\,2$ for
$\nu_0=0.5,\,0$ respectively. $\delta m\times10^3$ then falls
to $+0.5$ in the viscous case but to $\lesssim 0$ 
in the inviscid case. In the latter case, material 
flowing into the co-orbital region  removes the co-orbital mass deficit and 
 type III is  suppressed; migration experiences a more abrupt stall. $\delta m$  increases
again for $\nu_0=0$ while it remains approximately constant for
$\nu_0=0.5$ and decreases towards the end. Type
III migration can re-start in the inviscid disc but  such  behaviour is  not observed  for large viscosity.
Type III is  not operating in the late stages
of  the viscous case, in contrast to the inviscid evolution where
fast type III
migration  is recurrent, faster
than in the cases with applied viscosity,
and is associated with large values of $\delta m$. 

The above discussion shows that the magnitude of the
applied  viscosity  is  significant in
determining the character of the migration.
This is because the form of the flow through the co-orbital
region is sensitive to the choice of viscosity.  In particular, for high viscosity,  this flow is smooth and
there is less disruption of the co-orbital region. 
%\begin{landscape}
%\end{landscape}
%\begin{figure}
%\center
%\subfigure[$\nu_0=0$\label{angmom_dens14}]{\includegraphics[scale=.25]{angmom_dens14.ps}}
%\subfigure[$\nu_0=0$\label{angmom_mass14}]{\includegraphics[scale=.25]{angmom_mass14.ps}}
%\subfigure[$\nu_0=0.5$\label{angmom_dens15}]{\includegraphics[scale=.25]{angmom_dens15.ps}}
%\caption{ $\langle\Sigma rv_\phi\rangle$ and total mass of the co-orbital region
 % (middle), the ring just inside (inner) and the ring just outside
 %it.}
%\end{figure}
\section{Vortex-planet interaction}\label{inviscid}
We now focus on the inviscid case where the role
of vortices significantly affects the migration.
 We consider three phases  apparent from  (Fig. \ref{orbit2_visc}).
These are: 
a) $t\lesssim 45$ (slow migration); $b) 60 \lesssim t\lesssim 70$
(rapid migration); c)  $t\sim 75$ (stalling) and d) $75\lesssim t\lesssim
110$ (second phase of slow migration). 
  Typical migration rates at various times are
$\adot(25)\sim-3\times10^{-3}$; $\adot(65)\sim -2\times 10^{-2}$
and $\adot(85)\sim -2\times 10^{-3}$.
The vortex-induced rapid migration (phase b)) 
is almost an order of magnitude faster than
the phases a) and d). Hence we refer to the latter as slow migration, but
more accurately they are migration associated with gap formation. They
are not necessarily slow in comparison to type I or type II migration.

\begin{figure}
\center
\subfigure{\includegraphics[width=0.5\linewidth]{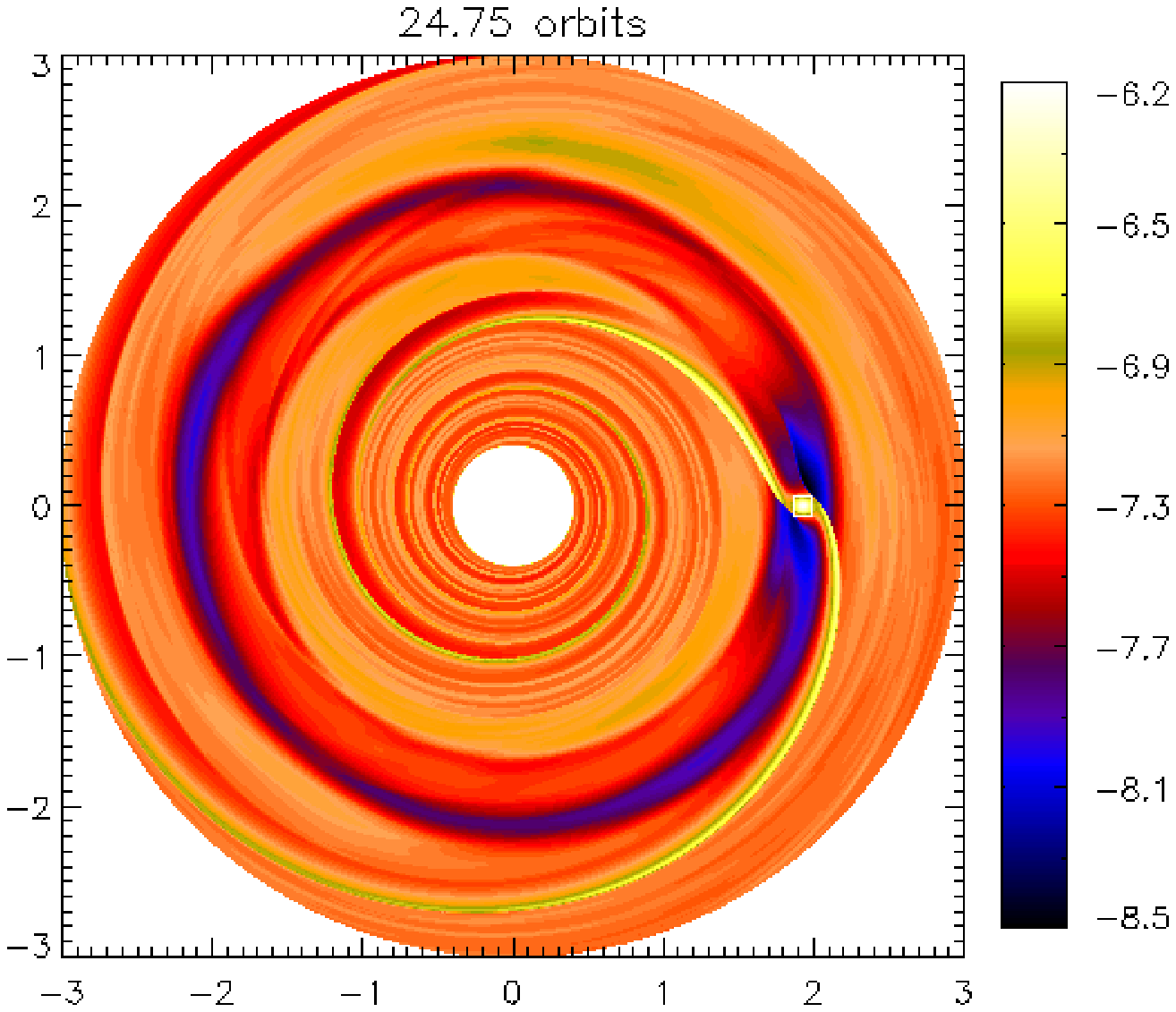}}\subfigure{\includegraphics[width=0.5\linewidth]{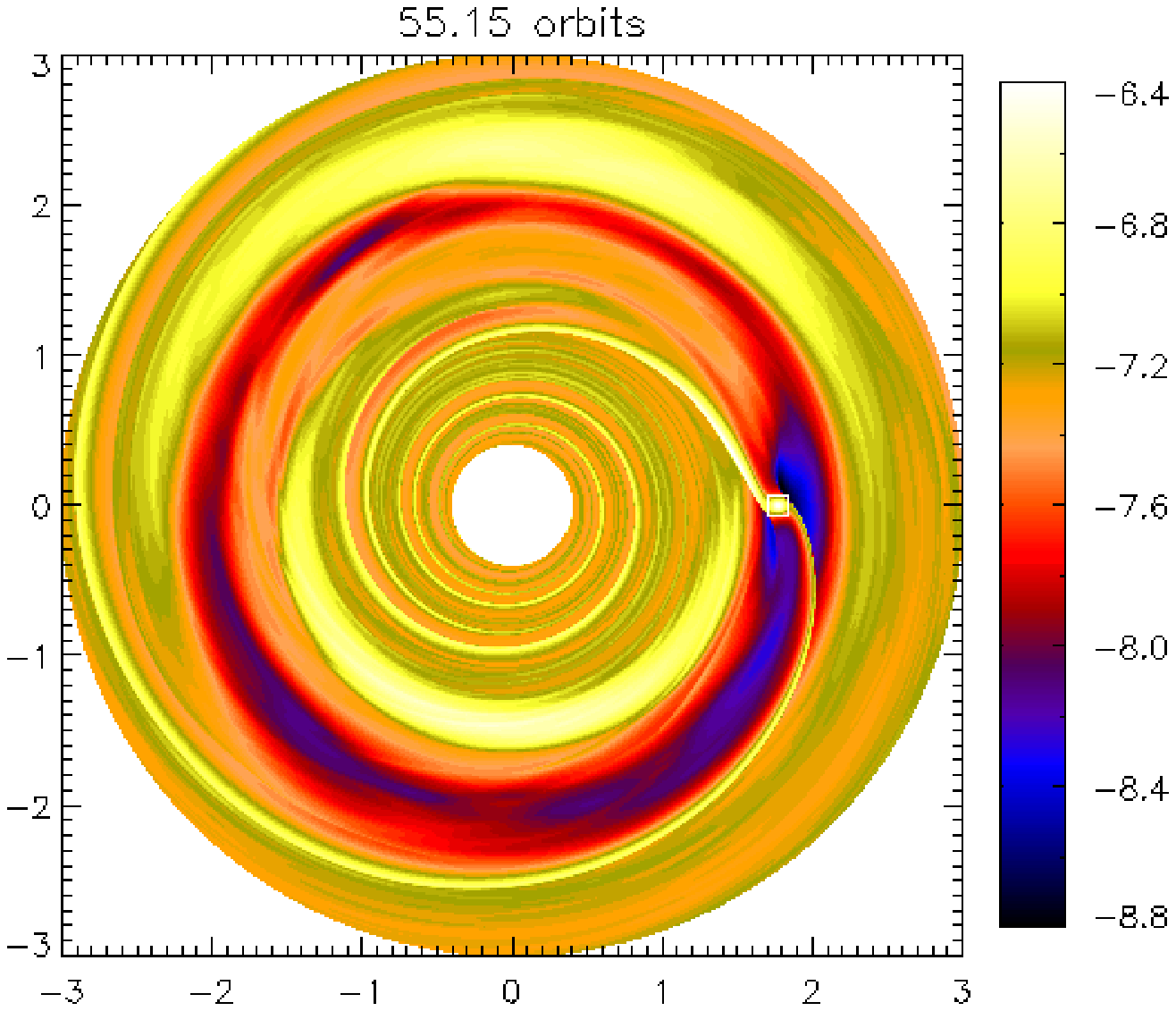}}\\
\subfigure{\includegraphics[width=0.5\linewidth]{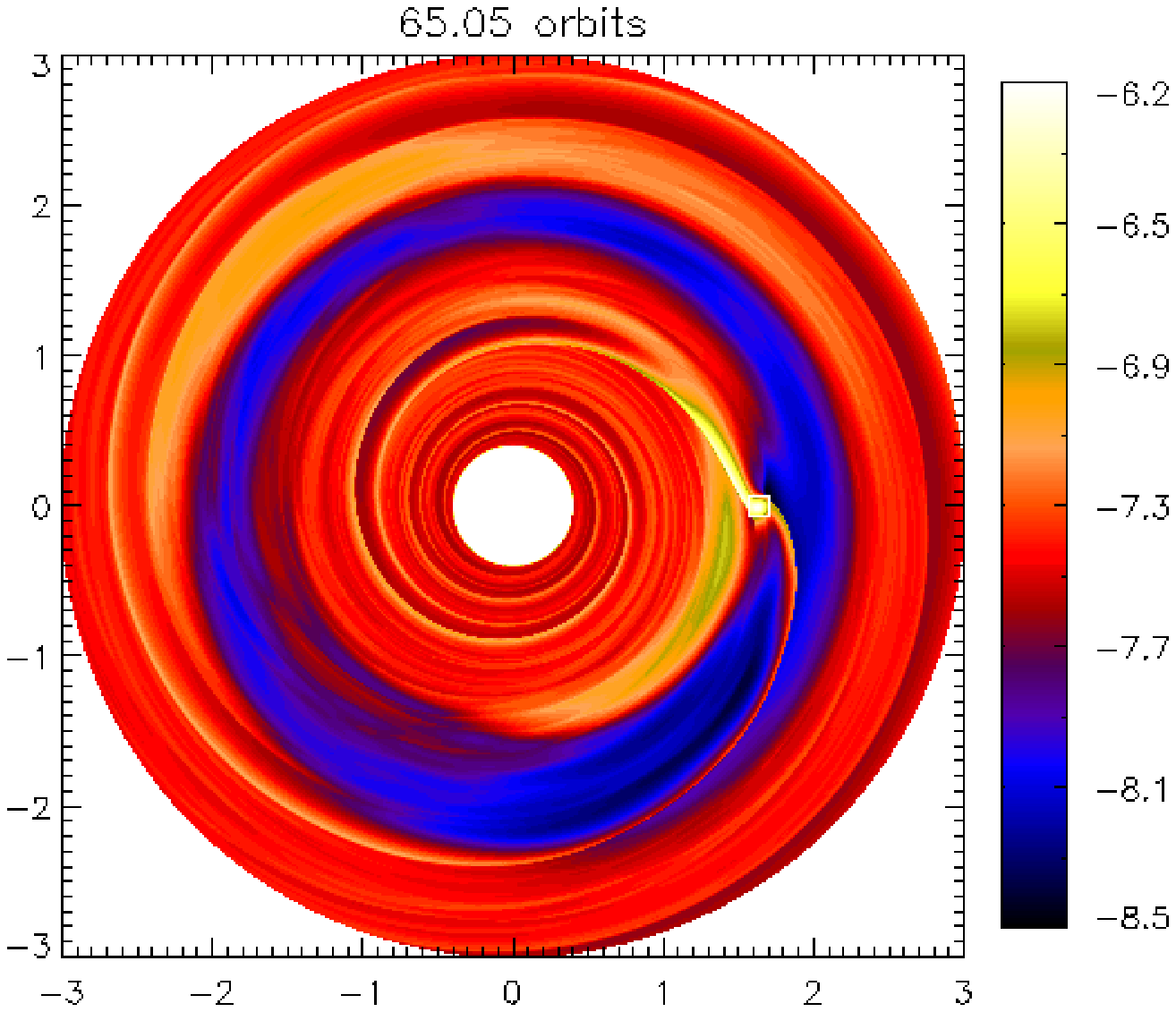}}\subfigure{\includegraphics[width=0.5\linewidth]{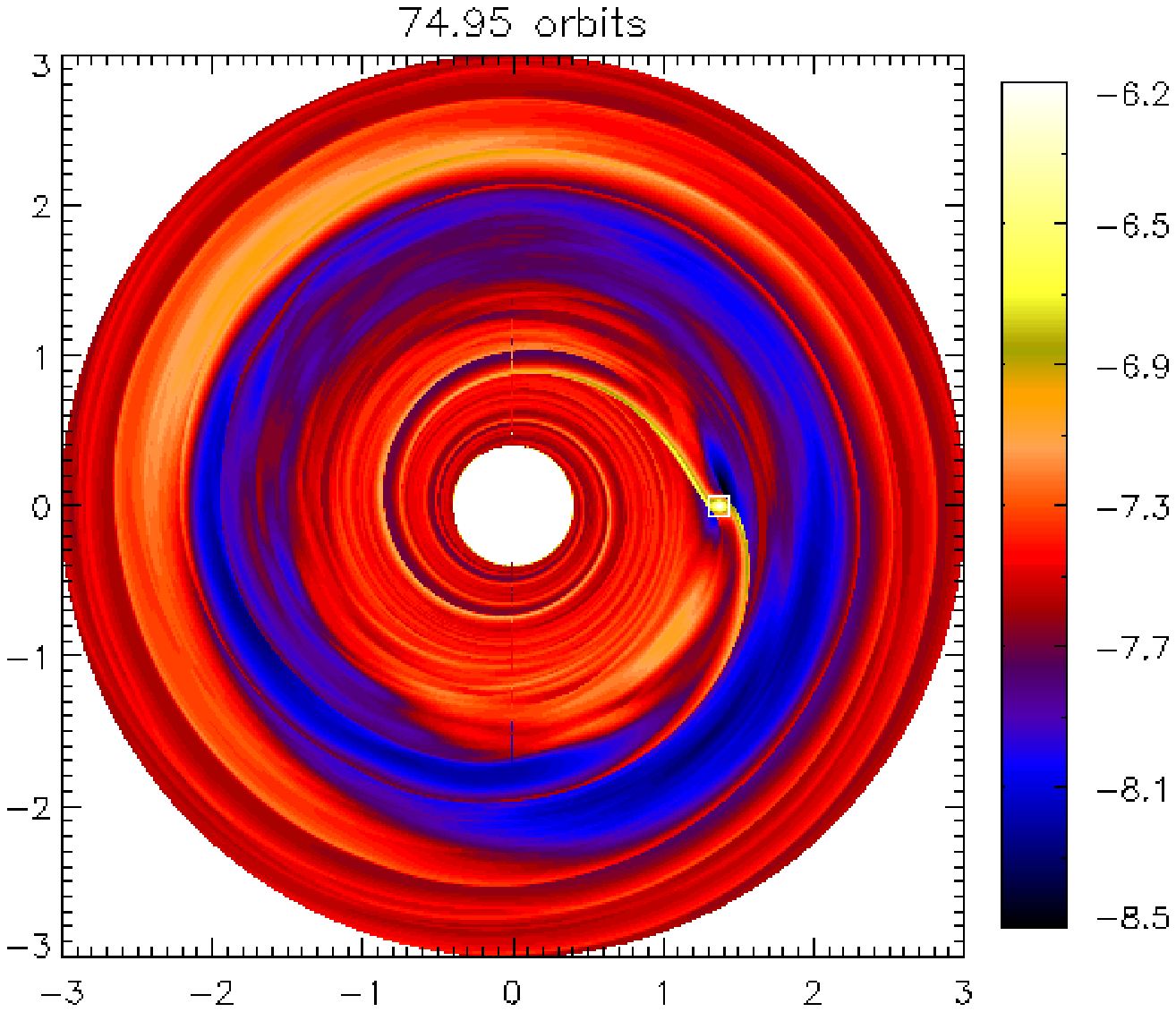}}
%% \subfigure{\includegraphics[scale=.5]{hyades/polar_dens240.jpg}}\subfigure{\includegraphics[scale=.5]{hyades/polar_dens284.jpg}}
\caption{ The evolution of $\ln{\Sigma}$ from early slow
  migration to  stalling  ($\sim 75$ orbits).\
\label{overview_density}}
 \end{figure}

Different  migration phases correspond to different disc
structures. Fig. \ref{overview_density} shows global the global surface
density evolution. 
At $t=24.75,$ during the  slow migration phase a), the planet resides in a
partial gap ($r\simeq r_p\pm2.3r_h$) 
with surface density $\sim 20\%$ lower than $\Sigma_0.$ 
The gap is circular  but
non-axisymmetric with a low surface  density arc along the outer gap edge
trailing  the planet.
 A partial gap is necessary for the  Type III  migration 
mode \citep{papaloizou06b} but not sufficient. 
A surface density asymmetry ahead/behind the 
planet is needed to provide the net co-orbital torque \citep{artymowicz04}.
At this stage this is too weak.
 Further azimuthal density asymmetry  has developed  in the gap by
 $t=55$ and there  is a factor of $\sim 3$   variation in the  gap  surface density, 
but   the asymmetry is 
still limited. 

Strong asymmetry can be provided  by large-scale vortices near the gap edges 
of azimuthal extent $\Delta \phi\sim\pi$ (Fig. \ref{overview_density}). 
The outer (inner) vortex rotates clockwise (counter-clockwise)
relative to the planet so they are not co-orbital.  Their origin was explained
as a natural consequence of the instability of the 
vortensity rings (see section \ref{Dynstab} above), their occurrence near gap
edges has a strong influence on the type III  migration mode.

\begin{figure}
\center
\subfigure{\includegraphics[width=0.5\linewidth]{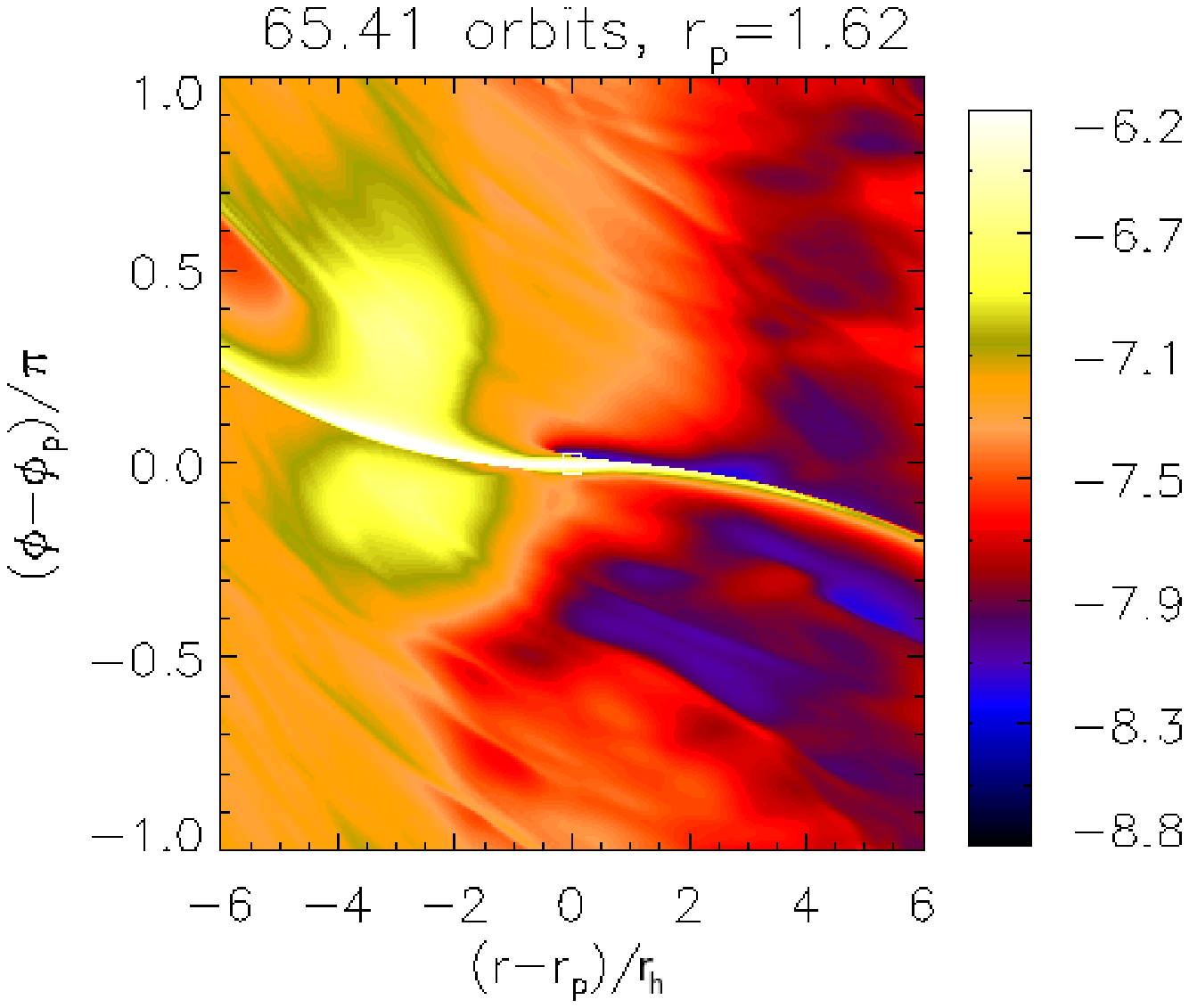}}\subfigure{\includegraphics[width=0.5\linewidth]{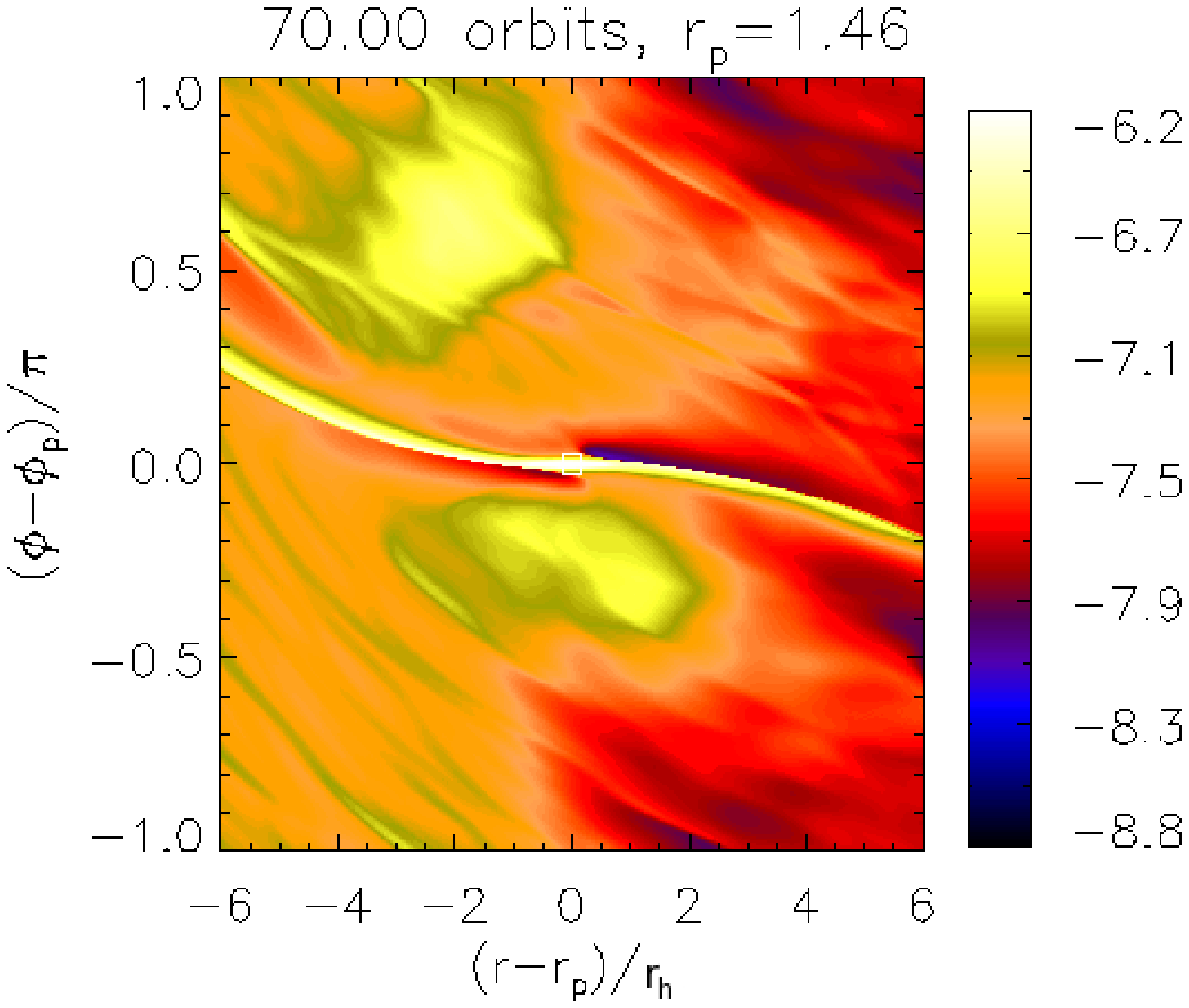}}
\subfigure{\includegraphics[width=0.5\linewidth]{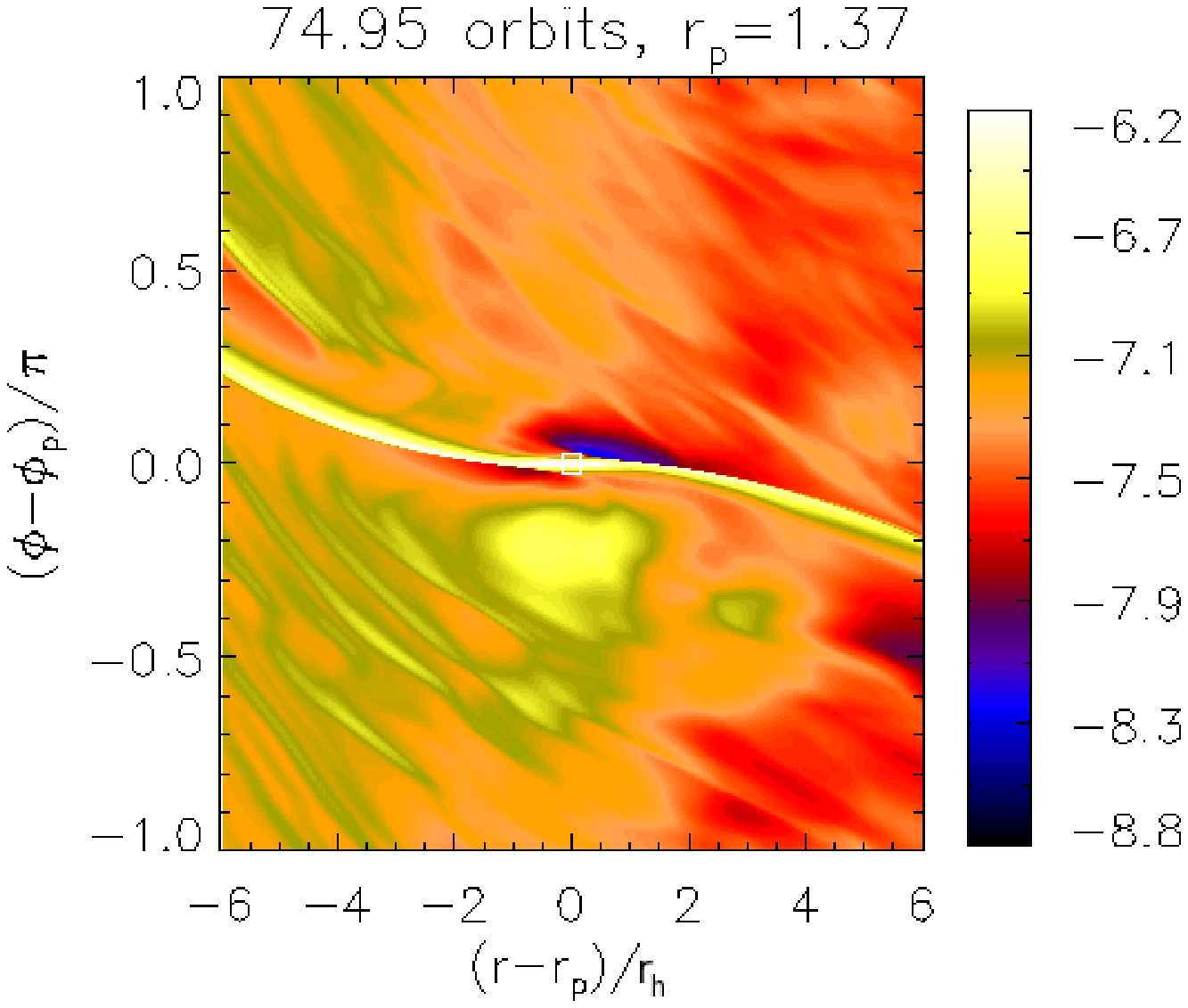}}
 \caption{ Illustration of the surface density evolution
 from the start of  rapid migration to
   stalling at $t=75.$ Maps of $\ln{\Sigma}$ are plotted.
 \label{stop_2d}}
 \end{figure}

 At $t=65.05$  Fig. \ref{overview_density} shows that 
 the planet is just inside the inner gap edge. 
The surface  density contrast in the neighbourhood of  the planet 
is largest as the  inner vortex enters the  co-orbital region from behind
the planet. 
It exerts a net negative torque  as the material crosses  the 
planet orbit and enters (is scattered into)  the exterior disc, and this snapshot corresponds to 
fast migration (phase b). At $t=75$ the migration stalls and the planet 
no longer resides inside a gap. 
The planet has effectively left its gap by scattering vortex material outwards. 
%% The co-orbital mass deficit decreases and the type III
%% mode cannot operate.
This completes a single vortex-planet episode, during which the outer vortex simply circulates around
the original outer gap edge
and does not influence the  co-orbital dynamics, though
it contributes  an oscillatory torque on the planet.\\
\indent The vortex-planet interaction is  magnified 
in Fig. \ref{stop_2d}. At $t=65$ the vortex
circulates at $\sim r_p - 3r_h$ and has radial extent $\sim 3r_h$. 
The gap depth is largest and  the migration is fast. 
As the planet
migrates  inwards, the vortex splits with some 
material entering  the co-orbital
 region while the rest continues  to circulate ($t=70$).
Vortex material becomes trapped just behind the planet at $t=75$, 
which would suggest a negative torque. However, the surface density 
distribution does not support the usual type III migration where
horseshoe material moves with planet. Furthermore,
the planet no longer resides in a gap. Part of the original horseshoe 
material is replaced with vortex material and gap-filling 
takes place. The co-orbital mass deficit is lost and the  migration stalls. 

\begin{figure}
\center
%% \subfigure{\includegraphics[scale=.33]{vort_070b.ps}}
\subfigure{\includegraphics[width=0.5\linewidth]{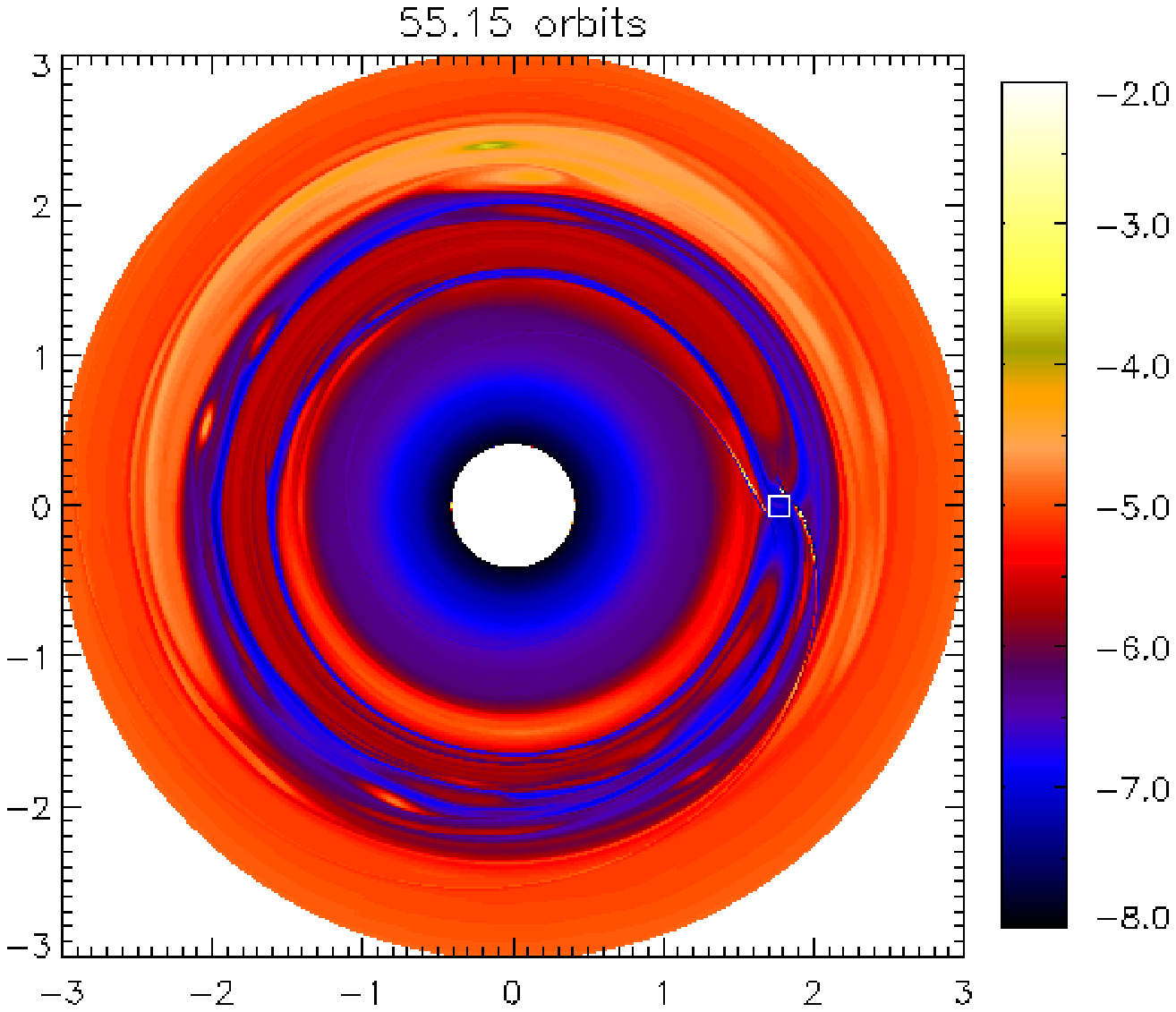}}\subfigure{\includegraphics[width=0.5\linewidth]{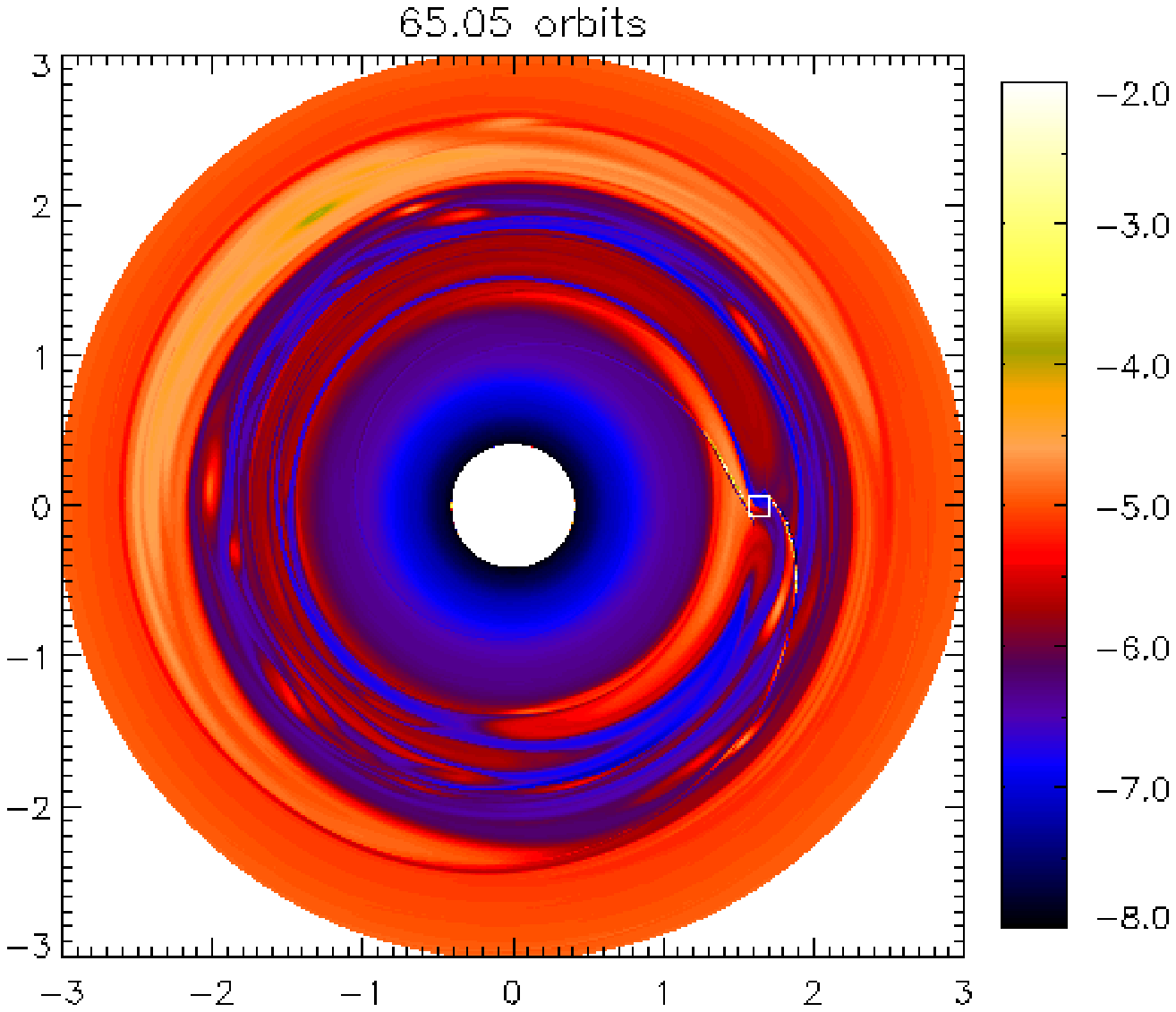}}\\
\subfigure{\includegraphics[width=0.5\linewidth]{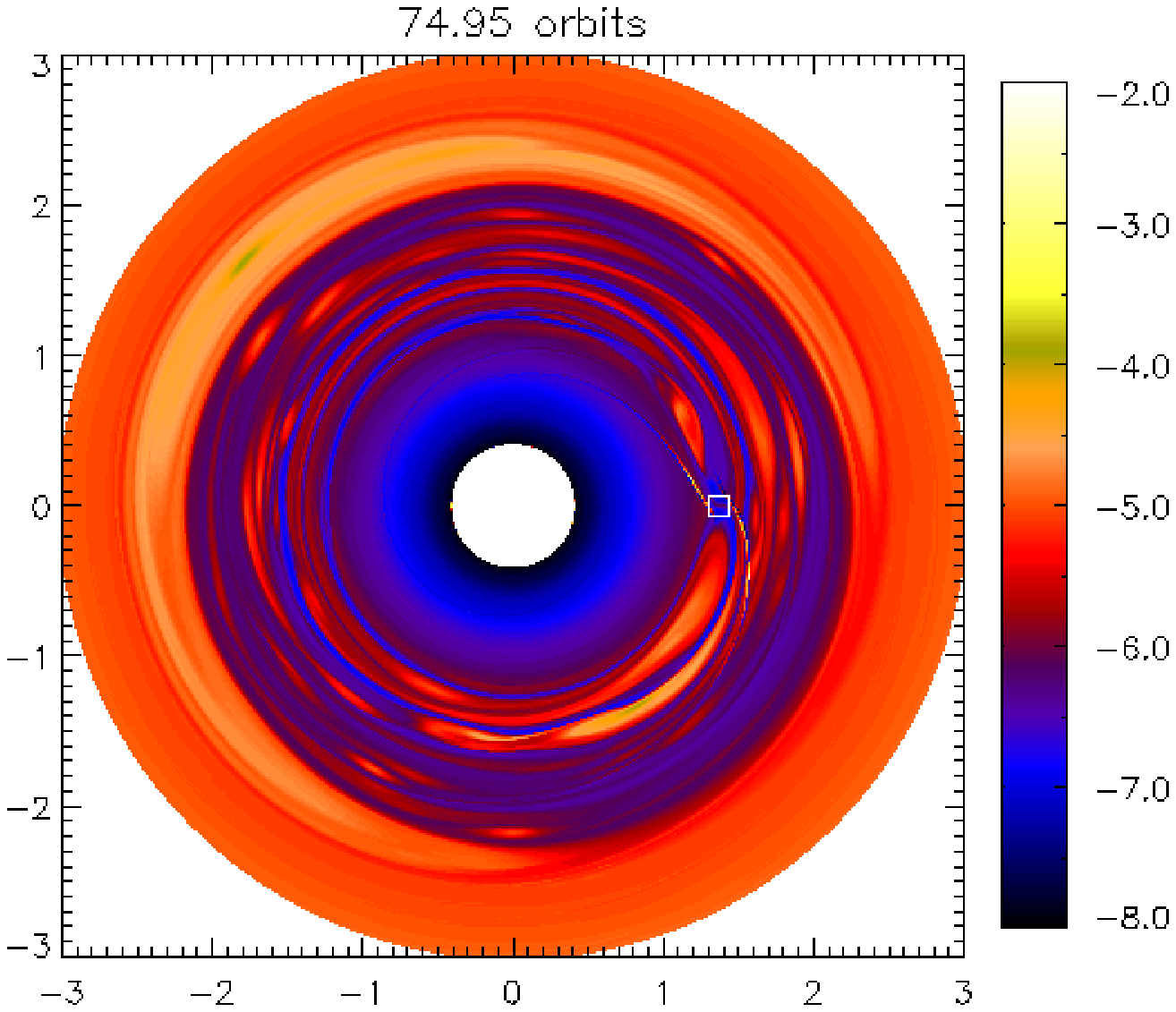}}\subfigure{\includegraphics[width=0.5\linewidth]{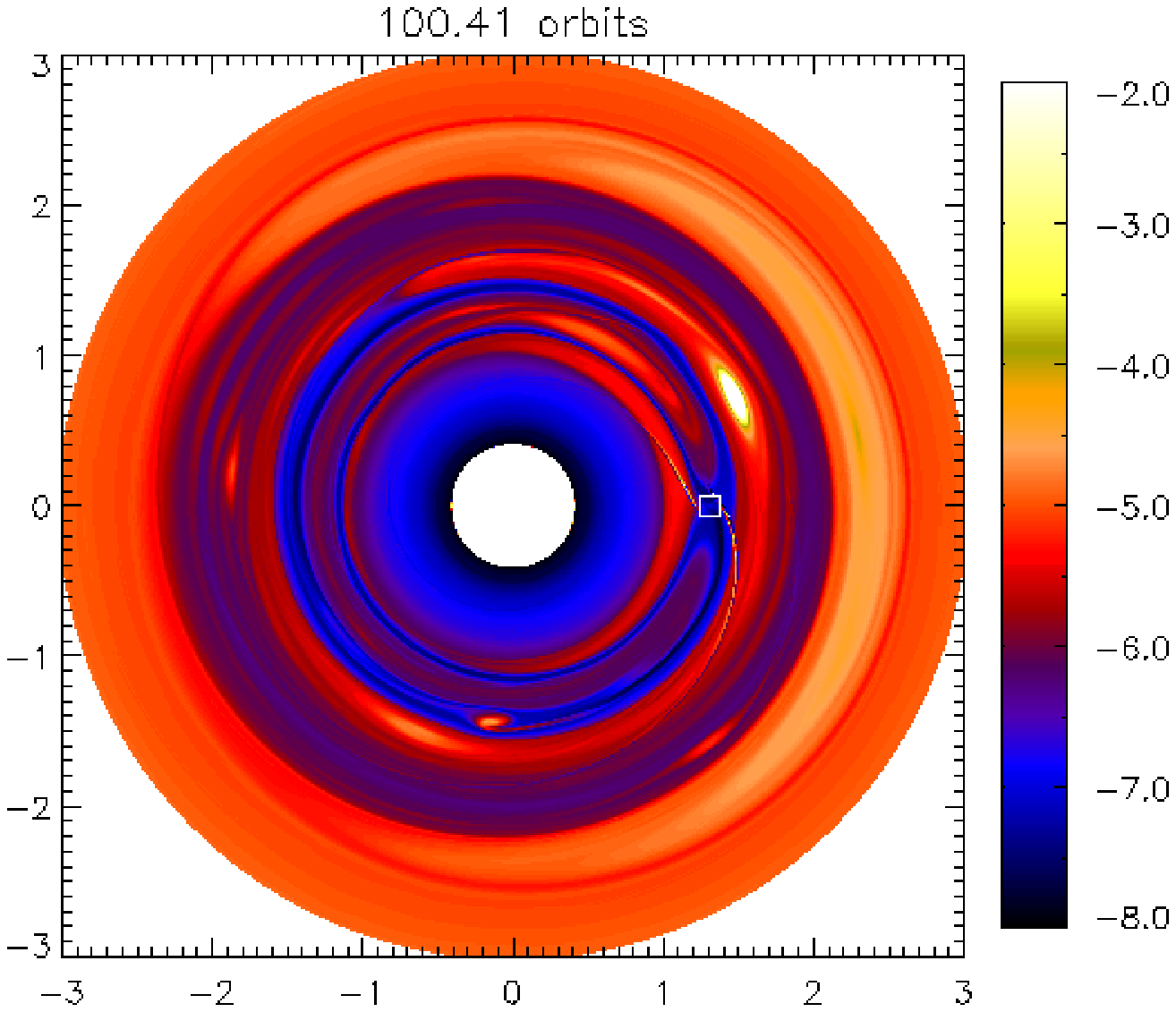}}
%% \subfigure{\includegraphics[scale=.33]{vort_240b.ps}}
\caption{ Vortensity evolution: maps of the
inverse vortensity  $\ln{(\Sigma/\omega)}$ are plotted spanning the
time interval   from early slow
  migration to  stalling ($\sim 75$ orbits) to the  second phase  of slow
  migration.\label{invort_evol}}
 \end{figure}

It is useful to examine the evolution of $\Sigma/\omega$, or inverse vortensity, since it defines 
co-orbital mass deficit $\delta m$ that drives the  type III torque \citep{masset03}.
This is illustrated in Fig. \ref{invort_evol}.
In inviscid discs $\Sigma/\omega$ is 
approximately conserved following a fluid so we can track material.
The vortensity ring basic state and its stability was discussed in section \ref{Dynstab}
(see Fig. \ref{vort_055}). By $t=55$ the inner vortex has formed via
non-linear evolution of the instability and begins to interact with
co-orbital region. Vortensity conservation implies that the `red blobs' near
 the outer ring were part of the inner vortex, consistent
with inward  migration  of the planet via Type III (see
Fig. \ref{invort_evol}). Each time this vortex passes by the planet,
some of its material crosses the planet orbit from behind, thereby
exerting a negative torque. Rapid migration at $t=65$ 
occurs when the main vortex body flows across  the 
co-orbital region. The inner ring is 
disrupted and no longer extends $2\pi$ in azimuth. 

This contrasts with  the usual type III scenario where material simply 
transfers from inner to outer disc leaving the co-orbital region unaffected. 
In the inviscid disc, disruption is \emph{necessary} due
to the existence of vortensity rings of much higher vortensity than the vortex. 
Under type III and vortensity conservation, vortex material must
cross the  planet orbit without changing its vortensity. 
This would not be possible if a  ring structure is maintained. 
In this sense, the vortensity rings oppose the type III mode. 
Hence, migration is slow until significant ring disruption occurs 
that is associated with the vortex flowing  across.

When migration stalls at $t=75$, Fig. \ref{invort_evol} shows
that the vortensity rings are much less
pronounced compared to initial phase. 
The vortex splits into several smaller patches circulating in 
the original gap. 
At the planet's new
radius, material of high $\Sigma/\omega$ fills the new co-orbital region, 
corresponding to lower co-orbital mass deficit. 
 However, by $t=85$ (not shown) new vortensity rings are setup   near the new 
orbital radius and are qualitatively similar to those present in
the ring basic state. The vortex material which passed to the outer disc during 
the first rapid migration phase is now irrelevant, 
much like the outermost vortex. 
During the build up to the  second rapid migration phase
at $t=100$, there is a single vortex associated with inner gap edge and two associated with the outer ring. 
This  simply leads to  a repeat of the first rapid migration phase.
 However, the presence of a vortex inside the co-orbital region, from 
the first rapid migration phase  reduces the co-orbital mass deficit 
(Fig. \ref{nu0}) and hence the migration rate for the second rapid
phase (at $t\sim 140$).

\subsection{The effect of changing the disc mass}\label{varmass}
 {\bf Here} we consider type III migration in inviscid discs 
of different masses obtained by scaling the parameter
 ($\Sigma_0$). Simulations here had a resolution $N_r\times N_\phi =
512\times1536$. %and the computation domain $r=[ The planet potential was switched on over 5 orbital
%periods. 

\begin{figure}
\center
\includegraphics[scale=.33]{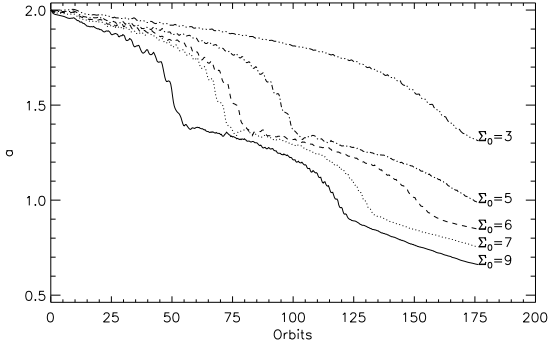}
\caption{Vortex-induced migration in discs with 
 different scalings of the initial surface density profile. The extent of 
rapid migration is independent of the  initial surface density
scaling.\label{orbit2_varmass}}
 \end{figure}

Fig. \ref{orbit2_varmass}
shows migration as a function of $\Sigma_0$
For $\Sigma_0=5$---9
the planet migrates by the same amount during the first rapid migration 
phase and stalls at the same radius. 
This is also observed for a
second rapid migration phase for $\Sigma_0=6$---9. 
Although the highest density case is unphysical due to the 
lack of inclusion  of self-gravity,
results are consistent with the notion that 
rapid migration is initiated by sufficient contrast between co-orbital (gap) and circulating fluid 
(vortex), measured by $\delta m$. 

The jump in orbital radius, should it occur, is
independent of $\Sigma_0$. As the interaction involves the vortex
flowing from the gap edge across the co-orbital region; its change in
specific angular momentum is independent of density, because the
co-orbital region size is fixed by planet mass. The results then
suggest the vortex needs to grow to a mass $M_v$, only dependent
on the planet mass, in order to scatter
the planet. Since the vortex forms at the gap edge, $M_v$ can be
linked to $\delta m$ because the co-orbital mass deficit depends on the
edge surface density.

As the vortex originates from instabilities with growth rate
 independent of $\Sigma_0$, increasing $\Sigma_0$ means less time is needed 
for the vortex to build up to critical mass or density. Hence,
increasing surface density only shortens the `waiting time' before rapid migration. However, if the density is too low, e.g. 
$\Sigma_0=3$ then vortex-induced rapid migration may never occur.

Consider an inner vortex of mass 
$M_v$  formed by instability at $r_v=r_p - \beta r_h$ ($\beta>0$) with
width $\alpha r_h,$
where the Hill radius  $r_h = f_0 r_p,\,f_0\equiv(M_p/(3M_*))^{1/3}.$
$M_v$ is clearly limited by the amount of material 
that can be gather into the vortex, so that 
$M_v < 2\pi \alpha r_h r_v \Sigma$. 
Taking $M_v/M_p=3.5$ as critical for rapid migration \footnote{We estimate
the vortex mass by monitoring the decrease in total mass of an annulus interior to
the inner gap edge.}, this means
\begin{align}
\Sigma_0 > \frac{3.5M_p}{2\pi \alpha f_0(1-\beta f_0)r_p^2}\times10^{4}.
\end{align}
Taking representative values found in simulations 
of $\alpha=3,\,\beta=4$ and $r_p\sim2$ gives $\Sigma_0 > 3.5.$ 
For such cases rapid, vortex-induced migration was indeed
observed (Fig.\ref{orbit2_varmass}) but for $\Sigma_0=3$ it was not.
 This is similar to the usual requirement that in order for
 Type III migration to  
occur for  intermediate planets, the disc should be
 sufficiently massive \citep{masset03}.  In our case the limitation is
specifically due to the maximum possible vortex mass.

\subsubsection{Critical co-orbital mass deficit}
\begin{figure}
\center
%% \subfigure{\includegraphics[scale=.15]{deltam_sigma_9.ps}}\subfigure{\includegraphics[scale=.15]{deltam_sigma_7.ps}}
%% \subfigure{\includegraphics[scale=.15]{deltam_sigma_5.ps}}\subfigure{\includegraphics[scale=.15]{deltam_sigma_3.ps}}\\
\subfigure[$\Sigma_0=9$]{\includegraphics[width=0.5\linewidth]{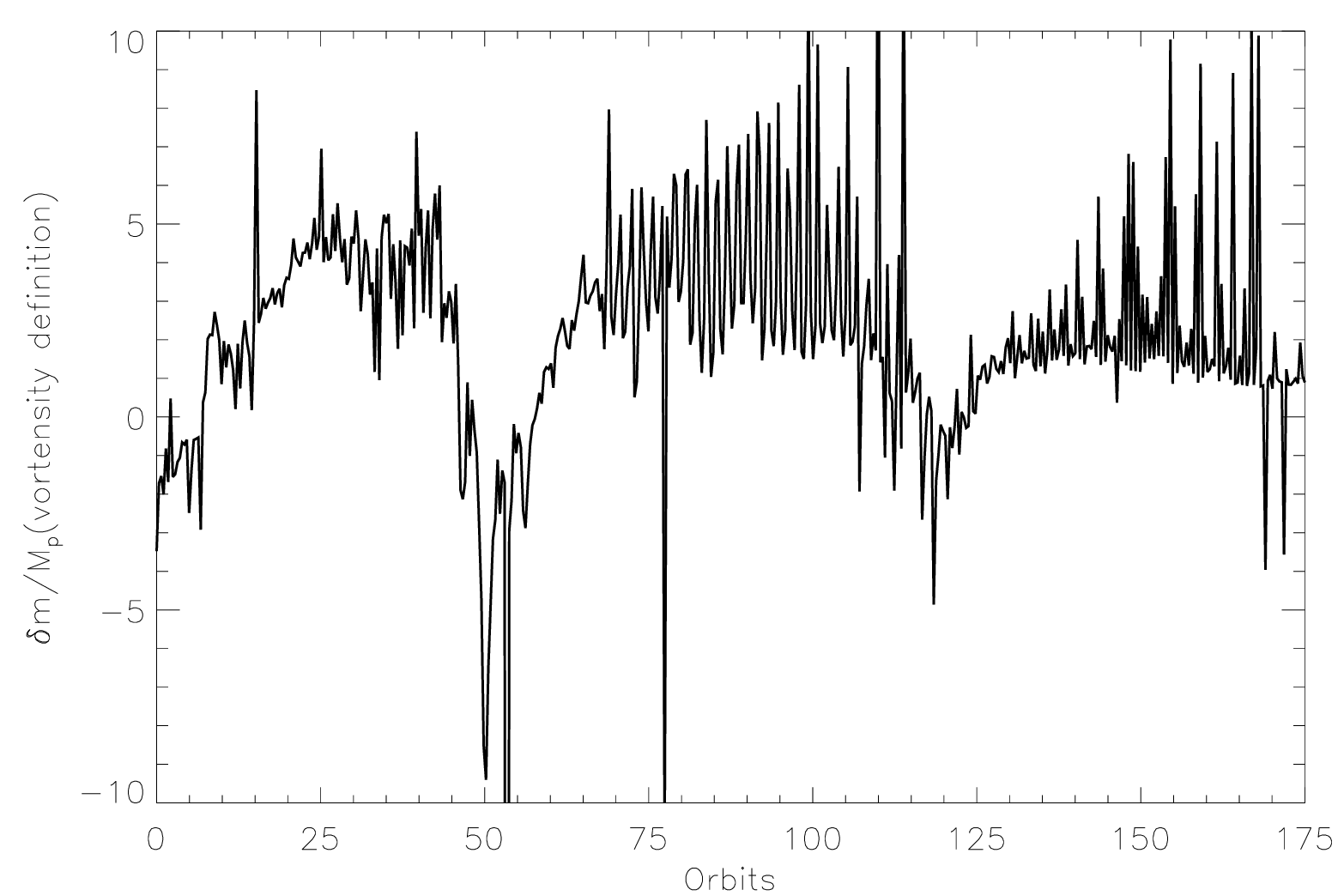}}\subfigure[$\Sigma_0=7$]{\includegraphics[width=0.5\linewidth]{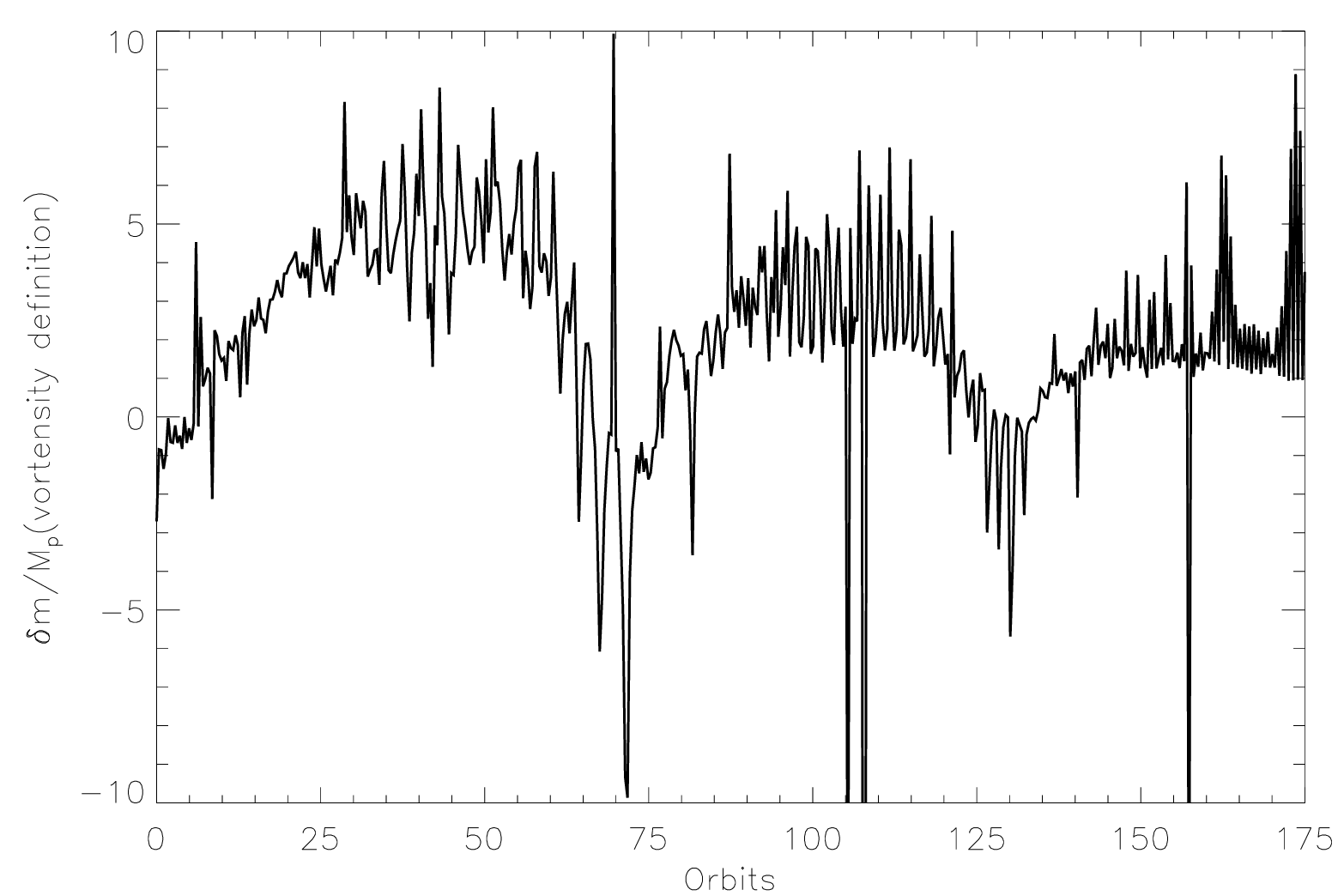}}
\subfigure[$\Sigma_0=5$]{\includegraphics[width=0.5\linewidth]{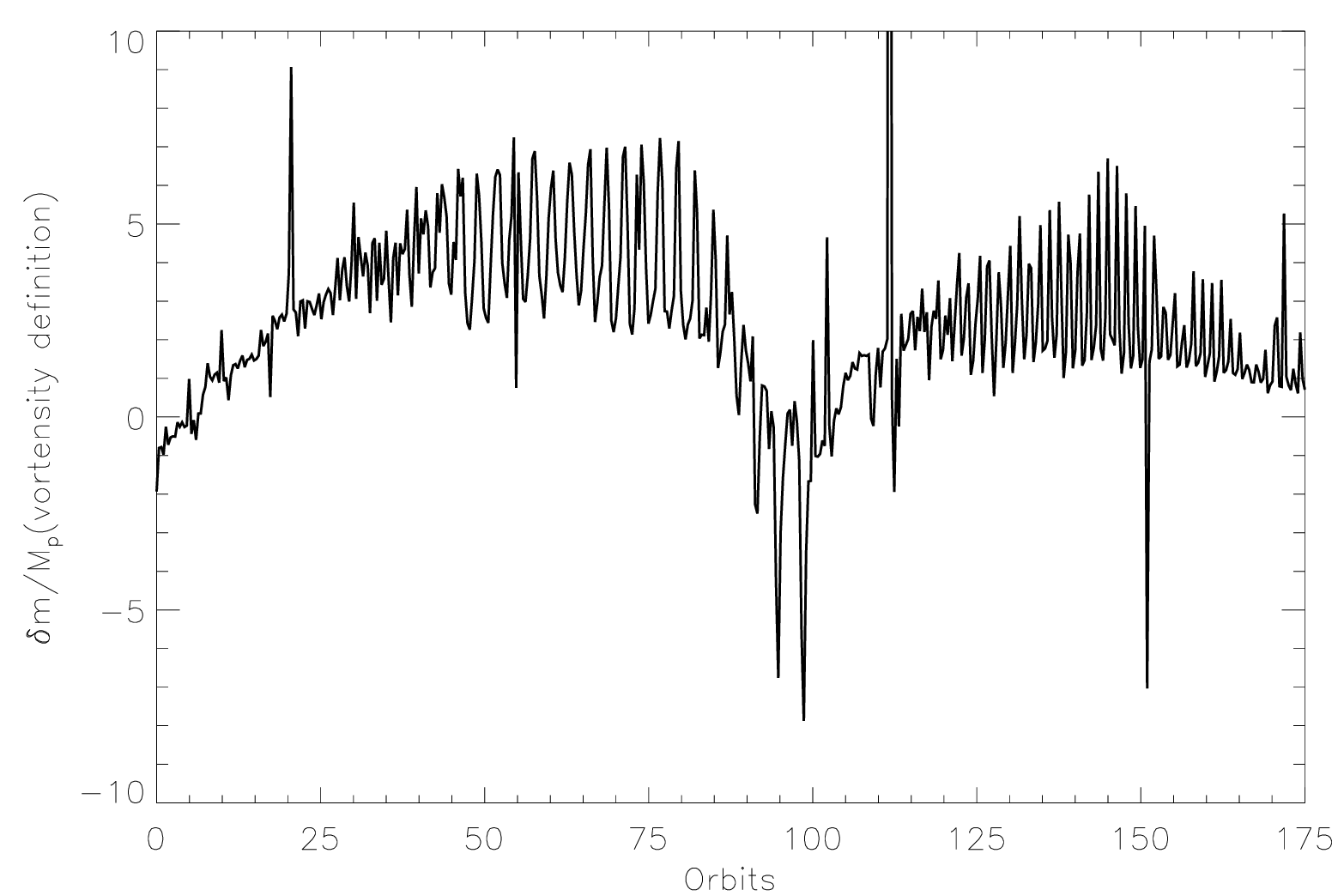}}\subfigure[$\Sigma_0=3$]{\includegraphics[width=0.5\linewidth]{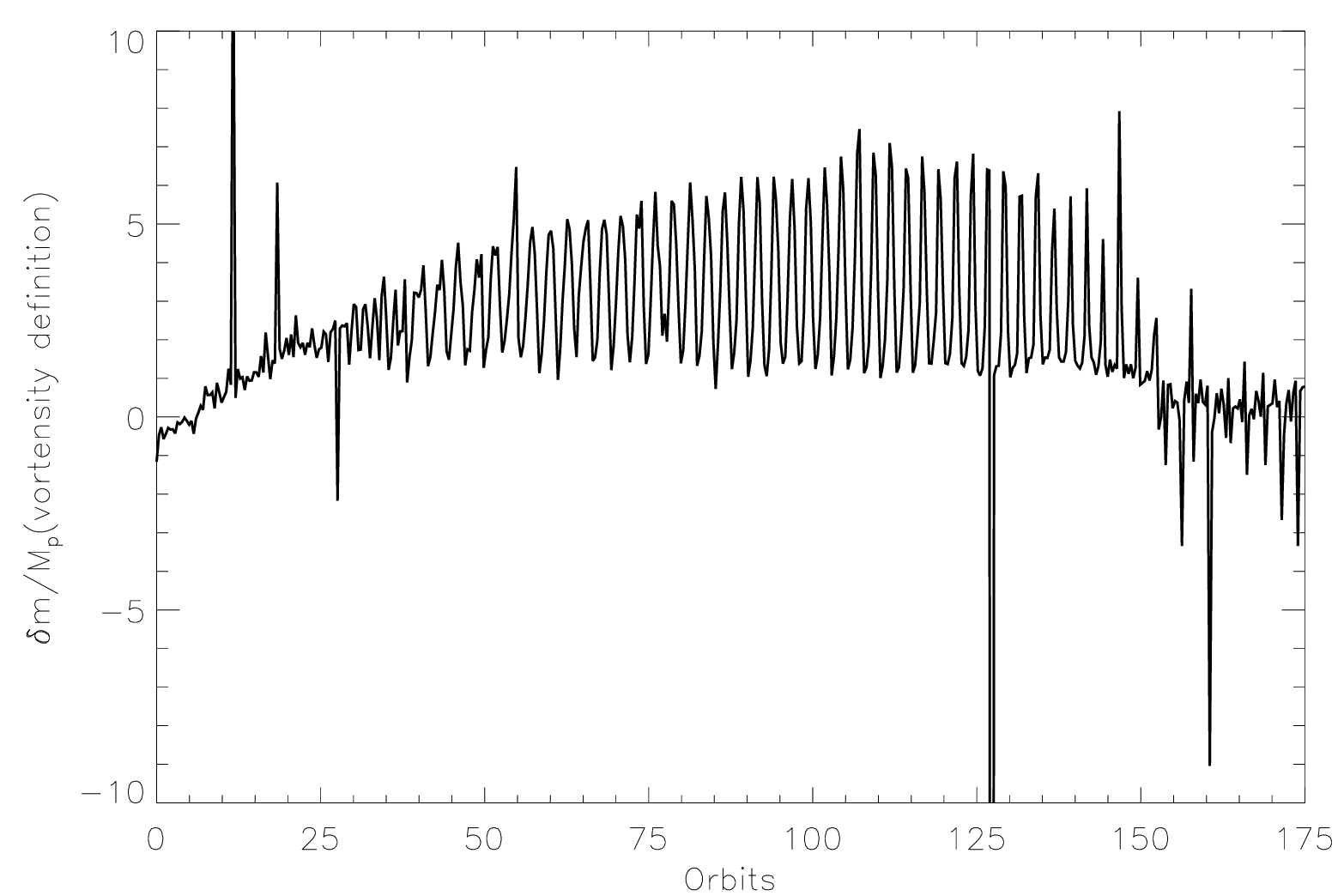}}
\caption{Evolution of the co-orbital mass deficit as defined
by inverse vortensity. The average
$\Sigma/\omega$ for  co-orbital material  is taken over 
$r-r_p=[-2.5r_h,0], \phi = [\phi_p,\phi_p-\pi/4]$; and that of circulating
material is taken over
$r-r_p=[-6,-2.5]r_h, \phi = [\phi_p,\phi_p-\pi/4]$. Very similar
behaviour was obtained when we
compared the co-orbital and circulating surface densities.\label{deltam}}
\end{figure}

In order to link type III migration and vortex-planet scattering, 
we measured the co-orbital mass deficit $\delta m$, which amounts to
comparing average inverse vortensity of co-orbital fluid just behind the planet, to that of circulating fluid just inside the 
inner separatrix but also behind the planet:
\begin{align*}
\delta m  = 2\pi x_s r_p^{-1/2} (\langle\Sigma/\omega\rangle_\mathrm{circ} - \langle\Sigma/\omega\rangle_\mathrm{coorb}).
%&\delta m _\mathrm{density} = 4\pi x_s r_p (\langle\Sigma\rangle_\mathrm{lib} - \langle\Sigma\rangle_\mathrm{coorb}).
\end{align*}
This is a simplified version of the definition in \cite{masset03}. 
Results are shown in Fig. \ref{deltam}. The oscillatory nature of
$\delta m $ reflects a vortex circulating at  
 the gap edge, $\delta m$ maximising when the vortex is
within the patch of fluid where averaging is done. As it grows, the high vortensity
vortex contributes to 
$\langle\Sigma/\omega\rangle_\mathrm{circ}$ hence favouring type III migration.
Cases with rapid migration
share the same evolution of $\delta m$. $\delta m$ increases up to $\sim
4$--5 before the vortex first induces fast migration. For $\Sigma_0=
3$, which does not show such a  rapid migration phase,
typically $\delta m\lesssim 5$ with smaller amplitude variation.   

During fast migration $\delta m$ rapidly decreases
and migration stalls when $\delta m \lesssim 0$. This is because as
the vortex flows across the co-orbital radius, it contributes to
$\langle\Sigma/\omega\rangle_\mathrm{coorb}$, lowering
$\delta m$. However, we comment that details depend on the
unperturbed ($t=0$) surface density profile. Discs discussed here initially have
uniform surface density. If a surface density $\Sigma\propto r^{-p},\,p>0$
were adopted, the planet can be scattered to a region of
higher background $\Sigma/\omega$ compared to the flat case, so
$\langle\Sigma/\omega\rangle_\mathrm{circ}$ may increase due to the
background. We have run simulations with $p=0.3,\,0.5$ and found that the periods of
stalling are of shorter duration, consistent with the discussion above regarding the variation with the
surface density scale.

%% The critical $\delta m/M_p$ is about 5 for rapid vortex induced migration; 
%% close to our above simple estimate  and independent of  initial
%% surface density. 

\subsection{Comparison to Jupiter-mass planet}

\begin{figure}
\center
\includegraphics[scale=.33]{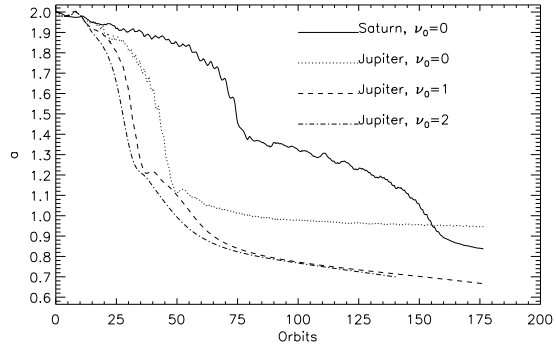}
\caption{Vortex-induced migration for a Jupiter-mass planet in an
  inviscid disc (dotted) and viscous disc ($\nu_0=1$, dashed;
  $\nu_0=2$, dashed-dot) compared to Saturn-mass planet in a inviscid
  disc (solid).\label{orbit2_jup}}
 \end{figure}

For completeness we briefly consider the case of a Jupiter-mass
planet. The setup is the same as the previous sub-section (with $\Sigma_0=7$),
but the planet potential is now switched on over 5 orbits since
Jupiter is 3 times more massive than Saturn. Fig. \ref{orbit2_jup}  compares the migration curves
$a(t)$ for $M_p=10^{-3}$ in inviscid and viscous ($\nu_0=1,\,2$) discs to a Saturn mass planet in a inviscid disc. The same vortex-planet
scattering occurs for the Jupiter-mass inviscid case. This was checked explicitly by examining the vortensity evolution and observing a vortex associated
with the inner gap edge build-up and flowing across the co-orbital region. Jupiter induces stronger shocks and therefore higher-amplitude vortensity rings,
so they are more unstable. The instability growth time-scale is therefore shorter than that of Saturn, so the vortex-induced migration occurs very soon after
the planet in introduced. 

The $a(t)$ for Jupiter in an inviscid disc
show two migration phases --- fast and slow, and that significant
orbital decay occur within $\sim 50P_0$.  These features were also
observed by \cite{peplinski08b}. However, the fast phase of
\citeauthor{peplinski08b} is almost a linear function of time, whereas
in our inviscid case migration is clearly accelerating inwards during
$t\lesssim50P_0$. In our case there is an abrupt transition to the
slow phase whereas that of \citeauthor{peplinski08b} is smoother and
no vortices were identified. Although
\citeauthor{peplinski08b} did not include physical viscosity,
their Cartesian code is more diffusive and vortex-formation is suppressed \citep{valborro07}.
In the late stages of our Jupiter in the inviscid disc,  thin
vortensity rings re-form and inhibit further migration. A new vortex
develops at this stage but it does not grow a sufficient size
to induce another episode within simulation time.

Interestingly, increasing to standard viscosity 
$\nu_0=1$ results in a similar behaviour for the fast phase, though
the transition to slow migration is smoother than zero viscosity.
The vortensity distribution for $\nu_0=1$ is also much smoother than
$\nu_0=0$ and individual vortices were not identified for the inner
gap edge. This is consistent with \cite{valborro07} who showed that
$\nu=O(10^{-5})$ suppress vortex formation. With $\nu_0=1$,
flow-through the co-orbital region is a smoother function of time,
unlike episodic behaviour of inviscid discs, and therefore do not
experience sudden stalling. 

A case with $\nu_0=2$ is also show in
Fig. \ref{orbit2_jup}. The orbital decay time-scale is again $\sim
50P_0$, consistent with type III migration, although there could be
complications from the fact that a Jovian-mass planet in a viscous
disc can undergo type II migration. The transition from fast to slow
is again smoothed (cf. $\nu_0=0,\,1$), the fast phase is now more
linear, qualitatively closer to \cite{peplinski08b}. These results
suggest that our inviscid cases have lower total viscosity than
those considered by \citeauthor{peplinski08b}.

%\indent $M_v$ is limited by the amount of material
%available in the annulus about $r_v$,
%\begin{align*}
%M_v \lesssim 2\pi r_v\times \alpha r_h \Sigma_0\equiv M_{v,\mathrm{max}},
%\end{align*}
%where $\alpha r_h$ is the annulus width and $\Sigma_0$. Then
%\begin{align*}
%\sqrt{\frac{r_p\prime}{r_p}} \gtrsim 1 + \frac{\sqrt{1-\beta f_0} -1}{1+M_p/M_{v,\mathrm{max}}}.
%\end{align*}
%Taking typical numbers $\beta=5,\,\alpha=3$
% ------------------------------------------------------------------------

%%% Local Variables: 
%%% mode: latex
%%% TeX-master: "../thesis"
%%% End: 

%% \\baselinestretch{1}
%% \chapter{Summary and future work}
%% \ifpdf
%%     \graphicspath{{Conclusions/ConclusionsFigs/PNG/}{Conclusions/ConclusionsFigs/PDF/}{Conclusions/ConclusionsFigs/}}
%% \else
    \graphicspath{{Conclusions/ConclusionsFigs/EPS/}{Conclusions/ConclusionsFigs/}}
%% \fi

%% \def\baselinestretch{1.66}
\section{Summary and discussion}\label{conc}
In this paper we have studied the role of large-scale vortices in
type III migration in low viscosity discs. We focused mainly on Saturn-mass planets because 
they open partial gaps, a configuration where type III migration can operate if the disc is massive \citep{masset03}. 
Type III migration would occur when the planet is Saturn mass before
growing to Jovian mass.  We first
demonstrated through numerical simulations and semi-analytic modelling
of inviscid discs, that vortensity rings originate from spiral shocks
induced by the planet. For Saturn or more massive planets, the rings
reside just inside the co-orbital region, while for less massive
planets they are not co-orbital features.  Vortensity rings are setup whether
the planet is introduced suddenly or switched on over several orbits
and instability occur in both approaches (the latter used by \cite{valborro07}).

We showed that the gap edges, associated with local vortensity minima,
are dynamically unstable to non-axisymmetric perturbations. Dominant
unstable modes are localised go on to develop into vortices which merge in
the non-linear regime, as verified by simulations.  The case of a
  Jupiter-mass planet held on a fixed orbit has already been discussed
by \cite{valborro07} which shows vortex-formation at gap edges.
The effect of vortices on migration is most significant in low
viscosity discs ($\nu\lesssim 0.25\times10^{-5}$) because the
instability is suppressed at higher viscosity.  In the inviscid
  limit only a small numerical viscosity is present. Suppose we
  consider a physical viscosity of $\nu=O(10^{-6})$ ,  it is still
  large compared to the numerical viscosity, so we expect the latter
  to be unimportant. However,  $\nu=O(10^{-6})$ is already unable to
  suppress the instability. Our conclusions are unaffected by
  numerical viscosity of this code.
The presence of
high density vortices at the gap edge produces  non-smooth migration,
with episodes of fast migration corresponding to the vortex-planet
interaction. This is analogous to planet-planet scattering, and the 
planet's orbital radius jumps by a few Hill radii in one episode. The vortex is also
responsible for stalling migration in discs with initially flat surface
density. In this case there can be repeated episodes of vortex-induced
migration. Viscosity smooths the flow across the co-orbital
radius, but has limited effect on the extent of orbital decay via the
type III mechanism.   

We also explored the role of vortices in inviscid discs of different
masses. The extent of orbital decay in a single episode of
fast migration is independent of initial surface density scaling. This
suggests a critical vortex mass or surface density is required to interact,
which can be linked to the concept of the co-orbital mass deficit
that drives type III migration. The case of a Jupiter-mass planet
in an inviscid disc also displayed vortex-planet interaction.

\subsection{Outstanding issues}
 One issue not considered in detail in this paper are boundary
  effects, although we used the damping boundary conditions
 \citep{valborro06} which is aimed to remove reflections.  
 The instabilities that lead to vortex development are localised near
 gap edges. As long as the planet, and hence
vortex-formation, is far from the disc edges, boundaries effects on
vortex-planet interaction should be limited. We have run a simulation where the domain is expanded to
$r=[0.2,6.0]$ and compared $a(t)$ to the fiducial disc
which extends $r=[0.4,4.0]$. The extended domain case still exhibited
non-smooth $a(t)$ with similar orbital decays
during the first two episodes of rapid migration. As the domain was extended inwards, there was a third phase
of rapid migration.  There is more discrepancy later in the evolution
as the planet is migrating towards the inner boundary. Boundary
effects need to be examined in more detail in the future, but should not
change the main conclusion of this paper, which is that in a low 
viscosity disc,  type III migration is induced by vortices and episodic.

Although we have demonstrated the effect of vortices on
migration, it is natural to question their existence as long-lived
structures in real discs with finite viscosity. Recently, \cite{li09}
studied migration in nearly-laminar discs in the type I regime. They
found effects at low viscosity, which tend to slow down migration,  begin to
appear for $\nu\leq10^{-7}$. Their simulations last $O(10^3)$
orbits. In our case, non-smooth migration can be
observed at $\nu=O(10^{-6})$. We adopted a more massive planet so vortex
formation occurs quickly, and since type III
migration time-scales are much shorter than type I, a
very small viscosity is not required
 in order to  allow the development of sharp features in the disc (which
the instability relies on). Vortices can thus interfere with migration
when viscosity is one order of magnitude smaller than the standard
value $\nu=10^{-5}$. 

We have used the simplest model and numerical setup to describe the
disc-planet system. One physical issue is the lack of inclusion of self-gravity (SG). 
It may be important when discussing type III migration since
this operates in massive discs \citep{masset03}. In order to have
self-consistent physics, self-gravity is essential. Although
\cite{li09} included SG, its role was not discussed. 
The effect of self-gravitating vortices on the migration of the larger mass planets considered here 
should be explored. We note in the standard viscous disc with $\nu=10^{-5}$,
vortices are transient features. However, it is conceivable that SG
may mitigate the effects of viscous diffusion because vortices have high surface density. Thus,
vortices may exist for higher viscosity values in SG discs. These issues will be the subject of a future study.

%%% ----------------------------------------------------------------------

% ------------------------------------------------------------------------

%%% Local Variables: 
%%% mode: latex
%%% TeX-master: "../thesis"
%%% End: 

%\bibliographystyle{Classes/CUEDbiblio}
%\bibliographystyle{Classes/jmb}
%\bibliographystyle{plainnat} %this works with package natbib
%\bibliographystyle{mn2e} % bibliography style
%\renewcommand{\bibname}{References} % changes default name Bibliography to References
%\bibliography{ref} % References file

\begin{thebibliography}{}

\bibitem[\protect\citeauthoryear{{Artymowicz}}{{Artymowicz}}{2004a}]{artymowic%
z04}
{Artymowicz} P.,  2004a, in {L.~Caroff, L.~J.~Moon, D.~Backman, \& E.~Praton}
  ed., Debris Disks and the Formation of Planets Vol.~324 of Astronomical
  Society of the Pacific Conference Series, {Dynamics of Gaseous Disks with
  Planets}.
pp 39--+

\bibitem[\protect\citeauthoryear{{Artymowicz}}{{Artymowicz}}{2004b}]{artymowic%
z04b}
{Artymowicz} P.,  2004b, in KITP Conference: Planet Formation: Terrestrial and
  Extra Solar {Migration Type III}

\bibitem[\protect\citeauthoryear{{Balmforth}, {Cunha}, {Dolez}, {Gough} \&
  {Vauclair}}{{Balmforth} et~al.}{2001}]{balmforth01}
{Balmforth} N.~J.,  {Cunha} M.~S.,  {Dolez} N.,  {Gough} D.~O.,    {Vauclair}
  S.,  2001, \mnras, 323, 362

\bibitem[\protect\citeauthoryear{{D'Angelo}, {Bate} \& {Lubow}}{{D'Angelo}
  et~al.}{2005}]{dangelo05}
{D'Angelo} G.,  {Bate} M.~R.,    {Lubow} S.~H.,  2005, \mnras, 358, 316

\bibitem[\protect\citeauthoryear{{de Val-Borro}, {Artymowicz}, {D'Angelo} \&
  {Peplinski}}{{de Val-Borro} et~al.}{2007}]{valborro07}
{de Val-Borro} M.,  {Artymowicz} P.,  {D'Angelo} G.,    {Peplinski} A.,  2007,
  \aap, 471, 1043

\bibitem[\protect\citeauthoryear{{de Val-Borro}}{{de
  Val-Borro}}{2006}]{valborro06}
{de Val-Borro} M.~e.,  2006, \mnras, 370, 529

\bibitem[\protect\citeauthoryear{{Goldreich} \& {Tremaine}}{{Goldreich} \&
  {Tremaine}}{1979}]{goldreich79}
{Goldreich} P.,  {Tremaine} S.,  1979, \apj, 233, 857

\bibitem[\protect\citeauthoryear{{Goldreich} \& {Tremaine}}{{Goldreich} \&
  {Tremaine}}{1980}]{goldreich80}
{Goldreich} P.,  {Tremaine} S.,  1980, \apj, 241, 425

\bibitem[\protect\citeauthoryear{Kevlahan}{Kevlahan}{1997}]{kevlahan97}
Kevlahan N.-R.,  1997, J. Fluid Mech., 341, 371

\bibitem[\protect\citeauthoryear{{Koller}, {Li} \& {Lin}}{{Koller}
  et~al.}{2003}]{koller03}
{Koller} J.,  {Li} H.,    {Lin} D.~N.~C.,  2003, \apjl, 596, L91

\bibitem[\protect\citeauthoryear{{Korycansky} \& {Papaloizou}}{{Korycansky} \&
  {Papaloizou}}{1995}]{korycansky95}
{Korycansky} D.~G.,  {Papaloizou} J.~C.~B.,  1995, \mnras, 274, 85

\bibitem[\protect\citeauthoryear{{Li}, {Colgate}, {Wendroff} \& {Liska}}{{Li}
  et~al.}{2001}]{li01}
{Li} H.,  {Colgate} S.~A.,  {Wendroff} B.,    {Liska} R.,  2001, \apj, 551, 874

\bibitem[\protect\citeauthoryear{{Li}, {Finn}, {Lovelace} \& {Colgate}}{{Li}
  et~al.}{2000}]{li00}
{Li} H.,  {Finn} J.~M.,  {Lovelace} R.~V.~E.,    {Colgate} S.~A.,  2000, \apj,
  533, 1023

\bibitem[\protect\citeauthoryear{{Li}, {Li}, {Koller}, {Wendroff}, {Liska},
  {Orban}, {Liang} \& {Lin}}{{Li} et~al.}{2005}]{li05}
{Li} H.,  {Li} S.,  {Koller} J.,  {Wendroff} B.~B.,  {Liska} R.,  {Orban}
  C.~M.,  {Liang} E.~P.~T.,    {Lin} D.~N.~C.,  2005, \apj, 624, 1003

\bibitem[\protect\citeauthoryear{{Li}, {Lubow}, {Li} \& {Lin}}{{Li}
  et~al.}{2009}]{li09}
{Li} H.,  {Lubow} S.~H.,  {Li} S.,    {Lin} D.~N.~C.,  2009, \apjl, 690, L52

\bibitem[\protect\citeauthoryear{{Lin} \& {Papaloizou}}{{Lin} \&
  {Papaloizou}}{1986}]{lin86}
{Lin} D.~N.~C.,  {Papaloizou} J.,  1986, \apj, 309, 846

\bibitem[\protect\citeauthoryear{{Lovelace}, {Li}, {Colgate} \&
  {Nelson}}{{Lovelace} et~al.}{1999}]{lovelace99}
{Lovelace} R.~V.~E.,  {Li} H.,  {Colgate} S.~A.,    {Nelson} A.~F.,  1999,
  \apj, 513, 805

\bibitem[\protect\citeauthoryear{{Lynden-Bell} \& {Pringle}}{{Lynden-Bell} \&
  {Pringle}}{1974}]{lyndenbell74}
{Lynden-Bell} D.,  {Pringle} J.~E.,  1974, \mnras, 168, 603

\bibitem[\protect\citeauthoryear{{Masset}}{{Masset}}{2000a}]{masset00a}
{Masset} F.,  2000a, \anas, 141, 165

\bibitem[\protect\citeauthoryear{{Masset}}{{Masset}}{2000b}]{masset00b}
{Masset} F.~S.,  2000b, in {Garz{\'o}n} G.,  {Eiroa} C.,  {de Winter} D.,
  {Mahoney} T.~J.,  eds, Disks, Planetesimals, and Planets Vol.~219 of
  Astronomical Society of the Pacific Conference Series, {FARGO: A Fast
  Eulerian Transport Algorithm for Differentially Rotating Disks}.
pp 75--+

\bibitem[\protect\citeauthoryear{{Masset}}{{Masset}}{2002}]{masset02}
{Masset} F.~S.,  2002, \aap, 387, 605

\bibitem[\protect\citeauthoryear{{Masset} \& {Papaloizou}}{{Masset} \&
  {Papaloizou}}{2003}]{masset03}
{Masset} F.~S.,  {Papaloizou} J.~C.~B.,  2003, \apj, 588, 494

\bibitem[\protect\citeauthoryear{{Mayor} \& {Queloz}}{{Mayor} \&
  {Queloz}}{1995}]{mayor95}
{Mayor} M.,  {Queloz} D.,  1995, \nat, 378, 355

\bibitem[\protect\citeauthoryear{{Ou}, {Ji}, {Liu} \& {Peng}}{{Ou}
  et~al.}{2007}]{ou07}
{Ou} S.,  {Ji} J.,  {Liu} L.,    {Peng} X.,  2007, \apj, 667, 1220

\bibitem[\protect\citeauthoryear{{Paardekooper}, {Baruteau}, {Crida} \&
  {Kley}}{{Paardekooper} et~al.}{2009}]{paardekooper09b}
{Paardekooper} S.,  {Baruteau} C.,  {Crida} A.,    {Kley} W.,  2009, ArXiv
  e-prints

\bibitem[\protect\citeauthoryear{{Paardekooper} \& {Papaloizou}}{{Paardekooper}
  \& {Papaloizou}}{2009}]{paardekooper09}
{Paardekooper} S.-J.,  {Papaloizou} J.~C.~B.,  2009, \mnras, 394, 2297

\bibitem[\protect\citeauthoryear{{Papaloizou}}{{Papaloizou}}{2005}]{papaloizou%
05}
{Papaloizou} J.~C.~B.,  2005, Celestial Mechanics and Dynamical Astronomy, 91,
  33

\bibitem[\protect\citeauthoryear{{Papaloizou} \& {Lin}}{{Papaloizou} \&
  {Lin}}{1989}]{papaloizou89}
{Papaloizou} J.~C.~B.,  {Lin} D.~N.~C.,  1989, \apj, 344, 645

\bibitem[\protect\citeauthoryear{{Papaloizou}, {Nelson}, {Kley}, {Masset} \&
  {Artymowicz}}{{Papaloizou} et~al.}{2007}]{papaloizou06}
{Papaloizou} J.~C.~B.,  {Nelson} R.~P.,  {Kley} W.,  {Masset} F.~S.,
  {Artymowicz} P.,  2007, Protostars and Planets V, pp 655--668

\bibitem[\protect\citeauthoryear{{Papaloizou}, {Nelson} \&
  {Snellgrove}}{{Papaloizou} et~al.}{2004}]{papaloizou04}
{Papaloizou} J.~C.~B.,  {Nelson} R.~P.,    {Snellgrove} M.~D.,  2004, \mnras,
  350, 829

\bibitem[\protect\citeauthoryear{{Papaloizou} \& {Pringle}}{{Papaloizou} \&
  {Pringle}}{1984}]{papaloizou84}
{Papaloizou} J.~C.~B.,  {Pringle} J.~E.,  1984, \mnras, 208, 721

\bibitem[\protect\citeauthoryear{{Papaloizou} \& {Pringle}}{{Papaloizou} \&
  {Pringle}}{1985}]{papaloizou85}
{Papaloizou} J.~C.~B.,  {Pringle} J.~E.,  1985, \mnras, 213, 799

\bibitem[\protect\citeauthoryear{{Papaloizou} \& {Pringle}}{{Papaloizou} \&
  {Pringle}}{1987}]{papaloizou87}
{Papaloizou} J.~C.~B.,  {Pringle} J.~E.,  1987, \mnras, 225, 267

\bibitem[\protect\citeauthoryear{{Papaloizou} \& {Terquem}}{{Papaloizou} \&
  {Terquem}}{2006}]{papaloizou06b}
{Papaloizou} J.~C.~B.,  {Terquem} C.,  2006, Reports on Progress in Physics,
  69, 119

\bibitem[\protect\citeauthoryear{{Pepli{\'n}ski}, {Artymowicz} \&
  {Mellema}}{{Pepli{\'n}ski} et~al.}{2008a}]{peplinski08a}
{Pepli{\'n}ski} A.,  {Artymowicz} P.,    {Mellema} G.,  2008a, \mnras, 386, 164

\bibitem[\protect\citeauthoryear{{Pepli{\'n}ski}, {Artymowicz} \&
  {Mellema}}{{Pepli{\'n}ski} et~al.}{2008b}]{peplinski08b}
{Pepli{\'n}ski} A.,  {Artymowicz} P.,    {Mellema} G.,  2008b, \mnras, 386, 179

\bibitem[\protect\citeauthoryear{{Pepli{\'n}ski}, {Artymowicz} \&
  {Mellema}}{{Pepli{\'n}ski} et~al.}{2008c}]{peplinski08c}
{Pepli{\'n}ski} A.,  {Artymowicz} P.,    {Mellema} G.,  2008c, \mnras, 387,
  1063

\bibitem[\protect\citeauthoryear{{Press}, {Teukolsky}, {Vetterling} \&
  {Flannery}}{{Press} et~al.}{1992}]{press92}
{Press} W.~H.,  {Teukolsky} S.~A.,  {Vetterling} W.~T.,    {Flannery} B.~P.,
  1992, {Numerical recipes in FORTRAN. The art of scientific computing}

\bibitem[\protect\citeauthoryear{{Stone} \& {Norman}}{{Stone} \&
  {Norman}}{1992}]{stone92}
{Stone} J.~M.,  {Norman} M.~L.,  1992, \apjs, 80, 753

\bibitem[\protect\citeauthoryear{{Tanaka}, {Takeuchi} \& {Ward}}{{Tanaka}
  et~al.}{2002}]{tanaka02}
{Tanaka} H.,  {Takeuchi} T.,    {Ward} W.~R.,  2002, \apj, 565, 1257

\bibitem[\protect\citeauthoryear{{Ward}}{{Ward}}{1997}]{ward97}
{Ward} W.~R.,  1997, Icarus, 126, 261

\end{thebibliography}
%\addcontentsline{toc}{chapter}{References} %adds References to contents page

\appendix
\section{Upper limit on the horseshoe width}\label{xslimit}

\indent We deduce an upper limit on the horseshoe width $x_s$  that is valid  in the limit
of  either zero  or constant pressure. Consider a local Cartesian
frame that  has origin at the planet and hence co-rotates with it with the Keplerian
angular velocity
$\Omega_p =\sqrt{GM_*/r_p^3}.$ Here $M_*$ is the mass of the central object and
$r_p$ is the orbital radius of the planet.
In this frame  fluid elements approach the
planet from $(x=x_0>0,\,y=\infty)$ with velocity $(0,-3\Omega_px_0/2).$
 Suppose such a fluid element  executes a horseshoe
turn crossing the co-orbital radius  at $q=(0,y)$ with velocity $(v_x,0)$.

 We wish to
determine an upper bound on the value of $x_0$  for which such motion occurs.
 The  equation of motion governing a fluid element or particle is
\begin{align}\label{horseshoe}
\frac{D\mathbf{v}}{Dt}+2\Omega_p\hat{\mathbf{z}}\wedge\mathbf{v} = 
 -\nabla\Phi_\mathrm{eff}, 
\end{align}
where
\begin{align}
 \Phi_\mathrm{eff}= -\frac{GM_p}{\sqrt{x^2+y^2}}-\frac{3}{2}\Omega_p^2x^2.
\end{align}
Here the {\it effective potential} $ \Phi_\mathrm{eff}$  contains contributions from the gravitational
potential of the planet of mass $M_p$ and  the tidal potential associated with
the central object.

From  equation  (\ref{horseshoe}) it follows that  the Jacobi invariant  
\begin{align}\label{jacobi}
J = \frac{1}{2}\mathbf{v}^2 + \Phi_\mathrm{eff}
\end{align}
is constant along a particle path. Equating $J$  evaluated at $(x,y)=(x_0,\infty)$
to $J$ evaluated at  $q$
gives
\begin{align}\label{eq1}
-\frac{3}{8}\Omega_p^2x_0^2 = \frac{1}{2}v_x^2 -\frac{GM_p}{y}.
\end{align}
The steady-state Euler equation of motion for $v_y$ evaluated at $q$ is
\begin{align}\label{eq2}
v_{x}\p_x v_y +2\Omega_p v_x = -GM_p/y^2.
\end{align}
Since in the neighbourhood of $q$ for the type of  streamline
we consider,  $v_x<0$ and    $ \p_x v_y<0$ we deduce that
\begin{align}\label{vxlimit}
|v_x|>\frac{GM_p}{2\Omega_p y^2}.
\end{align}
Combining equation  (\ref{eq1}) and equation  (\ref{vxlimit}) we find that
\begin{align}\label{limit}
-\frac{3}{8}\Omega_p^2x_0^2 > \frac{1}{2}\left(\frac{GM_p}{2\Omega_p y^2}\right)^2
 -\frac{GM_p}{y}.
\end{align}
Writing $x_0 = \hat{x}_0 r_h,$ where the Hill radius $r_h = (M_p/(3M_*))^{1/3}r_p$ and
similarly  setting $y = \hat{y} r_h $, we obtain
\begin{align}
  \hat{x}_0 < \sqrt{\frac{8}{\hat{y}} - \frac{3}{\hat{y}^4}} \lesssim 2.3.
\end{align}
Thus because  the maximum possible  value of the right hand side of the above is $2.3,$
we deduce that  particles executing a U-turn could not have originated
further than $2.3r_h.$ This is comparable to the value of $2.5r_h$ 
that has been estimated from hydrodynamic simulations \citep{artymowicz04b,paardekooper09}.
% ------------------------------------------------------------------------

%%% Local Variables: 
%%% mode: latex
%%% TeX-master: "../thesis"
%%% End: 

\section{Vorticity jump across a  steady isothermal shock}
\indent Consider an isothermal  shock
that is stationary in a frame  rotating with 
 angular velocity  $\Omega_p\hat{\mathbf{z}}$. 
In order to evaluate the vorticity jump across the shock it is convenient
to use a two dimensional orthogonal coordinate system  $(x_1,x_2)$ 
defined in the disc mid-plane such that  one 
of the curves $x_2 = {\rm constant} = x_s$  coincides with the shock.
The curves  $x_1 = {\rm constant}$  will then be normal  to the shock where they
intersect it.  In addition the coordinates  are set up so  that
$(x_1, x_2, z)$ is a right handed system.
 %% (see Fig.\ref{vortgen_shock} )
The $\hat{\mathbf{z}}$-component of relative vorticity $\omega_r$  can then be written as
\begin{align}\label{vorticity_def}
\omega_r=\frac{1}{h_1h_2}\left(\frac{\p (u_2 h_2)}{\p x_1}
-\frac{\p (u_1h_1)}{\p x_2}\right),
\end{align}
where $(h_1,h_2)$  are the components of the  coordinate  scale factor.

We note that on $x_2= x_s,$ $u_1$ is the velocity tangential  to the shock.
 The normal component $u_2$ and other state variables  undergo a jump
from pre-shock values to post-shock values on normally traversing  the curve $x_2= x_s.$
For an isothermal shock
\begin{align}\label{shockc}
\frac{u_{2,\mathrm{post}}}{u_2}=M^{-2}=\frac{\rho}{\rho _{\mathrm{post}}},
\end{align}
where $M=u_2/c_s$ is the pre-shock perpendicular Mach number,  and here and in similar
expressions below connecting pre-shock and post-shock quantities,  we have denoted 
post-shock values with a subscript post while leaving pre-shock 
quantities without a corresponding subscript.
\noindent Thus the jump in relative vorticity is
%\begin{align}\label{jump1}
%[\omega_r]=\frac{1}{h_1}\left(\frac{\p u_2}{\p x_1}\right)_\mathrm{post}
%+\frac{u_{2,\mathrm{post}}}{h_1 h_2}\frac{\p h_2}{\p x_1}
%-\frac{u_{1,\mathrm{post}}}{h_1 h_2}\frac{\p h_1}{\p x_2}
%-\frac{1}{h_2}\left(\frac{\p u_1}{\p x_2}\right)_\mathrm{post}-\omega_r.
%\end{align} 
\begin{align}\label{jump1}
[\omega_r]=\omega_{r,\mathrm{post}} -\omega_r
\end{align}

Quite generally the  $x_1$ component of the  equation of motion for a steady state flow  is
\begin{align}
&\frac{u_1}{h_1}\frac{\p u_1}{\p x_1}+\frac{u_2}{h_2}\frac{\p u_1}{\p x_2} -u_2\left(
\frac{u_2}{h_1 h_2}\frac{\p h_2}{\p x_1} -\frac{u_1 }{h_1 h_2}\frac{\p
  h_1}{\p x_2}\right)\notag\\
&=
-\frac{1}{\rho h_1}\frac{\p P}{\p x_1}-\frac{1}{h_1}\frac{\p\Phi}{\p x_1}+2\Omega_p u_2,\label{eqm1}
\end{align}
or equivalently
\begin{align}
\frac{u_1}{h_1}\frac{\p u_1}{\p x_1}+\frac{u_2}{h_1}\frac{\p u_2}{\p x_1} -(2\Omega_p+\omega_r)u_2
=
-\frac{1}{\rho h_1}\frac{\p P}{\p x_1}-\frac{1}{h_1}\frac{\p\Phi}{\p x_1},\label{eqm2}
\end{align}
where $P$ is the pressure and $\Phi$ the total potential (the latter quantity being continuous
across the shock).
From (\ref{eqm2}) we obtain an expression for the relative vorticity
in the form  
\begin{align}\label{vorticity_alt}
\omega_r=\frac{1}{h_1}\frac{\p u_2}{\p x_1}+\frac{u_1}{u_2 h_1}\frac{\p u_1}{\p x_1}
+\frac{1}{\rho u_2  h_1}\frac{\p P}{\p x_1}+\frac{1}{u_2 h_1}\frac{\p\Phi}{\p x_1}-2\Omega_p.
\end{align}
 Applying equation  (\ref{vorticity_alt}) to give an expression for the post shock
relative vorticity, and using    equation  (\ref{shockc})  
to express  the post-shock normal velocity and density
in terms of the corresponding pre-shock quantities, the vorticity jump  can be written in the form
\begin{align}\label{vorticity_alt1}
[\omega_r]=&\frac{1}{h_1}\frac{\p (M^{-2}u_2)}{\p x_1}+\frac{M^2 u_1}{u_2 h_1}
\frac{\p u_1}{\p x_1}
+\frac{M^2}{u_2 h_1}\frac{\p\Phi}{\p x_1}\notag\\+
&\frac{1}{\rho u_2  h_1}\frac{\p P_\mathrm{post}}{\p x_1}-(\omega_r+2\Omega_p).
\end{align}
Adopting  a locally isothermal equation of state, we have  $P_\mathrm{post}=M^2 P.$
Substituting this into equation  (\ref{vorticity_alt1}), while  making use of equation  (\ref{vorticity_alt}) together with 
the relation $u_2=c_s M,$ we obtain
\begin{align}\label{vorticity_alt2}
[\omega_r]\equiv [\omega] =&-\frac{c_s(M^2-1)^2}{M^2h_1}\frac{\p M}{\p
  x_1}+(M^2-1)\omega \notag\\
&-\frac{(M^4-1)}{Mh_1}\frac{\p c_s}{\p x_1},
\end{align}
where $\omega= 2\Omega_p+\omega_r$ is the absolute vorticity.

Setting $h_1dx_1 =dS$ with $dS$ being the corresponding element of  distance measured parallel
to the shock, this takes the form
\begin{align}\label{vorticity_alt3}
[\omega_r]\equiv [\omega] =&-\frac{c_s(M^2-1)^2}{M^2}\frac{\p M}{\p
  S}+(M^2-1)\omega\notag\\
&-\frac{(M^4-1)}{M}\frac{\p c_s}{\p S},
\end{align}
This gives the vorticity jump across a shock in terms of  pre-shock quantities 
measured in the rotating frame 
in which it appears steady. It is important to note that
the expression (\ref{vorticity_alt3}) applies specifically  in the right handed coordinate system
we have adopted with shock location given by $x_2=x_s.$ 
If we  instead adopt $x_1=x_s$ for this location, the signs of the derivative terms
would be reversed as in the expression (2.23) given in \cite{kevlahan97}.
 Note too that the last term on the right hand side
that is proportional to the gradient of $c_s$ along the shock arises from
the assumption of a {\it locally}   isothermal equation of state and is not present
in the treatment given by \cite{kevlahan97}.

% ------------------------------------------------------------------------

%%% Local Variables: 
%%% mode: latex
%%% TeX-master: "../thesis"
%%% End: 

\end{document}